\documentclass[%
 reprint,
 amsmath,amssymb,
 aps,
prx,
superscriptaddress,
]{revtex4-2}

\usepackage{graphicx}
\usepackage{dcolumn}
\usepackage{bm}
\usepackage{hyperref}

\usepackage{multirow}
\usepackage{textgreek}
\usepackage{upgreek}
\usepackage{bigints}
\usepackage{chemmacros}

\newcommand*\diff{\mathop{}\!\mathrm{d}}

\usepackage[flushleft]{threeparttable}

\usepackage{changes}

\definechangesauthor[name={MS}, color=blue]{Edits-MS}

\begin{document}

\preprint{APS/123-QED}

\title{Raman Spectroscopy Reveals Photobiomodulation-Induced \textalpha-Helix to \textbeta-Sheet Transition in Tubulins: Potential Implications for Alzheimer's and Other Neurodegenerative Diseases}

\author{Elisabetta Di Gregorio}
\thanks{These two authors contributed equally to this work}
\email{michael.staelens@ific.uv.es}
 \affiliation{Department of Physics, Faculty of Science, University of Alberta, Edmonton, AB T6G 2E1, Canada}
\affiliation{Department of Mechanical and Aerospace Engineering (DIMEAS), Faculty of Biomedical Engineering, Polytechnic University of Turin, 10129 Turin, Italy}
\affiliation{Department of Physics, Freie Universit\"at Berlin, 14195 Berlin, Germany}

\author{Michael Staelens}
\thanks{These two authors contributed equally to this work}
\email{michael.staelens@ific.uv.es}
\affiliation{Department of Physics, Faculty of Science, University of Alberta, Edmonton, AB T6G 2E1, Canada}
\affiliation{Instituto de F\'isica Corpuscular, CSIC--Universitat de Val\`encia, Carrer Catedr\`atic Jos\'e Beltr\'an 2, 46980 Paterna (Val\`encia), Spain}

\author{Nazanin Hosseinkhah}
\author{Mahroo Karimpoor}
\author{Janine Liburd}
\author{Lew Lim}
\affiliation{Vielight Inc., Toronto, ON M4Y 2G8, Canada}

\author{Karthik Shankar}
\affiliation{Department of Electrical and Computer Engineering, Faculty of Engineering, University of Alberta, Edmonton, AB T6G 1H9, Canada}

\author{Jack A. Tuszy\'nski}
\affiliation{Department of Physics, Faculty of Science, University of Alberta, Edmonton, AB T6G 2E1, Canada}
\affiliation{Department of Mechanical and Aerospace Engineering (DIMEAS), Faculty of Biomedical Engineering, Polytechnic University of Turin, 10129 Turin, Italy}
\affiliation{Department of Data Science and Engineering, Silesian University of Technology, 44-100 Gliwice, Poland}

\date{\today}

\begin{abstract}
In small clinical studies, the application of photobiomodulation (PBM), which typically delivers near-infrared (NIR) to treat the brain, has presented some remarkable results in the treatment of dementia and several neurodegenerative diseases. However, while the literature is rich with the mechanisms of action underlying PBM outcomes, the underlying mechanisms affecting a neurodegenerative disease are not entirely clear. While large clinical trials are warranted to validate these findings, evidence of the mechanisms can explain and hence provide credible support for PBM as a potential treatment for these diseases. Tubulin and its polymerized state of microtubules have been known to play important roles in the pathology of Alzheimer's and other neurodegenerative diseases. We investigated the effects of PBM on these structures in the quest for answers. In this study, we employed a Raman spectroscopic analysis of the amide~I band of polymerized samples of tubulin exposed to pulsed low-intensity NIR radiation ($810$~nm, $10$~Hz, $22.5$~J/cm$^{2}$ dose). Peaks in the Raman fingerprint region ($300$--$1900$~cm$^{-1}$), in particular, in the amide~I band ($1600$--$1700$~cm$^{-1}$), can be used to quantify the percentage of protein secondary structures. Under this band, hidden signals of $\mathrm{C}$=$\mathrm{O}$ stretching, belonging to different structures, are superimposed---producing a complex signal as a result. An accurate decomposition of the amide~I band is therefore required for the reliable analysis of the conformation of proteins, which we achieved through a straightforward method employing a Voigt profile. This approach was validated through secondary structure analyses of unexposed control samples, for which comparisons with other values available in the literature could be conducted. Subsequently, using this validated method, we present novel findings of statistically significant alterations in the secondary structures of NIR-exposed tubulin, characterized by a notable decrease in \textalpha-helix content and a concurrent increase in \textbeta-sheets compared to the control samples. The \textalpha-helix to \textbeta-sheet transition suggests that PBM reduces microtubule stability and introduces dynamism to allow for the remodeling and, consequently, refreshing of microtubule structures. This newly discovered mechanism could have implications for reducing the risks associated with brain aging, including neurodegenerative diseases like Alzheimer's disease.
\end{abstract}

\maketitle

\section{\label{sec:Intro}Introduction}
Healthy cellular function and structure are intrinsically linked to the integrity of tubulins. Tubulins are abundant, hydrophilic, and highly conserved cytoskeletal proteins found in all eukaryotic cells, which play a critical role in the structure and function of microtubules (MTs). Eukaryotic cells typically contain $\sim3$--$4\%$ tubulin~\cite{HILLER1978795,OAKLEY2000,FOURESTLIEUVIN2006183,VERDIERPINARD2009197,Widlund2012,KHAMIS20181}. Notably, however, mammalian brain tissue is particularly rich in tubulin content, consisting of $\sim10\%$ or more of the total protein content~\cite{HILLER1978795,FOURESTLIEUVIN2006183,VERDIERPINARD2009197,Widlund2012}.

Tubulin has a heterodimeric structure composed of two closely related monomeric subunits, \textalpha- and \textbeta-tubulin, which combine via protein folding and dimerization processes. Both monomers have molecular weights of $\sim55$~kDa each, share an amino acid sequence homology of $\sim40$--$55\%$~\cite{Luduena1973,Valenzuela1981, Little1981,Nogales1998}, and comprise a pair of \textbeta-sheets surrounded by \textalpha-helices~\cite{Nogales1998}. Their secondary structure compositions are dominated by \textalpha-helices (which is generally true for globular proteins~\cite{Haimov2016}).

Regarding the functionality of tubulin, nucleation and the polymerization rate are attributed to \textalpha-helices, whereas \textbeta-sheets play a dual role in regulating these functions and contributing to the stability of this highly dynamic protein~\cite{Audenaert1989}. Additionally, \textalpha\textbeta-tubulin heterodimers comprise oppositely charged ends, with the negative and positive ends formed by \textalpha-tubulin and \textbeta-tubulin, respectively. In contrast with its ordered structure, tubulin also presents a disordered portion: the negatively charged C-terminal tails, which play an important role in the interaction between tubulin and microtubule-associated proteins~\cite{Wall2016}.

Members of the tubulin protein family are known to possess unique electrostatic properties~\cite{Marracino2019} that are fundamental to their ability to form MTs. \textalpha\textbeta-tubulin dimers polymerize head-to-tail into intrinsically polar linear protofilaments that can further assemble into metastable MTs through lateral tubulin--tubulin interactions; generally, MTs comprise $13$ protofilaments arranged in a tubular lattice configuration. MTs are dynamic structures that play crucial roles in many cellular processes, such as cell division and chromosome segregation~\cite{Higuchi2005,Laband2017}; cell movement and motility~\cite{Manneville2004,Garcin2019}; maintaining cell structure and rigidity~\cite{Ingber2003}; and the transport of vesicles and organelles via kinesin and dynein motor proteins~\cite{Vale2003}.

In the formation of MTs, \textalpha\textbeta-tubulin binds guanosine triphosphate (GTP) at two different binding sites, one exchangeable (in \textbeta-tubulin) and one non-exchangeable (\textalpha-tubulin). Hydrolyzation of GTP at the exchangeable site allows tubulin assembly~\cite{Nogales1998, DavidPfeuty1977, Sierra-Fonseca17} into a mainly GDP-tubulin microtubule, with a final portion of GTP-bound tubulin known as the GTP cap. It is the presence of this cap that makes MT polymerization possible~\cite{Sierra-Fonseca17, Mitchison1984}. When this piece of MTs is lost, the catastrophe phenomenon occurs, causing MTs to shrink instead of grow~\cite{Mitchison1984}. Growth will only resume after the GTP cap is reacquired. This process is known as rescue~\cite{Walker1988, Gundersen2004}. Thus, MTs have a dynamic behavior, alternating between phases of shrinking and growing, permitting them to be reshaped in cells. This particular characteristic, known as dynamic instability~\cite{Mitchison1984, Desai1997}, is pivotal for MT integrity and, if lost, can alter cell division properties. In healthy cells, time and space are important factors in the regulation of MT dynamics, even across the cytoplasm~\cite{Pasquier2008}. In particular, during mitosis, interphase MTs disassemble to form the mitotic spindles, which are about $100$ times faster at assembling/disassembling~\cite{Albahde2021}. The mitotic spindle is responsible for chromosome segregation. After the cell is completely divided, MTs forming the mitotic spindle reassemble into cytoplasmic MTs~\cite{Albahde2021}.

MTs possess a variety of interesting and distinct electrical properties (reviewed in detail in Ref.~\cite{Kalra2020}), such as electrical conductance and impedance~\cite{nano10020265}, as well as a highly negatively charged surface due to the large negative electrostatic charge of \textalpha\textbeta-tubulin dimers ($Q_{\mathrm{eff}}\sim-23\:e$ for a dimer in an MT~\cite{Heuvel2007}); thus, they have been considered as bionanowires that, in addition to supporting ionic transport~\cite{Sataric2009,Sekulic2011}, could be responsible for intracellular signaling~\cite{GUNDERSEN199981,CRADDOCK2010,Dent2014,BARVITENKO2018191}. Given these unique electrical properties and the highly polar nature of MTs, they have been considered a potential target for electromagnetic field (EMF)-based therapies. Numerous investigations have documented diverse impacts of EMFs on MTs in solution, such as the alignment of MTs in the presence of electric fields~\cite{STRACKE2002602,Minoura2006,Heuvel2007,Uppalapati2008,Dujovne2008,Isozaki2015}, the disassembly of MTs by intense terahertz pulses~\cite{Hough2021}, and effects on MT polymerization induced by low-intensity near-infrared (NIR) light radiation~\cite{Staelens2022}.

In neurons, MTs are responsible for the maintenance of neuron shape and structure, neuronal soma migration~\cite{Baas2016}, the growth and structure of axons~\cite{LewisJr2013}, protein transport in axons and dendrites~\cite{Yogev2016}, and the support of morphological changes in dendrites potentially associated with neuroplasticity~\cite{Baas2016}. MTs in brain cells may exhibit varying levels of stability compared to other cell types, depending on the specific context and cellular functions they are involved in; however, on average, MTs are more stable in neurons compared to other cells~\cite{Baas2016}. Additionally, in axons and dendrites, neuronal MTs are found in unique and curious configurations as uniform parallel aligned arrays~\cite{Baas2016,KELLIHER201939}. A plethora of studies have reported MT loss and dysfunction in connection with the onset and progression of neurodegenerative diseases (NDs), such as Alzheimer's disease (AD)~\cite{McMurray2000,Jean2013,MATAMOROS2016217,Sferra2020,Fernandez2020,BOIARSKA2021604,Peris2022}.

Photobiomodulation (PBM) uses low-intensity, non-thermal, and non-ionizing sources of electromagnetic (EM) radiation, typically in the visible red and NIR regions of the EM frequency spectrum, to induce positive physiological changes and health outcomes. Several small clinical studies of PBM for NDs have demonstrated remarkable results~\cite{Saltmarche2017,Chao2019,Salehpour2019,Liebert2021}. For example, statistically significant improvements in patients' Alzheimer's Disease Assessment Scale-Cognitive Subscale (ADAS-Cog~\cite{Kueper2018}) scores were reported in two studies employing $12$-week transcranial--intranasal $810$~nm PBM ($-6.73$ points vs. baseline after $12$~weeks, $p < 0.023$~\cite{Saltmarche2017}; $-5.2$ points vs. baseline after $12$~weeks, $p = 0.007$~\cite{Chao2019}). Notably, both studies observed mean improvements in ADAS-Cog scores that were markedly greater than those reported in a phase III clinical trial with $10$~mg/day donepezil therapy ($\sim -2$ points vs. baseline after $12$~weeks, $p < 0.0001$)~\cite{Rogers136}, which for a long time has been the standard of treatment for AD. Moreover, with PBM, patients with mild-to-moderately severe dementia experienced noteworthy enhancements, including improved sleep, reduced anxiety, and increased functional ability, without any negative adverse effects~\cite{Saltmarche2017}.

Additionally, the neuroprotective effects of PBM for AD have been demonstrated both in vivo~\cite{MICHALIKOVA2008480,DeTaboada2011,GRILLO201313,Purushothuman2014,PURUSHOTHUMAN2015155,daLuzEltchechem2017} and in vitro~\cite{Sommer2012}. This has led to an increased interest in such therapies, and the number of studies on their efficacy in treating NDs has seen a substantial increase (see Ref. ~\cite{Salehpour2021} for a review); several clinical trials for treating NDs are currently ongoing~\cite{NCT03484143,NCT04018092,NCT03551392}. Despite the promises, the literature recognizes that the mechanisms of action underlying the observed efficacy of PBM are still not entirely clear, and studies to understand the biophysical and subcellular effects at the molecular level are lacking~\cite{Freitas2016}. Thus, more research on the molecular and biophysical mechanisms of action is highly warranted.

In this work, we present the results of such a study aimed at investigating the effects of the pulsed NIR light employed in PBM therapy on the secondary structures of tubulin and MTs---fundamental components of the cytoskeleton in eukaryotic cells. It significantly involves the transitioning of \textalpha-helical and \textbeta-sheet arrangements. Findings in this area would contribute to explaining PBM efficacy.

We used Raman spectroscopy to compare the secondary structure compositions of polymerized tubulins in buffer solutions when exposed or unexposed to NIR light. Raman spectroscopy and other spectroscopic techniques are typically employed to study the secondary structures of proteins~\cite{Miura1995,Lefevre2007}. The other spectroscopy techniques include Fourier-transform infrared (FTIR) spectroscopy~\cite{Barth2007,DeMeutter2021} and far-ultraviolet (UV) circular dichroism (CD) spectroscopy~\cite{LEE19784,Greenfield2006}, among others. Raman spectroscopy is a scattering-based technique that measures the inelastic scattering of light by molecules, which has several advantages over other spectroscopic techniques. One of the advantages of Raman spectroscopy over infrared spectroscopy is that $\mathrm{H}_{2}\mathrm{O}$ vibrations have less influence on Raman spectra, eliminating the need for $\mathrm{D}_{2}\mathrm{O}$ and reducing the error related to background subtractions~\cite{Miura1995, Li2019}. Additionally, Raman spectroscopy is more feasible for studying turbid solutions, such as solutions of polymerized tubulin, whereas methods such as CD are not suitable due to potential distortion in the measured signal caused by the turbidity. We will be comparing our observations for unexposed tubulins with Raman spectroscopy against other observations in the literature with Raman spectroscopy, CD, and FTIR as a check that our technique is in line with others in the literature.

In summary, Raman spectroscopy is a powerful, label-free method that has demonstrated utility for the chemical analysis of biological and non-biological samples. With this technique, we could observe macromolecule conformation modifications, which translate to shifts in the frequency bands acquired through this methodology~\cite{Miura1995}. Three Raman bands, amide~I, II, and III ($1600$--$1700$~cm$^{-1}$~\cite{Kuhar2021}, $1510$--$1580$~cm$^{-1}$~\cite{Sadat_2020}, and $1220$--$1310$~cm$^{-1}$~\cite{doi:10.1021/ac503647x}, respectively), are particularly useful for evaluating proteins and peptide structures~\cite{Rygula2013}. $\mathrm{C}$=$\mathrm{O}$ stretching vibrations account for around $80$\% of the amide~I band~\cite{Kuhar2021}. The remainder is related to $\mathrm{C}$--$\mathrm{N}$ out-of-plane stretching~\cite{Kuhar2021}. In contrast, the amide~II band is less sensitive to alterations in protein conformation~\cite{Sadat_2020, Krimm1986}. It accounts primarily for in-plane bending of $\mathrm{N}$--$\mathrm{H}$ groups ($40$--$60\%$) and vibrations related to the stretching of the $\mathrm{C}$--$\mathrm{N}$ groups ($18$--$40\%$)~\cite{Kuhar2021, Sadat_2020, Krimm1986}, while $\mathrm{C}$=$\mathrm{O}$ bending and $\mathrm{C}$--$\mathrm{C}$ stretching have little influence on this band~\cite{Kuhar2021}. Finally, amide~III peaks are related to the bending of in-plane $\mathrm{N}$--$\mathrm{H}$ groups and $\mathrm{C}$--$\mathrm{N}$ stretching~\cite{doi:10.1021/ac503647x}.

Only a couple of studies have performed Raman spectroscopic analyses of the secondary structures of tubulin~\cite{Audenaert1989} and MTs~\cite{Audenaert1989,Simi_Krsti__1991}. In this study, we exploit Raman spectroscopy to determine how tubulin changes its internal structure when exposed to pulsed low-intensity NIR light. As far as we are aware, this is the first such study to report changes in the secondary structures of tubulin induced by NIR radiation.

The remainder of this article is organized as follows. In Sec.~\ref{sec:MatsMethods}, we briefly describe the materials, equipment, and methodology employed in this study. The results are presented in Sec.~\ref{sec:Results}, beginning with a comparison of our results obtained for the secondary structure composition of unexposed tubulin (control samples) with values obtained in several other studies in the literature both to validate our methodology and to resolve some of the tension between the different values reported by these studies. Thereafter, we present our results for the secondary structures of the NIR-exposed tubulin samples and contrast these results against those obtained for the unexposed samples. We discuss these results and present several ensuing hypotheses regarding their possible connection to the reported efficacy of PBM for treating AD in Sec.~\ref{sec:Disc}. Lastly, conclusions and future outlooks are provided in Sec.~\ref{sec:Conc}.

\section{\label{sec:MatsMethods}Experimental Methodology}
\subsection{\label{subsec:TubRecon}Reconstitution of Tubulin Samples}
Unlabeled ultra-pure tubulin derived from porcine brain, purchased from Cytoskeleton, Inc. (T240), was employed in the following experiments. T240 samples were stored at $4$~$^{\circ}$C and later resuspended to $2.5$~mg/mL tubulin by adding to each vial $360$~\textmu L of ice-cold G-PEM buffer (GTP-supplemented PEM buffer: $80$ mM PIPES pH $6.9$, $0.5$ mM EGTA, and $2$ mM $\mathrm{Mg}\mathrm{Cl}_{2}$) and $40$~\textmu L of Microtubule Cushion Buffer (PEM buffer in $60\%$ v/v glycerol), and then placing the protein sample on ice. The G-PEM buffer was prepared immediately prior by adding $990$~\textmu L of PEM buffer to $10$~\textmu L of $100$~mM GTP; thus, the final GTP concentration of the G-PEM buffer was $1$~mM. After reconstitution, the samples were aliquoted into experimental amounts of $1$~mL and snap-frozen through immersion in liquid nitrogen to avoid protein denaturation. Finally, the samples were stored at $-80$~$^{\circ}$C until use in the experiments.

\subsection{\label{subsec:NIRexp}Near-Infrared Exposure of Tubulin}
The exposure of reconstituted tubulin samples was performed with the intranasal LED applicator of the Vielight Neuro Alpha transcranial--intranasal brain PBM device; its parameters are reported in Table~\ref{tab:table1}. On its own, the intranasal applicator has shown potential as a treatment method for neurological disorders~\cite{Salehpour2020,Yoo2021}. Tubulin samples collected from the $-80$~$^{\circ}$C freezer were exposed for $30$~minutes inside a $4$~$^{\circ}$C fridge to prevent polymerization of the samples during exposure. To avoid movement of the sample with respect to the LED, it was fixed directly to the intranasal applicator and finally to the inside of a cardboard cryo box. The box was also utilized to keep the sample in the dark during exposure to avoid light diffusion and reflection. From the power density of the LED, we can calculate both the delivered energy and the approximate strength of the electric field  generated. For a $30$-minute exposure, the total energy delivered amounts to $22.5$~J (i.e., a net dose of $22.5$~J/cm$^{2}$); the electric field strength is approximately $4$~V/cm. Separate tubulin samples that were not subjected to any NIR exposure were preserved for use as control samples. After exposure and prior to performing Raman spectroscopy, the tubulin samples were polymerized into MTs by placing them in a $37$~$^{\circ}$C incubator for $60$ minutes. Two independent experiments and subsequent measurements were performed.

\begin{table}[htb]
\caption{\label{tab:table1} Characteristic parameters of the intranasal LED applicator of the Vielight Neuro Alpha brain PBM device.}
\begin{ruledtabular}
\begin{tabular}{lc}
\textbf{Parameters}	& \textbf{Neuro Alpha Intranasal LED}\\
\colrule
Wavelength (nm)	& $810$\\
Power density (mW/cm$^{2}$) & $25$\\
Pulse frequency (Hz) & $10$\\
Pulse duty cycle & $50\%$\\
Beam spot size (cm$^{2}$) & $1$\\
\end{tabular}
\end{ruledtabular}
\end{table}

\subsection{\label{subsec:RamSpec}Raman Spectroscopy}
\begin{figure}[htb]
\includegraphics[width = 8.5 cm]{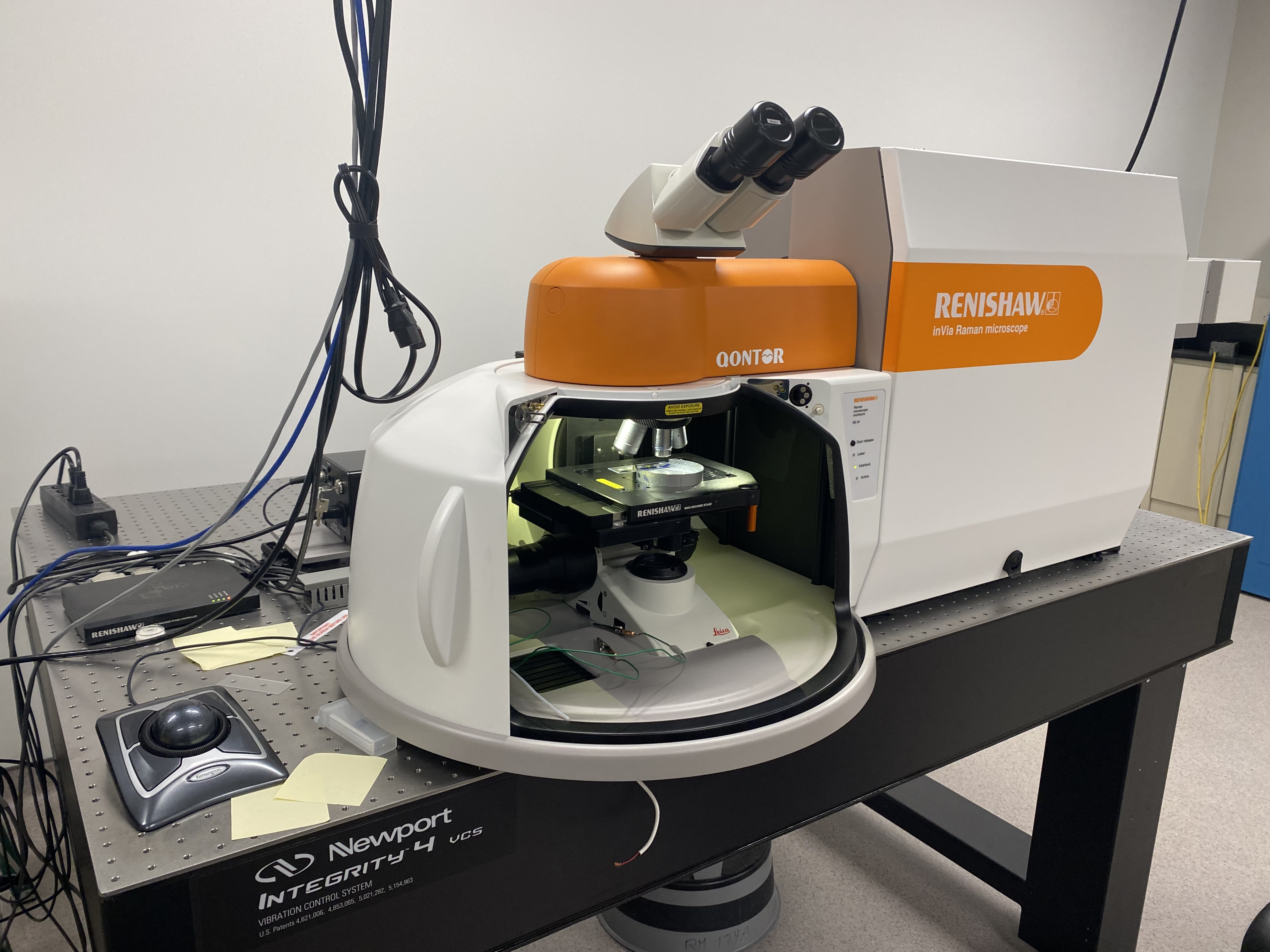}
\caption{\label{fig:Fig1} Digital photograph of the Renishaw inVia™ confocal Raman microscope used in our experiments.}
\end{figure}

A $5$~\textmu L droplet of polymerized tubulin solution, either untreated or NIR-exposed, was deposited onto a glass slide for measurement. Raman spectra were acquired at room temperature with a $532$~nm laser, with $1200$~lines per mm grating, $100\%$ power, and an exposure time of $1$~s, using a Renishaw inVia™ confocal Raman microscope. A photograph of the device is provided in Fig.~\ref{fig:Fig1}. The resulting spectra reported here derive from at least two samples measured. Several points of the same sample---focused through the $50\times$ magnification---were measured, capturing multiple acquisitions ($4$--$5$) for each position. The multiple acquisitions obtained were automatically averaged by Renishaw's WiRE™ software, which manages the collection of Raman data.

\subsection{\label{subsec:DataProc}Data Processing and Spectral Decomposition}
The data were processed using \textsc{Matlab}\textsuperscript{\textregistered} R2022a (v. 9.12). After the acquisition of spectra, range reductions (to $350$--$2700$~cm$^{-1}$) and baseline corrections were implemented. An asymmetric least-squares smoothing with a $0.01$ threshold, a smoothing factor of $5$, and $10$ iterations was employed for the baseline corrections~\cite{Eilers2005}. The data were smoothed using Savitzky--Golay filtering~\cite{Savitzky1964}, available in the \textsc{Matlab}\textsuperscript{\textregistered} Signal Processing Toolbox, based on a second-order polynomial and with a $17$-point window. Additional measurements were conducted for both the blank glass slide and the buffer solution, and in both cases, the resulting spectra in the amide~I region exhibited no band structure contributions, eliminating the need for any background subtractions associated with these potential contributions. For spectral deconvolution, all the measured spectra were restricted to the amide~I band and normalized between $0$ and $1$. Peak finding was performed by analyzing the second derivative of the spectra. Previous research has shown that the biophysics of protein folding processes, which secondary structures are a consequence of, can be effectively described by a Voigt profile~\cite{Maisuradze2021}. Accordingly, peak deconvolution of the measured Raman amide~I spectra was performed with a Voigt profile distribution, which is defined as a convolution of Lorentzian and Gaussian distributions:

\begin{eqnarray}
    f_{\mathrm{L}}(x)=\frac{2A}{\pi}\frac{w_{\mathrm{L}}}{{4(x-x_{\mathrm{c}}})^2+w_{\mathrm{L}}^2},
\end{eqnarray}
\begin{eqnarray}
    f_{\mathrm{G}}(x) = \sqrt{\frac{4\ln 2}{\pi w_{\mathrm{G}}^2}} \; \exp{\left(-\frac{4\ln 2}{w_{\mathrm{G}}^2}x^2\right)},
\end{eqnarray}
where $A$ represents the area, $x_{\mathrm{c}}$ represents the center, and $w_{\mathrm{L}}$ and $w_{\mathrm{G}}$ are parameters specifying the Lorentzian and Gaussian full width at half maximum, respectively. Explicitly, it can then be written as~\cite{DiRocco2001} 
\begin{eqnarray}
    y(x) = y_{0}+f_{\mathrm{L}}(x)\ast f_{\mathrm{G}}(x) \nonumber
\end{eqnarray}
\begin{eqnarray}
& = y_{0}+ \displaystyle{A\frac{2\ln 2}{\pi^{3/2}}\frac{w_{\mathrm{L}}}{w_{\mathrm{G}}^2}} \nonumber \\
& \times \displaystyle\int_{-\infty}^\infty  \left(\frac{\mathrm{e}^{-t^2}}{\left(\sqrt{\ln 2}\frac{w_{\mathrm{L}}}{w_{\mathrm{G}}}\right)^2+\left(2\sqrt{\ln 2}\frac{x-x_{\mathrm{c}}}{w_{\mathrm{G}}}-t\right)^2}\right)\diff t .
\end{eqnarray}

Curve fitting was performed using the Levenberg--Marquardt minimization algorithm function available in \textsc{Matlab}\textsuperscript{\textregistered}~\cite{LEVENBERG1944,Marquardt1963}. The coefficient of determination, $R^{2}$, was used to evaluate the goodness of the fits. Optimal values for this parameter were greater than $0.99$. Following the fitting procedure, the amide~I vibrational peak areas were used to evaluate the secondary structure content of each protein sample. This analysis was performed by adding the areas of all the amide~I peaks obtained that contribute to the secondary structures and calculating the individual contribution of each peak associated with a particular secondary structure: \textalpha-helices, \textbeta-sheets, random coils, and \textbeta-turns~\cite{Ngarize2004,Sadat_2020}. This is based on the assumption that the Raman cross-section for the mentioned structures is the same, as discussed by Surewicz et al.~\cite{Surewicz1993} and Sane et al.~\cite{Sane1999}. The peaks associated with each secondary structure were assigned to the specific sub-bands of the amide~I envelope summarized in Table~\ref{tab:table2}~\cite{WANG2020105149}; however, there are no universally agreed-upon definitions for these characteristic bands, and hence, the particular values stated differ slightly throughout the literature. Finally, for both groups of tubulin samples studied, the mean percentages of secondary structures were calculated.

\begin{table}[htb]
\caption{\label{tab:table2} Secondary structure peaks contribution bands, from Ref.~\cite{Ngarize2004}.}
\begin{ruledtabular}
\begin{tabular}{lc}
\textrm{\textbf{Secondary Structure}}&
\textrm{\textbf{Amide~I Band (cm$^{-1}$)}}\\
\colrule
\textbeta-sheet & $1620$--$1640$, $1670$--$1680$\\ 
\textalpha-helix & $1650$--$1660$\\ 
Random coil & $1660$--$1670$\\ 
\textbeta-turn & $1680$--$1699$
\end{tabular}
\end{ruledtabular}
\end{table}

\subsection{\label{subsec:Stats}Statistical Analyses}
Statistical hypothesis testing was performed to compare the resulting mean secondary structure compositions of the unexposed tubulin samples with other results reported in the literature, as well as with the results obtained for the NIR-exposed tubulin. Under a normality assumption, statistically significant differences between results were established using Welch's unequal variances $t$-tests~\cite{Welch1938,Welch1947} with a significance level of $\alpha=0.05$. All statistical analyses were performed using \textsc{Matlab}\textsuperscript{\textregistered} R2023a (v. 9.14).

\section{\label{sec:Results}Results}
We first report the results of our Raman spectroscopic analyses of the secondary structure compositions of the unexposed control samples. These results are compared with those from other studies available in the literature that analyzed the conformation of tubulin and microtubules. This is followed by a presentation of our secondary structure results obtained for the NIR-exposed tubulin samples and how they compare to the results for the control samples.

\subsection{\label{subsec:ExpResults}Raman Spectra and Secondary Structures of Polymerized Unexposed Tubulin}
The secondary structure compositions obtained for our control samples in the three independent experiments performed are presented in Table~\ref{tab:table3}. A sample of one of the control spectra obtained in these experiments and the resulting spectral deconvolution of the amide~I band is shown in Fig.~\ref{fig:Fig2} (the remaining spectra are available in the Supplementary Material file, Figs.~S1--S3~\cite{RamanSuppMat}). As previously stated, each peak was assigned to a characteristic secondary structure to estimate its total percentage. In all three experiments, we find that the secondary structure compositions of the unexposed samples are dominated by \textalpha-helices, which is typically the case for globular proteins. In two of the three experiments, the results indicate that the second-most abundant structures are \textbeta-sheets. The average ($\pm$ SD) results, reported in the first row of Table~\ref{tab:table4}, are consistent with this trend. In particular, we find an average secondary structure composition dominated by $36.0 \pm 4.2\%$ \textalpha-helices and $26.7 \pm 7.2\%$ \textbeta-sheets.

\begin{table}[htb]
\caption{\label{tab:table3} Secondary structure composition results obtained for the control (unexposed) polymerized tubulin samples. The last two rows correspond to separate measurements of two different points of the same sample.}
\begin{ruledtabular}
\begin{tabular}{cccccccc}
			\textbf{Sample} & \textbf{\textalpha-Helix}    & \textbf{\textbeta-Sheet}	&  \textbf{\textbeta-Turn}    &  \textbf{Random Coil} \\
			\colrule

           Control~1 & $32.7\%$ & $26.0\%$ & $21.6\%$ &  $19.7\%$ \\

           Control~2-1 & $40.7\%$ &  $19.9\%$ & $16.0\%$ &  $23.5\%$ \\

           Control~2-2 & $34.5\%$ & $34.2\%$ & $11.6\%$ & $19.8\%$ \\

\end{tabular}
\end{ruledtabular}
\end{table}

\begin{figure*}[htb]
\includegraphics[width = 16 cm]{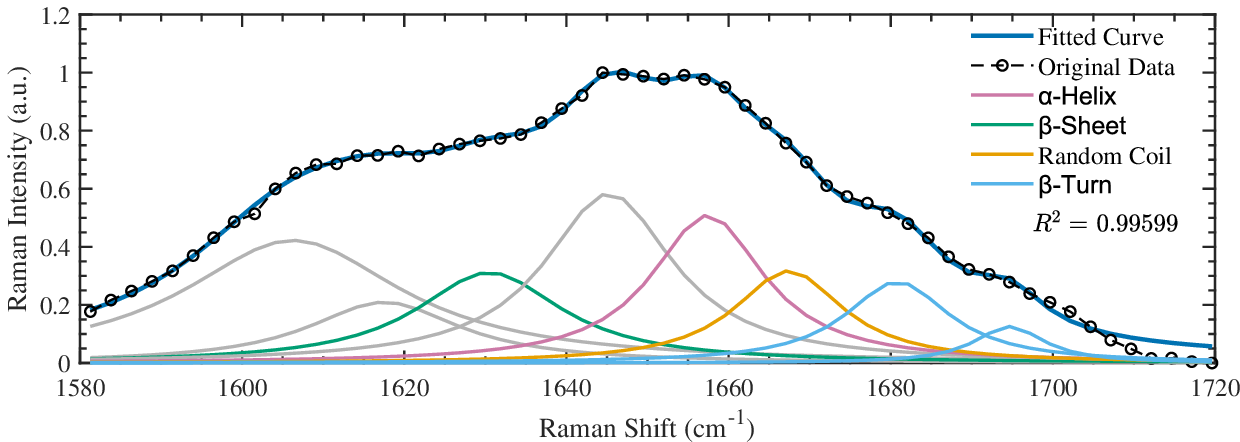}
\caption{\label{fig:Fig2} A representative example of one of the amide~I Raman spectra obtained for the control (unexposed) polymerized tubulin samples (labeled as Control~1 in Table~\ref{tab:table3}). Grey curves represent peaks obtained from the spectral deconvolution that are unassociated with any secondary structures.}
\end{figure*}

\begin{table*}[htb]
\caption{\label{tab:table4} Comparison of the secondary structure percentages for MTs and tubulin (dimer).}
\begin{ruledtabular}
\begin{tabular}{ccccccc}

    \textbf{Material} & \textbf{Source}	& \textbf{Method}	& \textbf{\textalpha-Helix}      & \textbf{\textbeta-Sheet}	& \textbf{\textbeta-Turn}      & \textbf{Random Coil} \\
	\colrule
     & & & & & &  \\

    \multirow{6}*{MTs} & This study		& Amide~I analysis of Raman spectra		&  $36.0 \pm 4.2\%$	& $26.7 \pm 7.2\%$  & 	$16.4 \pm 5.0\%$	&  $21.0 \pm 2.2\%$ \\
    & & & & & \\

    ~ & \multirow{2}*{Audenaert et al.~\cite{Audenaert1989}}		& \multirow{2}*{Amide~I analysis of Raman spectra}		& $21 \pm 3 \%$	& $48 \pm 2 \%$ 	& $19 \pm 1 \%$	& $12 \pm 1 \%$ \\

    ~ & ~ & ~ & (**) & (*) & (ns) & (*) \\
    
    & & & & & \\

    ~ & \multirow{2}*{Simi\'c-Krsti\'c et al.~\cite{Simi_Krsti__1991}}		& \multirow{2}*{Amide~I analysis of Raman spectra}		& $33\%$	& 	$27\%$ & $24\%$ & $16\%$ \\

   ~ & ~ & ~ & (ns) & (ns) & (ns) & (ns) \\
    
    & & & & & \\

    \hline

    & & & & & &  \\
    \multirow{11}*{Tubulin} & \multirow{2}*{Ventilla et al.~\cite{Ventilla1972}}		& \multirow{2}*{Far-UV CD spectroscopy}			&	$22\%$		& 	\multicolumn{2}{c}{$30\%$}			& $48\%$	 \\

    ~ & ~ & ~ & (*) & \multicolumn{2}{c}{(ns)} & (**) \\
    
    & & & & & &  \\
    
    ~ & \multirow{5}*{de Pereda et al.~\cite{de_Pereda_1996}}	& \multirow{2}*{Far-UV CD spectroscopy}			& 	$33\pm7\%$		& $21\pm5\%$ 			& $21\pm6\%$ & $25\pm6\%$ \\
    
    ~ & ~ & ~ & (ns) & (ns) & (ns) & (ns) \\
       
    & & & & & \\
        ~ &	 & \multirow{2}*{FTIR spectroscopy} & 	$37$\footnote{Maximum.}$\pm1\%$		& $24\pm1\%$ 	& $20\pm1\%$ & $18$\footnote{Minimum.}$\pm1\%$ \\
        
    ~ & ~ & ~ & (ns) & (ns) & (ns) & (ns) \\
           
    & & & & & &  \\

    ~ & \multirow{2}*{Afrasiabi et al.~\cite{Afrasiabi2013}}		& \multirow{2}*{Far-UV CD spectroscopy}			&	$38.02\%$		& 	$15.22\%$		& 	--		&  $46.76\%$ \\
    
       ~ & ~ & ~ & (ns) & (ns) & & (**) \\

\end{tabular}
    \begin{tablenotes}
      \small
      \item ** Highly statistically significant, $0.001 \leq p < 0.01$; * statistically significant, $0.01 \leq p < \alpha$; ns, not significant, $p \geq \alpha$.
    \end{tablenotes}
\end{ruledtabular}
\end{table*}

Two studies that analyzed the conformation of polymerized tubulin are available in the literature, both of which employed Raman spectroscopy and subsequent deconvolution of the amide~I band~\cite{Audenaert1989,Simi_Krsti__1991}. Their average results are presented in the upper portion of Table~\ref{tab:table4} (uncertainties are also displayed for every study that explicitly reported such values). Welch's unequal variances $t$-tests indicate that our results differ significantly from those reported by Audenaert et al.~\cite{Audenaert1989} for all secondary structures analyzed except \textbeta-turns. Notably, however, we find excellent agreement with the results obtained by Simi\'c-Krsti\'c et al.~\cite{Simi_Krsti__1991}, with no significant differences obtained for any of the structures analyzed, suggesting that our techniques are consistent. Moreover, for both \textalpha-helices and \textbeta-sheets, we find agreement with their results within one error interval.

Several other studies exist in the literature that analyzed the conformation of tubulin dimers using different spectroscopic techniques. We compare our results with three such studies: Ventilla et al.~\cite{Ventilla1972}, which employed far-UV CD spectroscopy; de Pereda et al.~\cite{de_Pereda_1996}, which included two independent analyses using both far-UV CD spectroscopy and FTIR spectroscopy; and finally, Afrasiabi et al.~\cite{Afrasiabi2013}, which also utilized far-UV CD spectroscopy. Their results on the secondary structure composition of tubulin dimers are presented in the lower portion of Table~\ref{tab:table4}. We find agreement with the results of both analyses reported by de Pereda et al.~\cite{de_Pereda_1996}, as well as the results of the study by Afrasiabi et al.~\cite{Afrasiabi2013}. On the other hand, our results largely disagree with those reported by Ventilla et al.~\cite{Ventilla1972}.

In summary, for the main secondary structures---\textalpha-helices and \textbeta-sheets---we find the best agreement with the values reported by de Pereda et al. (FTIR, $p=0.717$) and Simi\'c-Krsti\'c et al. ($p=0.949$), respectively. The comparable observations of ours with the literature suggest that our Raman spectroscopy technique was in line with others. Therefore, our observations of the unexposed tubulins are credible as baselines for comparing the effects of NIR irradiation. Detailed results from all these statistical tests can be found in Sec.~II of the Supplementary Material, Tables~S1--S6~\cite{RamanSuppMat}. Relevant information on the different experiments and analyses performed by each group is also provided (see Table~S7 for a comparative overview). Although it is not entirely clear what particular factors might be responsible for some of the observed disagreements between results, these experimental and methodological differences are notable and likely account for some of the variances.

\subsection{\label{subsec:UnexpResults}Raman Spectra and Secondary Structures of Polymerized NIR-Exposed Tubulin}
\begin{figure*}[htb]
\includegraphics[width = 16 cm]{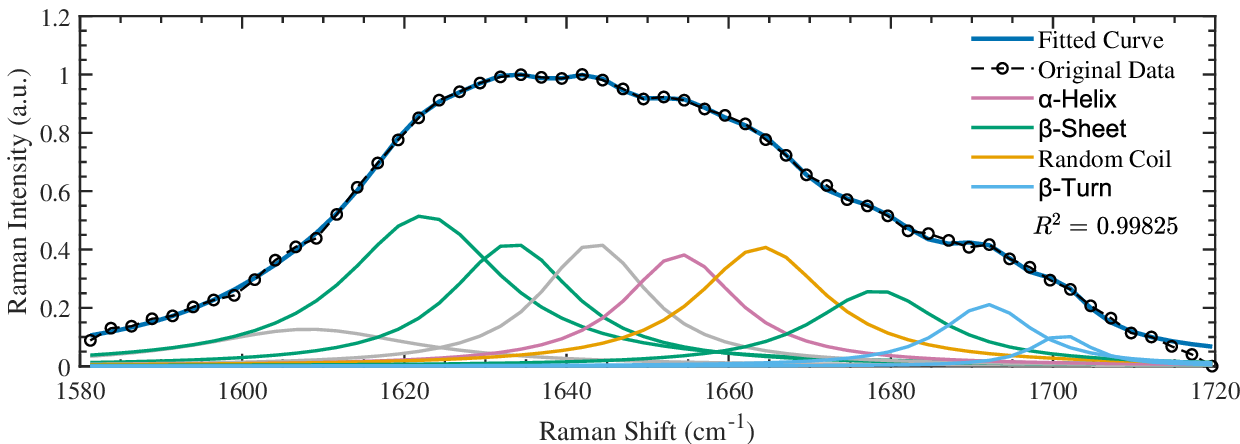}
\caption{\label{fig:Fig3} A representative example of one of the amide~I Raman spectra obtained for the NIR-exposed polymerized tubulin samples (labeled as Exposed~1 in Table~\ref{tab:table5}). Grey curves represent peaks obtained from the spectral deconvolution that are unassociated with any secondary structures.}
\end{figure*}

We now turn to our results for the secondary structure compositions of polymerized NIR-exposed tubulin obtained by deconvoluting their measured Raman spectra using the same procedure applied to the control samples. The secondary structure compositions derived from the exposed samples in the three independent experiments are detailed in Table~\ref{tab:table5}. Fig.~\ref{fig:Fig3} presents an example of one of the exposed spectra gathered during these experiments, along with the consequent spectral deconvolution of the amide~I band (additional spectra can be found in the Supplementary Material, Figs.~S4--S6~\cite{RamanSuppMat}). Notably, our results for the three different NIR-exposed samples indicate a significant change in conformation, in which the \textbeta-sheets dominate the secondary structure composition. The average results ($\pm$ SD) are reported in Table~\ref{tab:table6} alongside the mean results obtained for the control group and the statistical test results.

\begin{table}[htb]
\caption{\label{tab:table5} Secondary structure composition results obtained for the NIR-exposed polymerized tubulin samples. The last two rows correspond to separate measurements of two different points of the same sample.}
\begin{ruledtabular}
\begin{tabular}{cccccccc}
			\textbf{Sample} & \textbf{\textalpha-Helix}  &  \textbf{\textbeta-Sheet}	&  \textbf{\textbeta-Turn}    & \textbf{Random Coil} \\
			\colrule

           Exposed~1 & $15.0\%$ & $57.9\%$ & $8.7\%$ &  $18.4\%$ \\

           Exposed~2-1 & $13.8\%$ & $45.5\%$ &  $24.0\%$ & $16.6\%$ \\

           Exposed~2-2 & $13.0\%$ & $59.5\%$ & $17.2\%$ &  $10.3\%$ \\

\end{tabular}
\end{ruledtabular}
\end{table}

\begin{table*}[htb]
\caption{\label{tab:table6} Comparison of secondary structure percentages obtained for control versus NIR-exposed and polymerized tubulin samples. Statistically significant differences between the mean percentages of secondary structures in both groups were determined by Welch's unequal variances $t$-tests.}
\begin{ruledtabular}
\begin{tabular}{ccccc}
			\textbf{Group Analysis}	& \textbf{\textalpha-Helix} 	& \textbf{\textbeta-Sheet}     & \textbf{\textbeta-Turn}  & \textbf{Random Coil} \\
			\colrule

			Control	&	$36.0 \pm 4.2\%$		&  $26.7 \pm 7.2\%$ & 	$16.4 \pm 5.0\%$		& $21.0 \pm 2.2\%$  \\

			Exposed tubulin		&  $13.9 \pm 1.0\%$		& $54.3 \pm 7.7\%$ & 	$16.6 \pm 7.7\%$		& $15.1 \pm 4.3\%$ \\

   			$p$-value &  $0.00887$ & $0.0105$ & $0.967$ &  $0.123$ \\

         	Significance &  ** & * & ns & ns \\

\end{tabular}
    \begin{tablenotes}
      \small
      \item ** Highly statistically significant, $0.001 \leq p < 0.01$; * statistically significant, $0.01 \leq p < \alpha$; ns, not significant, $p \geq \alpha$.
    \end{tablenotes}
\end{ruledtabular}
\end{table*}

Compared to the unexposed samples, marked differences can be observed: the NIR-exposed tubulin samples present average \textalpha-helix and \textbeta-sheet contents of $13.9 \pm 1.0\%$ and $54.3 \pm 7.7\%$, respectively. The results of Welch's unequal variances $t$-tests indicate statistically significant differences in the mean values of \textbeta-sheets ($p=0.0105$) and \textalpha-helices ($p=0.00887$). The differences in \textbeta-turn and random coil (undefined) structures were not found to be statistically significant. The complete results of these statistical tests can be found in Table~S8 of the Supplementary Material~\cite{RamanSuppMat}. In conclusion, we found that in vitro exposure of tubulin to pulsed low-level NIR radiation led to a highly statistically significant reduction in \textalpha-helices and a concurrent statistically significant increase in \textbeta-sheets.

\section{\label{sec:Disc}Discussion}
The Raman spectroscopy results of this study appear to indicate that low-intensity, non-ionizing NIR radiation interacts with tubulin at the molecular level, modifying its secondary structures, as evidenced by significant changes in the amide~I band. This particular Raman mode is highly responsive to alterations in the hydrogen bonding strength between $\mathrm{N}$--$\mathrm{H}$ and $\mathrm{C}$=$\mathrm{O}$ groups~\cite{Kuhar2021}. In this region, the Raman spectra of the NIR-exposed and subsequently polymerized tubulin samples illustrate the transformation from a conformation dominated by \textalpha-helical content to one dominated by \textbeta-sheets. Gautam et al.~\cite{doi:10.1021/ac503647x}, in their Raman spectroscopic study of various gene mutants, accounted for such a loss of \textalpha-helical content in other proteins as ``structural unfolding and/or denaturation'' (where the mechanism underlying unfolding is likely the breaking of weak $\mathrm{H}$-bonds~\cite{Ackbarow2007}). At the same time, they justified \textbeta-sheet formation as the result of the interaction between exposed hydrophobic residues of different molecules with each other~\cite{doi:10.1021/ac503647x}.

Another example of the same protein unfolding--refolding behavior is reported by Perillo et al.~\cite{s20020497}. In their work concerning the impact of mechanical forces on protein structures, they witnessed a reduced Amide~I band signal intensity and a decrease in \textalpha-helix content in response to the applied strain forces~\cite{s20020497}. These effects were consistent with results obtained in a similar prior study conducted by the same authors~\cite{Camerlingo2014}. Independently, this behavior had also been previously demonstrated in another Raman spectroscopy study, reporting an \textalpha-helix to \textbeta-sheet transition in the proteins of strained keratin fibers, evidenced by a progressive increase in \textbeta-sheet content and decrease in \textalpha-helices as a function of the applied strain intensity~\cite{Paquin2007}.

Various studies in the literature also highlight interesting ways in which electromagnetic stimuli can induce conformational changes in proteins~\cite{Beyer2008_2,Afrasiabi2013,Bekard2014,Lundholm2015,Maugeri2018,Darwish2020}. For instance, using X-ray crystallography, Lundholm et al. observed terahertz-radiation-induced ($0.4$~THz, $62$~mW/cm$^{2}$) steady-state secondary structure changes in lysozymes characterized by \textalpha-helix compression, which the authors attributed to resonant interactions~\cite{Lundholm2015}. In another study, which investigated the effects of extremely low-frequency magnetic fields ($-2.4$--$2.4$~mT, $50$~Hz, $5.0$~min exposure) on cAMP response element-binding protein (CREB), FTIR spectroscopic analysis revealed lasting conformational changes evidenced by varying spectral band shifts in the amide~II, IV, and VI regions~\cite{Darwish2020}. Similarly, in our study, it is possible to observe the rearrangement in the NIR-exposed tubulin samples' secondary structure. Furthermore, our findings, based on non-simultaneous NIR exposure and acquisition of spectra, also indicate potential long-term effects or extended relaxation times. Interestingly, in a recent in vivo study on the effects of transcranial PBM on human subjects, significant changes in EEG were observed both $10$ and $30$~min after treatment (using the same device employed in this study)~\cite{Zomorrodi2021}.

While there is no other study in the literature investigating the effects of pulsed NIR light on the secondary structures of tubulin or MTs, we can draw comparisons with the Afrasiabi et al. study~\cite{Afrasiabi2013}, which, to the best of our knowledge, appears to be the most closely related one. Specifically, they performed far-UV CD spectroscopy on tubulin dimers ($2$~mg/mL) exposed in vitro for $30$~min to extremely low-frequency electromagnetic fields (ELF-EMFs) with frequencies of $50$, $100$, and $217$~Hz and an intensity of $0.2$~mT~\cite{Afrasiabi2013}. Consistent with our results for NIR-exposed tubulin with the same exposure time and a similar concentration ($2.5$~mg/mL), their corresponding secondary structures analyses of CD data revealed a reduction in \textalpha-helices and an increase in \textbeta-sheets for all three frequencies of ELF-EMFs studied~\cite{Afrasiabi2013}. In addition to their CD spectroscopy analyses, several other techniques were employed, including transmission electron microscopy and turbidity assays, to study MT polymerization. Results from both methods displayed a reduction in polymerization and an increase in the nucleation time (``lag'' phase) observed for the exposed tubulin samples for each of the ELF-EMFs studied, which the authors associated with the secondary structure alterations induced by exposure~\cite{Afrasiabi2013}.

These findings are consistent with the results of our previous study~\cite{Staelens2022}, in which we also employed turbidity measurements to explore how exposure of tubulin to the same PBM device used in this study affects its polymerization into MTs. For the same tubulin concentration and exposure conditions used in this study, we obtained a reduction in the polymerization rate and the final polymer mass of the exposed tubulin samples compared to the control samples, as well as an increase in the time required to produce $10\%$ of the maximal value of polymer (i.e., an increased nucleation phase)~\cite{Staelens2022}. This reduction in the polymerization rate is a key factor in determining if MTs will continue to grow or start to shrink. In particular, if the addition of new GTP-bound tubulin molecules occurs faster than the rate of GTP hydrolysis, the GTP cap is maintained, and MT growth will continue; conversely, if the polymerization rate falls below that of GTP hydrolysis, the tubulin--GTP subunit at the growing end of the MT will undergo hydrolysis, leading to the catastrophe phenomenon~\cite{Cooper2000}. Thus, the reduced rate of polymerization observed in the NIR-exposed tubulin samples could pose consequences for the stability of the ensuing MTs.

Additionally, in the Raman spectroscopic study of free and polymerized tubulin conducted by Audenaert et al., the GTP- and GDP-bound states of tubulin dimers were distinguished by a significant decrease in ordered \textalpha-helices and a concurrent increase in antiparallel \textbeta-sheets observed in the latter state~\cite{Audenaert1989}. In connection with this study, this suggests the possibility that the MTs assembled from NIR-exposed tubulin in our study might contain a larger proportion of GDP-bound tubulin in the MT lattice, again leading to increased MT instability.

These studies indicate a plausible relationship between NIR-exposure-induced changes in secondary structures and tubulin/MT polymerization dynamics. Based on these results, we present several hypotheses regarding this connection. First, we note that the reduced overall MT polymer mass measured for the NIR-exposed tubulin samples in our previous study has two plausible interpretations: 1) less polymerization occurred compared to the control samples, or 2) the MTs assembled from NIR-exposed tubulin were less stable, leading to increased MT disassembly and hence a lower final polymer mass measured. Likewise, in line with the first interpretation, we propose that the NIR-induced changes in secondary structural elements, in particular, the reduction in \textalpha-helices, cause a reduced polymerization rate and hindrance of nucleation that ultimately affect MT growth dynamics. Alternatively, in accordance with the second interpretation, we hypothesize that the induced conformational changes lead to reduced MT stability. A third likely possibility is that all of these effects occur simultaneously. Either way, these results have implications regarding the mechanisms underlying the efficacy of PBM in treating NDs such as AD.

In a recent study by Peris et al.~\cite{Peris2022}, both ex vivo and in vivo experiments demonstrated that a characteristic aspect of MT dysfunction in early AD is that they become overly stable, which hinders neuronal activity. The authors initially demonstrated this through post-mortem analyses of brain samples from AD patients, which exhibited increased levels of detyrosinated tubulin compared to samples collected from individuals without the disease~\cite{Peris2022}. Over time, the C-terminal end of \textalpha-tubulin in MTs naturally undergoes detyrosination; thus, high levels of tyrosinated tubulin are typically found in young, dynamic MTs, whereas detyrosinated tubulin is representative of aged, long-lived MTs. Further experiments performed using a heterozygous mouse model subjected to inhibited MT rejuvenation via downregulation of the tubulin tyrosine ligase gene---resulting in mice with an increased proportion of aged neuronal MTs---revealed consequential memory deficiencies and reduced synaptic content~\cite{Peris2022}.

Similarly, in the study by Muhia et al., neurons from memory- and learning-impaired mice with an inactivated kinesin family member 21B gene (responsible for encoding a kinesin motor protein involved in synaptic vesicle transport along neuronal MTs) also displayed evidence of impaired microtubule dynamics~\cite{MUHIA2016}. Together, their results suggest that while a certain amount of stable MTs is needed for cellular support, a significant proportion of dynamic MTs is also necessary for proper brain function, and a disruption in this equilibrium is detrimental. In fact, for the brain to encode new memories, a precise balance between dynamic and stable microtubules in neurons appears to be foundational. Thus, we hypothesize that through NIR PBM therapy in AD patients, where this balance is disrupted, induced changes in tubulin secondary structures leading to altered polymerization dynamics and reduced MT stability could promote MT depolymerization and encourage cytoskeletal remodeling by enabling the replacement of old, overly stable MTs with new dynamic MTs.

Typically, a pronounced increase in \textbeta-sheets would be cause for concern, as it might exacerbate the pathogenesis of NDs. This is grounded in the understanding that, in many functional proteins, the conversion of normal \textalpha-helix structures into \textbeta-sheets is often linked to protein misfolding and aggregation connected with the formation of amyloids~\cite{Ashraf2014}. Tubulins, however, have not been traditionally associated with amyloid formation, nor are they generally known for their propensity to form amyloids; thus, they are not currently recognized as typical amyloid-forming proteins. In this context, the highlighted observations regarding altered tubulin dynamics and PBM-induced delayed polymerization become particularly significant. The notable positive clinical outcomes from PBM suggest that the altered dynamics might contribute to the remodeling and renewal of the MT structures. This seems to mitigate the risks of misfolding and aggregation. Such an effect might be analogous to counteracting brain aging, offering potential benefits against AD and other NDs.

It is important to recognize that although PBM can modulate cellular processes, as demonstrated in this in vitro study, potentially leading to the renewal of microtubules, the outcomes in a living individual with a complex physiological backdrop might be much more variable and unpredictable. An important caveat is that, even with high power, NIR light's penetration into tissues is shallow, often under $2$~cm~\cite{Koster2022}, and its effects in deeper regions are largely dependent on indirect signaling pathways and other mechanisms that are not yet fully understood. While positive clinical outcomes have been reported, they stem from small studies. The mechanisms underlying PBM's impact on tubulin secondary structures, as well as these other influencing factors, still require more comprehensive investigations.

This study has several limitations, such as the relatively small number of samples analyzed and experiments performed, which led to sizable uncertainties in the mean values of secondary structures reported. Replication experiments and measurements of additional Raman modes, such as the amide~III band and $\mathrm{N}$--$\mathrm{H}$ stretching region, to confirm and further characterize the changes in secondary structures and probe the possible breaking of $\mathrm{H}$-bonds are warranted. Additionally, the buffer solution in which the resuspended tubulin was exposed is only an approximation to the intracellular environment. A key difference is that our experiments were conducted with tubulin in the absence of microtubule-associated proteins (MAPs), which appear to play a critical role in AD, especially MAP tau~\cite{IQBAL2005,Iqbal2010}. In particular, the inclusion of MAP--tubulin interactions may affect the observed changes in secondary structures induced by NIR radiation (and vice versa), which deserves future study. Moreover, in AD, there appears to be a notable acidification of the intracellular \pH\ (\pH$_{\mathrm{i}}$) ~\cite{LYROS2020, DECKER2021} associated with a decrease in mitochondrial respiration and connected with a reduction in neuronal activity~\cite{Schwartz2020}. Tubulin and MTs are highly sensitive to changes in \pH, and the \pH\ of the environment has a significant effect on their behavior and conformation~\cite{Ventilla1972, Audenaert1989}. Thus, it would be interesting and valuable to conduct further experiments that investigate how the NIR-induced changes in secondary structures observed in this study might vary as a function of \pH. Additionally, the persistence of the conformational changes reported in this study over longer timescales and the potential occurrence of protein refolding also merit further investigation. Lastly, single-cell Raman spectroscopy of NIR-exposed live neuronal cells aimed at studying further possible changes in protein activity at the single neural cell level presents an additional future direction for this research with potential value for the PBM community.

\section{\label{sec:Conc}Conclusions}
This study addressed the scarcity of data and conflicting reports in the literature regarding the secondary structures of tubulin in the polymerized state. Additionally, as the current literature is replete with increasing evidence of the effects of electromagnetic fields on relatively simple structures such as tubulin and microtubules, we sought to investigate potential conformational changes due to exposure to the low-intensity pulsed NIR radiation typically exploited in PBM. By employing Raman spectroscopy and subsequent spectral decomposition of the measured amide~I spectra of polymerized tubulin samples using a Voigt profile model, this research contributed to our understanding of the conformation of polymerized tubulin.

Although spectral deconvolution of Raman spectra based on a Voigt profile has been used previously in several studies to accurately quantify secondary structures of various proteins, this is the first time it has been applied to tubulin or MTs. Our secondary structure results obtained through Raman spectroscopy confirm the findings of a previous study and help to reconcile the disparities among reported values in the literature. Furthermore, the observed consistency in results obtained served to validate our methodology for this specific context.

Based on our analysis of the irradiated samples, we reported novel findings on the impact of pulsed low-intensity NIR radiation on the secondary structures of polymerized tubulin. We observed a statistically significant decrease in \textalpha-helix content and an increase in \textbeta-sheets following in vitro exposure to a net dose of $22.5$~J/cm$^{2}$ from an $810$~nm LED pulsed at a rate of $10$~Hz. While there are risks associated with an excessive increase in \textbeta-sheets in the context of neurodegenerative diseases, related clinical evidence supports an interpretation that the PBM-induced conformational changes in tubulins could lead to refreshed microtubule structures and possibly delay the aging process.

These structural alterations directly influence the polymerization kinetics of tubulin and microtubules, suggesting potential implications for the efficacy of NIR PBM, particularly when considering potential applications for Alzheimer's disease and related dementia. The observed remodeling appears to offer a refreshment or renewal of microtubules. This study serves not only to bridge key knowledge gaps in the existing literature but also to propose insights into the potential mechanisms by which NIR PBM might be beneficial, especially in the context of exploring therapeutic options for Alzheimer's disease and other neurodegenerative disorders. Further investigations into PBM's mechanisms are essential to better understand and possibly harness its therapeutic potential for neurodegenerative diseases.

\begin{acknowledgments}
Part of this work was conducted at the University of Alberta nanoFAB Centre. We wish to express our thanks to the funders and maintainers of this facility. We also acknowledge with thanks Prof. John Lewis for lab access and bench space and Katia Carmine-Simmen for her diligent management of the lab. Lastly, E.D.G. acknowledges Kazi Alam's assistance in utilizing the Raman microscope.
\\
\end{acknowledgments}

\paragraph*{Funding:}
E.D.G., J.A.T., and M.S. are grateful to Vielight Inc. for partial funding. J.A.T. thanks the Natural Sciences and Engineering Research Council of Canada (NSERC) for partial financial support under Discovery Grant No. RGPIN-2018-03837.
\\
\paragraph*{Supplementary Materials:}
The supplemental material document associated with this article is available online at \href{https://doi.org/10.6084/m9.figshare.24499825}{10.6084/m9.figshare.24499825}.
\\
\paragraph*{Author Contributions:}
Conceptualization, K.S. and J.A.T.; methodology, E.D.G. and K.S.; validation, E.D.G. and M.S.; formal analysis, E.D.G. and M.S.; investigation, E.D.G.; resources, N.H., M.K., J.L., L.L., K.S., and J.A.T.; data curation, E.D.G.; writing---original draft preparation, E.D.G. and M.S.; writing---review and editing, E.D.G., N.H., M.K., J.L., L.L., J.A.T., and M.S.; visualization, E.D.G. and M.S.; supervision, K.S. and J.A.T.; project administration, K.S. and J.A.T.; funding acquisition, L.L. and J.A.T. All authors have read and agreed to the published version of the manuscript. 
\\
\paragraph*{Conflicts of Interest:}
N.H., M.K., J.L., and L.L. are employed by Vielight Inc. This study was funded, in part, by Vielight Inc. This funder and the employees had limited involvement in the study; they provided the Neuro Alpha device and details regarding the specifications, setup, and usage of the device and contributed to the editing of the manuscript, particularly relating to PBM, AD, and neurodegeneration. This funder had no role in the design of the study, in the collection and curation of data, or in the subsequent analyses. The remaining authors declare that the research was conducted in the absence of any commercial or financial relationships that could be construed as potential conflicts of interest.
\\
\paragraph*{Data Availability:}
The raw data associated with this study are openly available through Figshare at the following DOI: \href{https://doi.org/10.6084/m9.figshare.24492100}{10.6084/m9.figshare.24492100}.

\bibliography{RamanTubulin}

\begin{thebibliography}{123}%
\makeatletter
\providecommand \@ifxundefined [1]{%
 \@ifx{#1\undefined}
}%
\providecommand \@ifnum [1]{%
 \ifnum #1\expandafter \@firstoftwo
 \else \expandafter \@secondoftwo
 \fi
}%
\providecommand \@ifx [1]{%
 \ifx #1\expandafter \@firstoftwo
 \else \expandafter \@secondoftwo
 \fi
}%
\providecommand \natexlab [1]{#1}%
\providecommand \enquote  [1]{``#1''}%
\providecommand \bibnamefont  [1]{#1}%
\providecommand \bibfnamefont [1]{#1}%
\providecommand \citenamefont [1]{#1}%
\providecommand \href@noop [0]{\@secondoftwo}%
\providecommand \href [0]{\begingroup \@sanitize@url \@href}%
\providecommand \@href[1]{\@@startlink{#1}\@@href}%
\providecommand \@@href[1]{\endgroup#1\@@endlink}%
\providecommand \@sanitize@url [0]{\catcode `\\12\catcode `\$12\catcode `\&12\catcode `\#12\catcode `\^12\catcode `\_12\catcode `\%12\relax}%
\providecommand \@@startlink[1]{}%
\providecommand \@@endlink[0]{}%
\providecommand \url  [0]{\begingroup\@sanitize@url \@url }%
\providecommand \@url [1]{\endgroup\@href {#1}{\urlprefix }}%
\providecommand \urlprefix  [0]{URL }%
\providecommand \Eprint [0]{\href }%
\providecommand \doibase [0]{https://doi.org/}%
\providecommand \selectlanguage [0]{\@gobble}%
\providecommand \bibinfo  [0]{\@secondoftwo}%
\providecommand \bibfield  [0]{\@secondoftwo}%
\providecommand \translation [1]{[#1]}%
\providecommand \BibitemOpen [0]{}%
\providecommand \bibitemStop [0]{}%
\providecommand \bibitemNoStop [0]{.\EOS\space}%
\providecommand \EOS [0]{\spacefactor3000\relax}%
\providecommand \BibitemShut  [1]{\csname bibitem#1\endcsname}%
\let\auto@bib@innerbib\@empty
\bibitem [{\citenamefont {Hiller}\ and\ \citenamefont {Weber}(1978)}]{HILLER1978795}%
  \BibitemOpen
  \bibfield  {author} {\bibinfo {author} {\bibfnamefont {G.}~\bibnamefont {Hiller}}\ and\ \bibinfo {author} {\bibfnamefont {K.}~\bibnamefont {Weber}},\ }\bibfield  {title} {\bibinfo {title} {{Radioimmunoassay for tubulin: a quantitative comparison of the tubulin content of different established tissue culture cells and tissues}},\ }\href {https://doi.org/10.1016/0092-8674(78)90335-5} {\bibfield  {journal} {\bibinfo  {journal} {Cell}\ }\textbf {\bibinfo {volume} {14}},\ \bibinfo {pages} {795} (\bibinfo {year} {1978})},\ \bibinfo {note} {{\href{https://pubmed.ncbi.nlm.nih.gov/688394/}{PMID: 688394}}}\BibitemShut {NoStop}%
\bibitem [{\citenamefont {Oakley}(2000)}]{OAKLEY2000}%
  \BibitemOpen
  \bibfield  {author} {\bibinfo {author} {\bibfnamefont {B.~R.}\ \bibnamefont {Oakley}},\ }\bibfield  {title} {\bibinfo {title} {{An abundance of tubulins}},\ }\href {https://doi.org/10.1016/S0962-8924(00)01857-2} {\bibfield  {journal} {\bibinfo  {journal} {Trends Cell Biol.}\ }\textbf {\bibinfo {volume} {10}},\ \bibinfo {pages} {537} (\bibinfo {year} {2000})},\ \bibinfo {note} {{\href{https://pubmed.ncbi.nlm.nih.gov/11121746/}{PMID: 11121746}}}\BibitemShut {NoStop}%
\bibitem [{\citenamefont {Fourest-Lieuvin}(2006)}]{FOURESTLIEUVIN2006183}%
  \BibitemOpen
  \bibfield  {author} {\bibinfo {author} {\bibfnamefont {A.}~\bibnamefont {Fourest-Lieuvin}},\ }\bibfield  {title} {\bibinfo {title} {{Purification of tubulin from limited volumes of cultured cells}},\ }\href {https://doi.org/10.1016/j.pep.2005.05.011} {\bibfield  {journal} {\bibinfo  {journal} {Protein Expr. Purif.}\ }\textbf {\bibinfo {volume} {45}},\ \bibinfo {pages} {183} (\bibinfo {year} {2006})},\ \bibinfo {note} {{\href{https://pubmed.ncbi.nlm.nih.gov/16027005/}{PMID: 16027005}}}\BibitemShut {NoStop}%
\bibitem [{\citenamefont {Verdier-Pinard}\ \emph {et~al.}(2009)\citenamefont {Verdier-Pinard}, \citenamefont {Pasquier}, \citenamefont {Xiao}, \citenamefont {Burd}, \citenamefont {Villard}, \citenamefont {Lafitte}, \citenamefont {Miller}, \citenamefont {Angeletti}, \citenamefont {Horwitz},\ and\ \citenamefont {Braguer}}]{VERDIERPINARD2009197}%
  \BibitemOpen
  \bibfield  {author} {\bibinfo {author} {\bibfnamefont {P.}~\bibnamefont {Verdier-Pinard}}, \bibinfo {author} {\bibfnamefont {E.}~\bibnamefont {Pasquier}}, \bibinfo {author} {\bibfnamefont {H.}~\bibnamefont {Xiao}}, \bibinfo {author} {\bibfnamefont {B.}~\bibnamefont {Burd}}, \bibinfo {author} {\bibfnamefont {C.}~\bibnamefont {Villard}}, \bibinfo {author} {\bibfnamefont {D.}~\bibnamefont {Lafitte}}, \bibinfo {author} {\bibfnamefont {L.~M.}\ \bibnamefont {Miller}}, \bibinfo {author} {\bibfnamefont {R.~H.}\ \bibnamefont {Angeletti}}, \bibinfo {author} {\bibfnamefont {S.~B.}\ \bibnamefont {Horwitz}},\ and\ \bibinfo {author} {\bibfnamefont {D.}~\bibnamefont {Braguer}},\ }\bibfield  {title} {\bibinfo {title} {{Tubulin proteomics: Towards breaking the code}},\ }\href {https://doi.org/10.1016/j.ab.2008.09.020} {\bibfield  {journal} {\bibinfo  {journal} {Anal. Biochem.}\ }\textbf {\bibinfo {volume} {384}},\ \bibinfo {pages} {197} (\bibinfo {year} {2009})},\ \bibinfo {note}
  {{\href{https://pubmed.ncbi.nlm.nih.gov/18840397/}{PMID: 18840397}}}\BibitemShut {NoStop}%
\bibitem [{\citenamefont {Widlund}\ \emph {et~al.}(2012)\citenamefont {Widlund}, \citenamefont {Podolski}, \citenamefont {Reber}, \citenamefont {Alper}, \citenamefont {Storch}, \citenamefont {Hyman}, \citenamefont {Howard},\ and\ \citenamefont {Drechsel}}]{Widlund2012}%
  \BibitemOpen
  \bibfield  {author} {\bibinfo {author} {\bibfnamefont {P.~O.}\ \bibnamefont {Widlund}}, \bibinfo {author} {\bibfnamefont {M.}~\bibnamefont {Podolski}}, \bibinfo {author} {\bibfnamefont {S.}~\bibnamefont {Reber}}, \bibinfo {author} {\bibfnamefont {J.}~\bibnamefont {Alper}}, \bibinfo {author} {\bibfnamefont {M.}~\bibnamefont {Storch}}, \bibinfo {author} {\bibfnamefont {A.~A.}\ \bibnamefont {Hyman}}, \bibinfo {author} {\bibfnamefont {J.}~\bibnamefont {Howard}},\ and\ \bibinfo {author} {\bibfnamefont {D.~N.}\ \bibnamefont {Drechsel}},\ }\bibfield  {title} {\bibinfo {title} {{One-step purification of assembly-competent tubulin from diverse eukaryotic sources}},\ }\href {https://doi.org/10.1091/mbc.e12-06-0444} {\bibfield  {journal} {\bibinfo  {journal} {Mol. Biol. Cell}\ }\textbf {\bibinfo {volume} {23}},\ \bibinfo {pages} {4393} (\bibinfo {year} {2012})},\ \bibinfo {note} {{\href{https://pubmed.ncbi.nlm.nih.gov/22993214/}{PMID: 22993214}}}\BibitemShut {NoStop}%
\bibitem [{\citenamefont {Khamis}\ and\ \citenamefont {Heikkila}(2018)}]{KHAMIS20181}%
  \BibitemOpen
  \bibfield  {author} {\bibinfo {author} {\bibfnamefont {I.}~\bibnamefont {Khamis}}\ and\ \bibinfo {author} {\bibfnamefont {J.~J.}\ \bibnamefont {Heikkila}},\ }\bibfield  {title} {\bibinfo {title} {{Effect of isothiocyanates, BITC and PEITC, on stress protein accumulation, protein aggregation and aggresome-like structure formation in Xenopus A6 kidney epithelial cells}},\ }\href {https://doi.org/10.1016/j.cbpc.2017.10.011} {\bibfield  {journal} {\bibinfo  {journal} {Comp. Biochem. Physiol. C Toxicol. Pharmacol.}\ }\textbf {\bibinfo {volume} {204}},\ \bibinfo {pages} {1} (\bibinfo {year} {2018})},\ \bibinfo {note} {{\href{https://pubmed.ncbi.nlm.nih.gov/29100952/}{PMID: 29100952}}}\BibitemShut {NoStop}%
\bibitem [{\citenamefont {Luduena}\ and\ \citenamefont {Woodward}(1973)}]{Luduena1973}%
  \BibitemOpen
  \bibfield  {author} {\bibinfo {author} {\bibfnamefont {R.~F.}\ \bibnamefont {Luduena}}\ and\ \bibinfo {author} {\bibfnamefont {D.~O.}\ \bibnamefont {Woodward}},\ }\bibfield  {title} {\bibinfo {title} {{Isolation and Partial Characterization of $\upalpha$- and $\upbeta$-Tubulin from Outer Doublets of Sea-Urchin Sperm and Microtubules of Chick-Embryo Brain}},\ }\href {https://doi.org/10.1073/pnas.70.12.3594} {\bibfield  {journal} {\bibinfo  {journal} {Proc. Natl. Acad. Sci. U.S.A.}\ }\textbf {\bibinfo {volume} {70}},\ \bibinfo {pages} {3594} (\bibinfo {year} {1973})},\ \bibinfo {note} {{\href{https://pubmed.ncbi.nlm.nih.gov/4519648/}{PMID: 4519648}}}\BibitemShut {NoStop}%
\bibitem [{\citenamefont {Valenzuela}\ \emph {et~al.}(1981)\citenamefont {Valenzuela}, \citenamefont {Quiroga}, \citenamefont {Zaldivar}, \citenamefont {Rutter}, \citenamefont {Kirschner},\ and\ \citenamefont {Cleveland}}]{Valenzuela1981}%
  \BibitemOpen
  \bibfield  {author} {\bibinfo {author} {\bibfnamefont {P.}~\bibnamefont {Valenzuela}}, \bibinfo {author} {\bibfnamefont {M.}~\bibnamefont {Quiroga}}, \bibinfo {author} {\bibfnamefont {J.}~\bibnamefont {Zaldivar}}, \bibinfo {author} {\bibfnamefont {W.~J.}\ \bibnamefont {Rutter}}, \bibinfo {author} {\bibfnamefont {M.~W.}\ \bibnamefont {Kirschner}},\ and\ \bibinfo {author} {\bibfnamefont {D.~W.}\ \bibnamefont {Cleveland}},\ }\bibfield  {title} {\bibinfo {title} {{Nucleotide and corresponding amino acid sequences encoded by $\alpha$ and $\beta$ tubulin mRNAs}},\ }\href {https://doi.org/10.1038/289650a0} {\bibfield  {journal} {\bibinfo  {journal} {Nature}\ }\textbf {\bibinfo {volume} {289}},\ \bibinfo {pages} {650} (\bibinfo {year} {1981})},\ \bibinfo {note} {{\href{https://pubmed.ncbi.nlm.nih.gov/7464932/}{PMID: 7464932}}}\BibitemShut {NoStop}%
\bibitem [{\citenamefont {Little}\ \emph {et~al.}(1981)\citenamefont {Little}, \citenamefont {Krauhs},\ and\ \citenamefont {Ponstingl}}]{Little1981}%
  \BibitemOpen
  \bibfield  {author} {\bibinfo {author} {\bibfnamefont {M.}~\bibnamefont {Little}}, \bibinfo {author} {\bibfnamefont {E.}~\bibnamefont {Krauhs}},\ and\ \bibinfo {author} {\bibfnamefont {H.}~\bibnamefont {Ponstingl}},\ }\bibfield  {title} {\bibinfo {title} {{Tubulin sequence conservation}},\ }\href {https://doi.org/10.1016/0303-2647(81)90031-9} {\bibfield  {journal} {\bibinfo  {journal} {Biosystems}\ }\textbf {\bibinfo {volume} {14}},\ \bibinfo {pages} {239} (\bibinfo {year} {1981})},\ \bibinfo {note} {{\href{https://pubmed.ncbi.nlm.nih.gov/7337807/}{PMID: 7337807}}}\BibitemShut {NoStop}%
\bibitem [{\citenamefont {Nogales}\ \emph {et~al.}(1998)\citenamefont {Nogales}, \citenamefont {Wolf},\ and\ \citenamefont {Downing}}]{Nogales1998}%
  \BibitemOpen
  \bibfield  {author} {\bibinfo {author} {\bibfnamefont {E.}~\bibnamefont {Nogales}}, \bibinfo {author} {\bibfnamefont {S.~G.}\ \bibnamefont {Wolf}},\ and\ \bibinfo {author} {\bibfnamefont {K.~H.}\ \bibnamefont {Downing}},\ }\bibfield  {title} {\bibinfo {title} {{Structure of the $\upalpha$$\upbeta$ tubulin dimer by electron crystallography}},\ }\href {https://doi.org/10.1038/34465} {\bibfield  {journal} {\bibinfo  {journal} {Nature}\ }\textbf {\bibinfo {volume} {391}},\ \bibinfo {pages} {199} (\bibinfo {year} {1998})},\ \bibinfo {note} {{\href{https://pubmed.ncbi.nlm.nih.gov/9428769/}{PMID: 9428769}}. Erratum in \href{https://doi.org/10.1038/30288}{Nature \textbf{393}, 191 (1998)}}\BibitemShut {NoStop}%
\bibitem [{\citenamefont {Haimov}\ and\ \citenamefont {Srebnik}(2016)}]{Haimov2016}%
  \BibitemOpen
  \bibfield  {author} {\bibinfo {author} {\bibfnamefont {B.}~\bibnamefont {Haimov}}\ and\ \bibinfo {author} {\bibfnamefont {S.}~\bibnamefont {Srebnik}},\ }\bibfield  {title} {\bibinfo {title} {{A closer look into the $\upalpha$-helix basin}},\ }\href {https://doi.org/10.1038/srep38341} {\bibfield  {journal} {\bibinfo  {journal} {Sci. Rep.}\ }\textbf {\bibinfo {volume} {6}},\ \bibinfo {pages} {38341} (\bibinfo {year} {2016})},\ \bibinfo {note} {{\href{https://pubmed.ncbi.nlm.nih.gov/27917894/}{PMID: 27917894}}}\BibitemShut {NoStop}%
\bibitem [{\citenamefont {Audenaert}\ \emph {et~al.}(1989)\citenamefont {Audenaert}, \citenamefont {Heremans}, \citenamefont {Heremans},\ and\ \citenamefont {Engelborghs}}]{Audenaert1989}%
  \BibitemOpen
  \bibfield  {author} {\bibinfo {author} {\bibfnamefont {R.}~\bibnamefont {Audenaert}}, \bibinfo {author} {\bibfnamefont {L.}~\bibnamefont {Heremans}}, \bibinfo {author} {\bibfnamefont {K.}~\bibnamefont {Heremans}},\ and\ \bibinfo {author} {\bibfnamefont {Y.}~\bibnamefont {Engelborghs}},\ }\bibfield  {title} {\bibinfo {title} {{Secondary structure analysis of tubulin and microtubules with Raman spectroscopy}},\ }\href {https://doi.org/10.1016/0167-4838(89)90102-7} {\bibfield  {journal} {\bibinfo  {journal} {Biochim. Biophys. Acta - Protein Struct. Mol. Enzymol.}\ }\textbf {\bibinfo {volume} {996}},\ \bibinfo {pages} {110} (\bibinfo {year} {1989})},\ \bibinfo {note} {{\href{https://pubmed.ncbi.nlm.nih.gov/2736254/}{PMID: 2736254}}}\BibitemShut {NoStop}%
\bibitem [{\citenamefont {Wall}\ \emph {et~al.}(2016)\citenamefont {Wall}, \citenamefont {Pagratis}, \citenamefont {Armstrong}, \citenamefont {Balsbaugh}, \citenamefont {Verbeke}, \citenamefont {Pearson},\ and\ \citenamefont {Hough}}]{Wall2016}%
  \BibitemOpen
  \bibfield  {author} {\bibinfo {author} {\bibfnamefont {K.~P.}\ \bibnamefont {Wall}}, \bibinfo {author} {\bibfnamefont {M.}~\bibnamefont {Pagratis}}, \bibinfo {author} {\bibfnamefont {G.}~\bibnamefont {Armstrong}}, \bibinfo {author} {\bibfnamefont {J.~L.}\ \bibnamefont {Balsbaugh}}, \bibinfo {author} {\bibfnamefont {E.}~\bibnamefont {Verbeke}}, \bibinfo {author} {\bibfnamefont {C.~G.}\ \bibnamefont {Pearson}},\ and\ \bibinfo {author} {\bibfnamefont {L.~E.}\ \bibnamefont {Hough}},\ }\bibfield  {title} {\bibinfo {title} {{Molecular Determinants of Tubulin’s C-Terminal Tail Conformational Ensemble}},\ }\href {https://doi.org/10.1021/acschembio.6b00507} {\bibfield  {journal} {\bibinfo  {journal} {{ACS} Chem. Biol.}\ }\textbf {\bibinfo {volume} {11}},\ \bibinfo {pages} {2981} (\bibinfo {year} {2016})},\ \bibinfo {note} {{\href{https://pubmed.ncbi.nlm.nih.gov/27541566/}{PMID: 27541566}}}\BibitemShut {NoStop}%
\bibitem [{\citenamefont {Marracino}\ \emph {et~al.}(2019)\citenamefont {Marracino}, \citenamefont {Havelka}, \citenamefont {Pr{\r{u}}{\v{s}}a}, \citenamefont {Liberti}, \citenamefont {Tuszynski}, \citenamefont {Ayoub}, \citenamefont {Apollonio},\ and\ \citenamefont {Cifra}}]{Marracino2019}%
  \BibitemOpen
  \bibfield  {author} {\bibinfo {author} {\bibfnamefont {P.}~\bibnamefont {Marracino}}, \bibinfo {author} {\bibfnamefont {D.}~\bibnamefont {Havelka}}, \bibinfo {author} {\bibfnamefont {J.}~\bibnamefont {Pr{\r{u}}{\v{s}}a}}, \bibinfo {author} {\bibfnamefont {M.}~\bibnamefont {Liberti}}, \bibinfo {author} {\bibfnamefont {J.}~\bibnamefont {Tuszynski}}, \bibinfo {author} {\bibfnamefont {A.~T.}\ \bibnamefont {Ayoub}}, \bibinfo {author} {\bibfnamefont {F.}~\bibnamefont {Apollonio}},\ and\ \bibinfo {author} {\bibfnamefont {M.}~\bibnamefont {Cifra}},\ }\bibfield  {title} {\bibinfo {title} {{Tubulin response to intense nanosecond-scale electric field in molecular dynamics simulation}},\ }\href {https://doi.org/10.1038/s41598-019-46636-4} {\bibfield  {journal} {\bibinfo  {journal} {Sci. Rep.}\ }\textbf {\bibinfo {volume} {9}},\ \bibinfo {pages} {10477} (\bibinfo {year} {2019})},\ \bibinfo {note} {{\href{https://pubmed.ncbi.nlm.nih.gov/31324834/}{PMID: 31324834}}}\BibitemShut {NoStop}%
\bibitem [{\citenamefont {Higuchi}\ and\ \citenamefont {Uhlmann}(2005)}]{Higuchi2005}%
  \BibitemOpen
  \bibfield  {author} {\bibinfo {author} {\bibfnamefont {T.}~\bibnamefont {Higuchi}}\ and\ \bibinfo {author} {\bibfnamefont {F.}~\bibnamefont {Uhlmann}},\ }\bibfield  {title} {\bibinfo {title} {{Stabilization of microtubule dynamics at anaphase onset promotes chromosome segregation}},\ }\href {https://doi.org/10.1038/nature03240} {\bibfield  {journal} {\bibinfo  {journal} {Nature}\ }\textbf {\bibinfo {volume} {433}},\ \bibinfo {pages} {171} (\bibinfo {year} {2005})},\ \bibinfo {note} {{\href{https://pubmed.ncbi.nlm.nih.gov/15650742/}{PMID: 15650742}}}\BibitemShut {NoStop}%
\bibitem [{\citenamefont {Laband}\ \emph {et~al.}(2017)\citenamefont {Laband}, \citenamefont {{Le Borgne}}, \citenamefont {Edwards}, \citenamefont {Stefanutti}, \citenamefont {Canman}, \citenamefont {Verbavatz},\ and\ \citenamefont {Dumont}}]{Laband2017}%
  \BibitemOpen
  \bibfield  {author} {\bibinfo {author} {\bibfnamefont {K.}~\bibnamefont {Laband}}, \bibinfo {author} {\bibfnamefont {R.}~\bibnamefont {{Le Borgne}}}, \bibinfo {author} {\bibfnamefont {F.}~\bibnamefont {Edwards}}, \bibinfo {author} {\bibfnamefont {M.}~\bibnamefont {Stefanutti}}, \bibinfo {author} {\bibfnamefont {J.~C.}\ \bibnamefont {Canman}}, \bibinfo {author} {\bibfnamefont {J.-M.}\ \bibnamefont {Verbavatz}},\ and\ \bibinfo {author} {\bibfnamefont {J.}~\bibnamefont {Dumont}},\ }\bibfield  {title} {\bibinfo {title} {{Chromosome segregation occurs by microtubule pushing in oocytes}},\ }\href {https://doi.org/10.1038/s41467-017-01539-8} {\bibfield  {journal} {\bibinfo  {journal} {Nat. Commun.}\ }\textbf {\bibinfo {volume} {8}},\ \bibinfo {pages} {1499} (\bibinfo {year} {2017})},\ \bibinfo {note} {{\href{https://pubmed.ncbi.nlm.nih.gov/29133801/}{PMID: 29133801}}}\BibitemShut {NoStop}%
\bibitem [{\citenamefont {Etienne-Manneville}(2004)}]{Manneville2004}%
  \BibitemOpen
  \bibfield  {author} {\bibinfo {author} {\bibfnamefont {S.}~\bibnamefont {Etienne-Manneville}},\ }\bibfield  {title} {\bibinfo {title} {{Actin and Microtubules in Cell Motility: Which One is in Control?}},\ }\href {https://doi.org/10.1111/j.1600-0854.2004.00196.x} {\bibfield  {journal} {\bibinfo  {journal} {Traffic}\ }\textbf {\bibinfo {volume} {5}},\ \bibinfo {pages} {470} (\bibinfo {year} {2004})},\ \bibinfo {note} {{\href{https://pubmed.ncbi.nlm.nih.gov/15180824/}{PMID: 15180824}}}\BibitemShut {NoStop}%
\bibitem [{\citenamefont {Garcin}\ and\ \citenamefont {Straube}(2019)}]{Garcin2019}%
  \BibitemOpen
  \bibfield  {author} {\bibinfo {author} {\bibfnamefont {C.}~\bibnamefont {Garcin}}\ and\ \bibinfo {author} {\bibfnamefont {A.}~\bibnamefont {Straube}},\ }\bibfield  {title} {\bibinfo {title} {{{Microtubules in cell migration}}},\ }\href {https://doi.org/10.1042/EBC20190016} {\bibfield  {journal} {\bibinfo  {journal} {Essays Biochem.}\ }\textbf {\bibinfo {volume} {63}},\ \bibinfo {pages} {509} (\bibinfo {year} {2019})},\ \bibinfo {note} {{\href{https://pubmed.ncbi.nlm.nih.gov/31358621/}{PMID: 31358621}}}\BibitemShut {NoStop}%
\bibitem [{\citenamefont {Ingber}(2003)}]{Ingber2003}%
  \BibitemOpen
  \bibfield  {author} {\bibinfo {author} {\bibfnamefont {D.~E.}\ \bibnamefont {Ingber}},\ }\bibfield  {title} {\bibinfo {title} {{{Tensegrity I. Cell structure and hierarchical systems biology}}},\ }\href {https://doi.org/10.1242/jcs.00359} {\bibfield  {journal} {\bibinfo  {journal} {J. Cell. Sci.}\ }\textbf {\bibinfo {volume} {116}},\ \bibinfo {pages} {1157} (\bibinfo {year} {2003})},\ \bibinfo {note} {{\href{https://pubmed.ncbi.nlm.nih.gov/12615960/}{PMID: 12615960}}}\BibitemShut {NoStop}%
\bibitem [{\citenamefont {Vale}(2003)}]{Vale2003}%
  \BibitemOpen
  \bibfield  {author} {\bibinfo {author} {\bibfnamefont {R.~D.}\ \bibnamefont {Vale}},\ }\bibfield  {title} {\bibinfo {title} {{The Molecular Motor Toolbox for Intracellular Transport}},\ }\href {https://doi.org/10.1016/S0092-8674(03)00111-9} {\bibfield  {journal} {\bibinfo  {journal} {Cell}\ }\textbf {\bibinfo {volume} {112}},\ \bibinfo {pages} {467} (\bibinfo {year} {2003})},\ \bibinfo {note} {{\href{https://pubmed.ncbi.nlm.nih.gov/12600311/}{PMID: 12600311}}}\BibitemShut {NoStop}%
\bibitem [{\citenamefont {David-Pfeuty}\ \emph {et~al.}(1977)\citenamefont {David-Pfeuty}, \citenamefont {Erickson},\ and\ \citenamefont {Pantaloni}}]{DavidPfeuty1977}%
  \BibitemOpen
  \bibfield  {author} {\bibinfo {author} {\bibfnamefont {T.}~\bibnamefont {David-Pfeuty}}, \bibinfo {author} {\bibfnamefont {H.~P.}\ \bibnamefont {Erickson}},\ and\ \bibinfo {author} {\bibfnamefont {D.}~\bibnamefont {Pantaloni}},\ }\bibfield  {title} {\bibinfo {title} {{Guanosinetriphosphatase activity of tubulin associated with microtubule assembly}},\ }\href {https://doi.org/10.1073/pnas.74.12.5372} {\bibfield  {journal} {\bibinfo  {journal} {Proc. Natl. Acad. Sci. U.S.A.}\ }\textbf {\bibinfo {volume} {74}},\ \bibinfo {pages} {5372} (\bibinfo {year} {1977})},\ \bibinfo {note} {{\href{https://pubmed.ncbi.nlm.nih.gov/202954/}{PMID: 202954}}}\BibitemShut {NoStop}%
\bibitem [{\citenamefont {Roychowdhury}\ and\ \citenamefont {Sierra-Fonseca}(2017)}]{Sierra-Fonseca17}%
  \BibitemOpen
  \bibfield  {author} {\bibinfo {author} {\bibfnamefont {S.}~\bibnamefont {Roychowdhury}}\ and\ \bibinfo {author} {\bibfnamefont {J.~A.}\ \bibnamefont {Sierra-Fonseca}},\ }\bibfield  {title} {\bibinfo {title} {{Heterotrimeric G Proteins and the Regulation of Microtubule Assembly}},\ }in\ \href {https://doi.org/10.5772/66929} {\emph {\bibinfo {booktitle} {Cytoskeleton}}},\ \bibinfo {editor} {edited by\ \bibinfo {editor} {\bibfnamefont {J.~C.}\ \bibnamefont {Jimenez-Lopez}}}\ (\bibinfo  {publisher} {IntechOpen},\ \bibinfo {address} {Rijeka},\ \bibinfo {year} {2017})\ Chap.~\bibinfo {chapter} {14}\BibitemShut {NoStop}%
\bibitem [{\citenamefont {Mitchison}\ and\ \citenamefont {Kirschner}(1984)}]{Mitchison1984}%
  \BibitemOpen
  \bibfield  {author} {\bibinfo {author} {\bibfnamefont {T.}~\bibnamefont {Mitchison}}\ and\ \bibinfo {author} {\bibfnamefont {M.}~\bibnamefont {Kirschner}},\ }\bibfield  {title} {\bibinfo {title} {{Dynamic instability of microtubule growth}},\ }\href {https://doi.org/10.1038/312237a0} {\bibfield  {journal} {\bibinfo  {journal} {Nature}\ }\textbf {\bibinfo {volume} {312}},\ \bibinfo {pages} {237} (\bibinfo {year} {1984})},\ \bibinfo {note} {{\href{https://pubmed.ncbi.nlm.nih.gov/6504138/}{PMID: 6504138}}}\BibitemShut {NoStop}%
\bibitem [{\citenamefont {Walker}\ \emph {et~al.}(1988)\citenamefont {Walker}, \citenamefont {O{\textquotesingle}Brien}, \citenamefont {Pryer}, \citenamefont {Soboeiro}, \citenamefont {Voter}, \citenamefont {Erickson},\ and\ \citenamefont {Salmon}}]{Walker1988}%
  \BibitemOpen
  \bibfield  {author} {\bibinfo {author} {\bibfnamefont {R.~A.}\ \bibnamefont {Walker}}, \bibinfo {author} {\bibfnamefont {E.~T.}\ \bibnamefont {O{\textquotesingle}Brien}}, \bibinfo {author} {\bibfnamefont {N.~K.}\ \bibnamefont {Pryer}}, \bibinfo {author} {\bibfnamefont {M.~F.}\ \bibnamefont {Soboeiro}}, \bibinfo {author} {\bibfnamefont {W.~A.}\ \bibnamefont {Voter}}, \bibinfo {author} {\bibfnamefont {H.~P.}\ \bibnamefont {Erickson}},\ and\ \bibinfo {author} {\bibfnamefont {E.~D.}\ \bibnamefont {Salmon}},\ }\bibfield  {title} {\bibinfo {title} {{Dynamic Instability of Individual Microtubules Analyzed by Video Light Microscopy: Rate Constants and Transition Frequencies}},\ }\href {https://doi.org/10.1083/jcb.107.4.1437} {\bibfield  {journal} {\bibinfo  {journal} {J. Cell Biol.}\ }\textbf {\bibinfo {volume} {107}},\ \bibinfo {pages} {1437} (\bibinfo {year} {1988})},\ \bibinfo {note} {{\href{https://pubmed.ncbi.nlm.nih.gov/3170635/}{PMID: 3170635}}}\BibitemShut {NoStop}%
\bibitem [{\citenamefont {Gundersen}\ \emph {et~al.}(2004)\citenamefont {Gundersen}, \citenamefont {Gomes},\ and\ \citenamefont {Wen}}]{Gundersen2004}%
  \BibitemOpen
  \bibfield  {author} {\bibinfo {author} {\bibfnamefont {G.~G.}\ \bibnamefont {Gundersen}}, \bibinfo {author} {\bibfnamefont {E.~R.}\ \bibnamefont {Gomes}},\ and\ \bibinfo {author} {\bibfnamefont {Y.}~\bibnamefont {Wen}},\ }\bibfield  {title} {\bibinfo {title} {{Cortical control of microtubule stability and polarization}},\ }\href {https://doi.org/10.1016/j.ceb.2003.11.010} {\bibfield  {journal} {\bibinfo  {journal} {Curr. Opin. Cell Biol.}\ }\textbf {\bibinfo {volume} {16}},\ \bibinfo {pages} {106} (\bibinfo {year} {2004})},\ \bibinfo {note} {{\href{https://pubmed.ncbi.nlm.nih.gov/15037313/}{PMID: 15037313}}}\BibitemShut {NoStop}%
\bibitem [{\citenamefont {Desai}\ and\ \citenamefont {Mitchison}(1997)}]{Desai1997}%
  \BibitemOpen
  \bibfield  {author} {\bibinfo {author} {\bibfnamefont {A.}~\bibnamefont {Desai}}\ and\ \bibinfo {author} {\bibfnamefont {T.~J.}\ \bibnamefont {Mitchison}},\ }\bibfield  {title} {\bibinfo {title} {{Microtubule polymerization dynamics}},\ }\href {https://doi.org/10.1146/annurev.cellbio.13.1.83} {\bibfield  {journal} {\bibinfo  {journal} {Annu. Rev. Cell Dev. Biol.}\ }\textbf {\bibinfo {volume} {13}},\ \bibinfo {pages} {83} (\bibinfo {year} {1997})},\ \bibinfo {note} {{\href{https://pubmed.ncbi.nlm.nih.gov/9442869/}{PMID: 9442869}}}\BibitemShut {NoStop}%
\bibitem [{\citenamefont {Pasquier}\ and\ \citenamefont {Kavallaris}(2008)}]{Pasquier2008}%
  \BibitemOpen
  \bibfield  {author} {\bibinfo {author} {\bibfnamefont {E.}~\bibnamefont {Pasquier}}\ and\ \bibinfo {author} {\bibfnamefont {M.}~\bibnamefont {Kavallaris}},\ }\bibfield  {title} {\bibinfo {title} {{Microtubules: A dynamic target in cancer therapy}},\ }\href {https://doi.org/10.1002/iub.25} {\bibfield  {journal} {\bibinfo  {journal} {{IUBMB} Life}\ }\textbf {\bibinfo {volume} {60}},\ \bibinfo {pages} {165} (\bibinfo {year} {2008})},\ \bibinfo {note} {{\href{https://pubmed.ncbi.nlm.nih.gov/18380008/}{PMID: 18380008}}}\BibitemShut {NoStop}%
\bibitem [{\citenamefont {{Albahde}}\ \emph {et~al.}(2021)\citenamefont {{Albahde}}, \citenamefont {Abdrakhimov}, \citenamefont {Li}, \citenamefont {Zhou}, \citenamefont {Zhou}, \citenamefont {Xu}, \citenamefont {Qian},\ and\ \citenamefont {Wang}}]{Albahde2021}%
  \BibitemOpen
  \bibfield  {author} {\bibinfo {author} {\bibfnamefont {M.~A.~H.}\ \bibnamefont {{Albahde}}}, \bibinfo {author} {\bibfnamefont {B.}~\bibnamefont {Abdrakhimov}}, \bibinfo {author} {\bibfnamefont {G.-Q.}\ \bibnamefont {Li}}, \bibinfo {author} {\bibfnamefont {X.}~\bibnamefont {Zhou}}, \bibinfo {author} {\bibfnamefont {D.}~\bibnamefont {Zhou}}, \bibinfo {author} {\bibfnamefont {H.}~\bibnamefont {Xu}}, \bibinfo {author} {\bibfnamefont {H.}~\bibnamefont {Qian}},\ and\ \bibinfo {author} {\bibfnamefont {W.}~\bibnamefont {Wang}},\ }\bibfield  {title} {\bibinfo {title} {{The Role of Microtubules in Pancreatic Cancer: Therapeutic Progress}},\ }\href {https://doi.org/10.3389/fonc.2021.640863} {\bibfield  {journal} {\bibinfo  {journal} {Front. Oncol.}\ }\textbf {\bibinfo {volume} {11}},\ \bibinfo {pages} {640863} (\bibinfo {year} {2021})},\ \bibinfo {note} {{\href{https://pubmed.ncbi.nlm.nih.gov/34094924/}{PMID: 34094924}}}\BibitemShut {NoStop}%
\bibitem [{\citenamefont {Kalra}\ \emph {et~al.}(2020{\natexlab{a}})\citenamefont {Kalra}, \citenamefont {Eakins}, \citenamefont {Patel}, \citenamefont {Ciniero}, \citenamefont {Rezania}, \citenamefont {Shankar},\ and\ \citenamefont {Tuszy\'nski}}]{Kalra2020}%
  \BibitemOpen
  \bibfield  {author} {\bibinfo {author} {\bibfnamefont {A.~P.}\ \bibnamefont {Kalra}}, \bibinfo {author} {\bibfnamefont {B.~B.}\ \bibnamefont {Eakins}}, \bibinfo {author} {\bibfnamefont {S.~D.}\ \bibnamefont {Patel}}, \bibinfo {author} {\bibfnamefont {G.}~\bibnamefont {Ciniero}}, \bibinfo {author} {\bibfnamefont {V.}~\bibnamefont {Rezania}}, \bibinfo {author} {\bibfnamefont {K.}~\bibnamefont {Shankar}},\ and\ \bibinfo {author} {\bibfnamefont {J.~A.}\ \bibnamefont {Tuszy\'nski}},\ }\bibfield  {title} {\bibinfo {title} {{All Wired Up: An Exploration of the Electrical Properties of Microtubules and Tubulin}},\ }\href {https://doi.org/10.1021/acsnano.0c06945} {\bibfield  {journal} {\bibinfo  {journal} {ACS Nano}\ }\textbf {\bibinfo {volume} {14}},\ \bibinfo {pages} {16301} (\bibinfo {year} {2020}{\natexlab{a}})},\ \bibinfo {note} {{\href{https://pubmed.ncbi.nlm.nih.gov/33213135/}{PMID: 33213135}}}\BibitemShut {NoStop}%
\bibitem [{\citenamefont {Kalra}\ \emph {et~al.}(2020{\natexlab{b}})\citenamefont {Kalra}, \citenamefont {Patel}, \citenamefont {Bhuiyan}, \citenamefont {Preto}, \citenamefont {Scheuer}, \citenamefont {Mohammed}, \citenamefont {Lewis}, \citenamefont {Rezania}, \citenamefont {Shankar},\ and\ \citenamefont {Tuszynski}}]{nano10020265}%
  \BibitemOpen
  \bibfield  {author} {\bibinfo {author} {\bibfnamefont {A.~P.}\ \bibnamefont {Kalra}}, \bibinfo {author} {\bibfnamefont {S.~D.}\ \bibnamefont {Patel}}, \bibinfo {author} {\bibfnamefont {A.~F.}\ \bibnamefont {Bhuiyan}}, \bibinfo {author} {\bibfnamefont {J.}~\bibnamefont {Preto}}, \bibinfo {author} {\bibfnamefont {K.~G.}\ \bibnamefont {Scheuer}}, \bibinfo {author} {\bibfnamefont {U.}~\bibnamefont {Mohammed}}, \bibinfo {author} {\bibfnamefont {J.~D.}\ \bibnamefont {Lewis}}, \bibinfo {author} {\bibfnamefont {V.}~\bibnamefont {Rezania}}, \bibinfo {author} {\bibfnamefont {K.}~\bibnamefont {Shankar}},\ and\ \bibinfo {author} {\bibfnamefont {J.~A.}\ \bibnamefont {Tuszynski}},\ }\bibfield  {title} {\bibinfo {title} {{Investigation of the Electrical Properties of Microtubule Ensembles under Cell-Like Conditions}},\ }\href {https://doi.org/10.3390/nano10020265} {\bibfield  {journal} {\bibinfo  {journal} {Nanomaterials}\ }\textbf {\bibinfo {volume} {10}},\ \bibinfo {pages} {265} (\bibinfo {year} {2020}{\natexlab{b}})},\
  \bibinfo {note} {{\href{https://pubmed.ncbi.nlm.nih.gov/32033331/}{PMID: 32033331}}}\BibitemShut {NoStop}%
\bibitem [{\citenamefont {{van den Heuvel}}\ \emph {et~al.}(2007)\citenamefont {{van den Heuvel}}, \citenamefont {{de Graaff}}, \citenamefont {Lemay},\ and\ \citenamefont {Dekker}}]{Heuvel2007}%
  \BibitemOpen
  \bibfield  {author} {\bibinfo {author} {\bibfnamefont {M.~G.~L.}\ \bibnamefont {{van den Heuvel}}}, \bibinfo {author} {\bibfnamefont {M.~P.}\ \bibnamefont {{de Graaff}}}, \bibinfo {author} {\bibfnamefont {S.~G.}\ \bibnamefont {Lemay}},\ and\ \bibinfo {author} {\bibfnamefont {C.}~\bibnamefont {Dekker}},\ }\bibfield  {title} {\bibinfo {title} {{Electrophoresis of individual microtubules in microchannels}},\ }\href {https://doi.org/10.1073/pnas.0608316104} {\bibfield  {journal} {\bibinfo  {journal} {Proc. Natl. Acad. Sci. U.S.A.}\ }\textbf {\bibinfo {volume} {104}},\ \bibinfo {pages} {7770} (\bibinfo {year} {2007})},\ \bibinfo {note} {{\href{https://pubmed.ncbi.nlm.nih.gov/17470799/}{PMID: 17470799}}}\BibitemShut {NoStop}%
\bibitem [{\citenamefont {Satari\'c}\ \emph {et~al.}(2009)\citenamefont {Satari\'c}, \citenamefont {Ili\'c}, \citenamefont {Ralevi\'c},\ and\ \citenamefont {Tuszynski}}]{Sataric2009}%
  \BibitemOpen
  \bibfield  {author} {\bibinfo {author} {\bibfnamefont {M.~V.}\ \bibnamefont {Satari\'c}}, \bibinfo {author} {\bibfnamefont {D.~I.}\ \bibnamefont {Ili\'c}}, \bibinfo {author} {\bibfnamefont {N.}~\bibnamefont {Ralevi\'c}},\ and\ \bibinfo {author} {\bibfnamefont {J.~A.}\ \bibnamefont {Tuszynski}},\ }\bibfield  {title} {\bibinfo {title} {{A nonlinear model of ionic wave propagation along microtubules}},\ }\href {https://doi.org/10.1007/s00249-009-0421-5} {\bibfield  {journal} {\bibinfo  {journal} {Eur. Biophys. J.}\ }\textbf {\bibinfo {volume} {38}},\ \bibinfo {pages} {637} (\bibinfo {year} {2009})},\ \bibinfo {note} {{\href{https://pubmed.ncbi.nlm.nih.gov/19259657/}{PMID: 19259657}}. Erratum in \href{https://doi.org/10.1007/s00249-009-0540-z}{Eur. Biophys. J. \textbf{38}, 1147 (2009)}}\BibitemShut {NoStop}%
\bibitem [{\citenamefont {Sekuli\'c}\ \emph {et~al.}(2011)\citenamefont {Sekuli\'c}, \citenamefont {Satari\'c}, \citenamefont {Tuszynski},\ and\ \citenamefont {Satari\'c}}]{Sekulic2011}%
  \BibitemOpen
  \bibfield  {author} {\bibinfo {author} {\bibfnamefont {D.~L.}\ \bibnamefont {Sekuli\'c}}, \bibinfo {author} {\bibfnamefont {B.~M.}\ \bibnamefont {Satari\'c}}, \bibinfo {author} {\bibfnamefont {J.~A.}\ \bibnamefont {Tuszynski}},\ and\ \bibinfo {author} {\bibfnamefont {M.~V.}\ \bibnamefont {Satari\'c}},\ }\bibfield  {title} {\bibinfo {title} {{Nonlinear ionic pulses along microtubules}},\ }\href {https://doi.org/10.1140/epje/i2011-11049-0} {\bibfield  {journal} {\bibinfo  {journal} {Eur. Phys. J. E}\ }\textbf {\bibinfo {volume} {34}},\ \bibinfo {pages} {49} (\bibinfo {year} {2011})},\ \bibinfo {note} {{\href{https://pubmed.ncbi.nlm.nih.gov/21604102/}{PMID: 21604102}}}\BibitemShut {NoStop}%
\bibitem [{\citenamefont {Gundersen}\ and\ \citenamefont {Cook}(1999)}]{GUNDERSEN199981}%
  \BibitemOpen
  \bibfield  {author} {\bibinfo {author} {\bibfnamefont {G.~G.}\ \bibnamefont {Gundersen}}\ and\ \bibinfo {author} {\bibfnamefont {T.~A.}\ \bibnamefont {Cook}},\ }\bibfield  {title} {\bibinfo {title} {{Microtubules and signal transduction}},\ }\href {https://doi.org/10.1016/S0955-0674(99)80010-6} {\bibfield  {journal} {\bibinfo  {journal} {Curr. Opin. Cell Biol.}\ }\textbf {\bibinfo {volume} {11}},\ \bibinfo {pages} {81} (\bibinfo {year} {1999})},\ \bibinfo {note} {{\href{https://pubmed.ncbi.nlm.nih.gov/10047525/}{PMID: 10047525}}}\BibitemShut {NoStop}%
\bibitem [{\citenamefont {Craddock}\ \emph {et~al.}(2010)\citenamefont {Craddock}, \citenamefont {Tuszynski}, \citenamefont {Priel},\ and\ \citenamefont {Freedman}}]{CRADDOCK2010}%
  \BibitemOpen
  \bibfield  {author} {\bibinfo {author} {\bibfnamefont {T.~J.~A.}\ \bibnamefont {Craddock}}, \bibinfo {author} {\bibfnamefont {J.~A.}\ \bibnamefont {Tuszynski}}, \bibinfo {author} {\bibfnamefont {A.}~\bibnamefont {Priel}},\ and\ \bibinfo {author} {\bibfnamefont {H.}~\bibnamefont {Freedman}},\ }\bibfield  {title} {\bibinfo {title} {{Microtubule ionic conduction and its implications for higher cognitive functions}},\ }\href {https://doi.org/10.1142/S0219635210002421} {\bibfield  {journal} {\bibinfo  {journal} {J. Integr. Neurosci.}\ }\textbf {\bibinfo {volume} {09}},\ \bibinfo {pages} {103} (\bibinfo {year} {2010})},\ \bibinfo {note} {{\href{https://pubmed.ncbi.nlm.nih.gov/20589950/}{PMID: 20589950}}}\BibitemShut {NoStop}%
\bibitem [{\citenamefont {Dent}\ and\ \citenamefont {Baas}(2014)}]{Dent2014}%
  \BibitemOpen
  \bibfield  {author} {\bibinfo {author} {\bibfnamefont {E.~W.}\ \bibnamefont {Dent}}\ and\ \bibinfo {author} {\bibfnamefont {P.~W.}\ \bibnamefont {Baas}},\ }\bibfield  {title} {\bibinfo {title} {{Microtubules in neurons as information carriers}},\ }\href {https://doi.org/10.1111/jnc.12621} {\bibfield  {journal} {\bibinfo  {journal} {J. Neurochem.}\ }\textbf {\bibinfo {volume} {129}},\ \bibinfo {pages} {235} (\bibinfo {year} {2014})},\ \bibinfo {note} {{\href{https://pubmed.ncbi.nlm.nih.gov/24266899/}{PMID: 24266899}}}\BibitemShut {NoStop}%
\bibitem [{\citenamefont {Barvitenko}\ \emph {et~al.}(2018)\citenamefont {Barvitenko}, \citenamefont {Lawen}, \citenamefont {Aslam}, \citenamefont {Pantaleo}, \citenamefont {Saldanha}, \citenamefont {Skverchinskaya}, \citenamefont {Regolini},\ and\ \citenamefont {Tuszynski}}]{BARVITENKO2018191}%
  \BibitemOpen
  \bibfield  {author} {\bibinfo {author} {\bibfnamefont {N.}~\bibnamefont {Barvitenko}}, \bibinfo {author} {\bibfnamefont {A.}~\bibnamefont {Lawen}}, \bibinfo {author} {\bibfnamefont {M.}~\bibnamefont {Aslam}}, \bibinfo {author} {\bibfnamefont {A.}~\bibnamefont {Pantaleo}}, \bibinfo {author} {\bibfnamefont {C.}~\bibnamefont {Saldanha}}, \bibinfo {author} {\bibfnamefont {E.}~\bibnamefont {Skverchinskaya}}, \bibinfo {author} {\bibfnamefont {M.}~\bibnamefont {Regolini}},\ and\ \bibinfo {author} {\bibfnamefont {J.~A.}\ \bibnamefont {Tuszynski}},\ }\bibfield  {title} {\bibinfo {title} {Integration of intracellular signaling: Biological analogues of wires, processors and memories organized by a centrosome 3d reference system},\ }\href {https://doi.org/10.1016/j.biosystems.2018.08.007} {\bibfield  {journal} {\bibinfo  {journal} {Biosystems}\ }\textbf {\bibinfo {volume} {173}},\ \bibinfo {pages} {191} (\bibinfo {year} {2018})},\ \bibinfo {note} {{\href{https://pubmed.ncbi.nlm.nih.gov/30142359/}{PMID:
  30142359}}}\BibitemShut {NoStop}%
\bibitem [{\citenamefont {Stracke}\ \emph {et~al.}(2002)\citenamefont {Stracke}, \citenamefont {B\"{o}hm}, \citenamefont {Wollweber}, \citenamefont {Tuszynski},\ and\ \citenamefont {Unger}}]{STRACKE2002602}%
  \BibitemOpen
  \bibfield  {author} {\bibinfo {author} {\bibfnamefont {R.}~\bibnamefont {Stracke}}, \bibinfo {author} {\bibfnamefont {K.~J.}\ \bibnamefont {B\"{o}hm}}, \bibinfo {author} {\bibfnamefont {L.}~\bibnamefont {Wollweber}}, \bibinfo {author} {\bibfnamefont {J.~A.}\ \bibnamefont {Tuszynski}},\ and\ \bibinfo {author} {\bibfnamefont {E.}~\bibnamefont {Unger}},\ }\bibfield  {title} {\bibinfo {title} {{Analysis of the migration behaviour of single microtubules in electric fields}},\ }\href {https://doi.org/10.1016/S0006-291X(02)00251-6} {\bibfield  {journal} {\bibinfo  {journal} {Biochem. Biophys. Res. Commun.}\ }\textbf {\bibinfo {volume} {293}},\ \bibinfo {pages} {602} (\bibinfo {year} {2002})},\ \bibinfo {note} {{\href{https://pubmed.ncbi.nlm.nih.gov/12054645/}{PMID: 12054645}}}\BibitemShut {NoStop}%
\bibitem [{\citenamefont {Minoura}\ and\ \citenamefont {Muto}(2006)}]{Minoura2006}%
  \BibitemOpen
  \bibfield  {author} {\bibinfo {author} {\bibfnamefont {I.}~\bibnamefont {Minoura}}\ and\ \bibinfo {author} {\bibfnamefont {E.}~\bibnamefont {Muto}},\ }\bibfield  {title} {\bibinfo {title} {{Dielectric Measurement of Individual Microtubules Using the Electroorientation Method}},\ }\href {https://doi.org/10.1529/biophysj.105.071324} {\bibfield  {journal} {\bibinfo  {journal} {Biophys. J.}\ }\textbf {\bibinfo {volume} {90}},\ \bibinfo {pages} {3739} (\bibinfo {year} {2006})},\ \bibinfo {note} {{\href{https://pubmed.ncbi.nlm.nih.gov/16500962/}{PMID: 16500962}}}\BibitemShut {NoStop}%
\bibitem [{\citenamefont {Uppalapati}\ \emph {et~al.}(2008)\citenamefont {Uppalapati}, \citenamefont {Huang}, \citenamefont {Jackson},\ and\ \citenamefont {Hancock}}]{Uppalapati2008}%
  \BibitemOpen
  \bibfield  {author} {\bibinfo {author} {\bibfnamefont {M.}~\bibnamefont {Uppalapati}}, \bibinfo {author} {\bibfnamefont {Y.-M.}\ \bibnamefont {Huang}}, \bibinfo {author} {\bibfnamefont {T.~N.}\ \bibnamefont {Jackson}},\ and\ \bibinfo {author} {\bibfnamefont {W.~O.}\ \bibnamefont {Hancock}},\ }\bibfield  {title} {\bibinfo {title} {{Microtubule Alignment and Manipulation Using AC Electrokinetics}},\ }\href {https://doi.org/10.1002/smll.200701088} {\bibfield  {journal} {\bibinfo  {journal} {Small}\ }\textbf {\bibinfo {volume} {4}},\ \bibinfo {pages} {1371} (\bibinfo {year} {2008})},\ \bibinfo {note} {{\href{https://pubmed.ncbi.nlm.nih.gov/18720434/}{PMID: 18720434}}}\BibitemShut {NoStop}%
\bibitem [{\citenamefont {Dujovne}\ \emph {et~al.}(2008)\citenamefont {Dujovne}, \citenamefont {{van den Heuvel}}, \citenamefont {Shen}, \citenamefont {{de Graaff}},\ and\ \citenamefont {Dekker}}]{Dujovne2008}%
  \BibitemOpen
  \bibfield  {author} {\bibinfo {author} {\bibfnamefont {I.}~\bibnamefont {Dujovne}}, \bibinfo {author} {\bibfnamefont {M.}~\bibnamefont {{van den Heuvel}}}, \bibinfo {author} {\bibfnamefont {Y.}~\bibnamefont {Shen}}, \bibinfo {author} {\bibfnamefont {M.}~\bibnamefont {{de Graaff}}},\ and\ \bibinfo {author} {\bibfnamefont {C.}~\bibnamefont {Dekker}},\ }\bibfield  {title} {\bibinfo {title} {{Velocity Modulation of Microtubules in Electric Fields}},\ }\href {https://doi.org/10.1021/nl801837j} {\bibfield  {journal} {\bibinfo  {journal} {Nano Lett.}\ }\textbf {\bibinfo {volume} {8}},\ \bibinfo {pages} {4217} (\bibinfo {year} {2008})},\ \bibinfo {note} {{\href{https://pubmed.ncbi.nlm.nih.gov/19367799/}{PMID: 19367799}}}\BibitemShut {NoStop}%
\bibitem [{\citenamefont {Isozaki}\ \emph {et~al.}(2015)\citenamefont {Isozaki}, \citenamefont {Ando}, \citenamefont {Nakahara}, \citenamefont {Shintaku}, \citenamefont {Kotera}, \citenamefont {Meyh\"{o}fer},\ and\ \citenamefont {Yokokawa}}]{Isozaki2015}%
  \BibitemOpen
  \bibfield  {author} {\bibinfo {author} {\bibfnamefont {N.}~\bibnamefont {Isozaki}}, \bibinfo {author} {\bibfnamefont {S.}~\bibnamefont {Ando}}, \bibinfo {author} {\bibfnamefont {T.}~\bibnamefont {Nakahara}}, \bibinfo {author} {\bibfnamefont {H.}~\bibnamefont {Shintaku}}, \bibinfo {author} {\bibfnamefont {H.}~\bibnamefont {Kotera}}, \bibinfo {author} {\bibfnamefont {E.}~\bibnamefont {Meyh\"{o}fer}},\ and\ \bibinfo {author} {\bibfnamefont {R.}~\bibnamefont {Yokokawa}},\ }\bibfield  {title} {\bibinfo {title} {{Control of microtubule trajectory within an electric field by altering surface charge density}},\ }\href {https://doi.org/10.1038/srep07669} {\bibfield  {journal} {\bibinfo  {journal} {Sci. Rep.}\ }\textbf {\bibinfo {volume} {5}},\ \bibinfo {pages} {7669} (\bibinfo {year} {2015})},\ \bibinfo {note} {{\href{https://pubmed.ncbi.nlm.nih.gov/25567007/}{PMID: 25567007}}}\BibitemShut {NoStop}%
\bibitem [{\citenamefont {Hough}\ \emph {et~al.}(2021)\citenamefont {Hough}, \citenamefont {Purschke}, \citenamefont {Bell}, \citenamefont {Kalra}, \citenamefont {Oliva}, \citenamefont {Huang}, \citenamefont {Tuszynski}, \citenamefont {Warkentin},\ and\ \citenamefont {Hegmann}}]{Hough2021}%
  \BibitemOpen
  \bibfield  {author} {\bibinfo {author} {\bibfnamefont {C.~M.}\ \bibnamefont {Hough}}, \bibinfo {author} {\bibfnamefont {D.~N.}\ \bibnamefont {Purschke}}, \bibinfo {author} {\bibfnamefont {C.}~\bibnamefont {Bell}}, \bibinfo {author} {\bibfnamefont {A.~P.}\ \bibnamefont {Kalra}}, \bibinfo {author} {\bibfnamefont {P.~J.}\ \bibnamefont {Oliva}}, \bibinfo {author} {\bibfnamefont {C.}~\bibnamefont {Huang}}, \bibinfo {author} {\bibfnamefont {J.~A.}\ \bibnamefont {Tuszynski}}, \bibinfo {author} {\bibfnamefont {B.~J.}\ \bibnamefont {Warkentin}},\ and\ \bibinfo {author} {\bibfnamefont {F.~A.}\ \bibnamefont {Hegmann}},\ }\bibfield  {title} {\bibinfo {title} {{Disassembly of microtubules by intense terahertz pulses}},\ }\href {https://doi.org/10.1364/boe.433240} {\bibfield  {journal} {\bibinfo  {journal} {Biomed. Opt. Express}\ }\textbf {\bibinfo {volume} {12}},\ \bibinfo {pages} {5812} (\bibinfo {year} {2021})},\ \bibinfo {note} {{\href{https://pubmed.ncbi.nlm.nih.gov/34692217/}{PMID: 34692217}}}\BibitemShut {NoStop}%
\bibitem [{\citenamefont {Staelens}\ \emph {et~al.}(2022)\citenamefont {Staelens}, \citenamefont {{Di Gregorio}}, \citenamefont {Kalra}, \citenamefont {Le}, \citenamefont {Hosseinkhah}, \citenamefont {Karimpoor}, \citenamefont {Lim},\ and\ \citenamefont {Tuszy\'nski}}]{Staelens2022}%
  \BibitemOpen
  \bibfield  {author} {\bibinfo {author} {\bibfnamefont {M.}~\bibnamefont {Staelens}}, \bibinfo {author} {\bibfnamefont {E.}~\bibnamefont {{Di Gregorio}}}, \bibinfo {author} {\bibfnamefont {A.~P.}\ \bibnamefont {Kalra}}, \bibinfo {author} {\bibfnamefont {H.~T.}\ \bibnamefont {Le}}, \bibinfo {author} {\bibfnamefont {N.}~\bibnamefont {Hosseinkhah}}, \bibinfo {author} {\bibfnamefont {M.}~\bibnamefont {Karimpoor}}, \bibinfo {author} {\bibfnamefont {L.}~\bibnamefont {Lim}},\ and\ \bibinfo {author} {\bibfnamefont {J.~A.}\ \bibnamefont {Tuszy\'nski}},\ }\bibfield  {title} {\bibinfo {title} {{Near-Infrared Photobiomodulation of Living Cells, Tubulin, and Microtubules \textit{In Vitro}}},\ }\href {https://doi.org/10.3389/fmedt.2022.871196} {\bibfield  {journal} {\bibinfo  {journal} {Front. Med. Technol.}\ }\textbf {\bibinfo {volume} {4}},\ \bibinfo {pages} {871196} (\bibinfo {year} {2022})},\ \bibinfo {note} {{\href{https://pubmed.ncbi.nlm.nih.gov/35600165/}{PMID: 35600165}}}\BibitemShut {NoStop}%
\bibitem [{\citenamefont {Baas}\ \emph {et~al.}(2016)\citenamefont {Baas}, \citenamefont {Rao}, \citenamefont {Matamoros},\ and\ \citenamefont {Leo}}]{Baas2016}%
  \BibitemOpen
  \bibfield  {author} {\bibinfo {author} {\bibfnamefont {P.~W.}\ \bibnamefont {Baas}}, \bibinfo {author} {\bibfnamefont {A.~N.}\ \bibnamefont {Rao}}, \bibinfo {author} {\bibfnamefont {A.~J.}\ \bibnamefont {Matamoros}},\ and\ \bibinfo {author} {\bibfnamefont {L.}~\bibnamefont {Leo}},\ }\bibfield  {title} {\bibinfo {title} {{Stability properties of neuronal microtubules}},\ }\href {https://doi.org/10.1002/cm.21286} {\bibfield  {journal} {\bibinfo  {journal} {Cytoskeleton}\ }\textbf {\bibinfo {volume} {73}},\ \bibinfo {pages} {442} (\bibinfo {year} {2016})},\ \bibinfo {note} {{\href{https://pubmed.ncbi.nlm.nih.gov/26887570/}{PMID: 26887570}}}\BibitemShut {NoStop}%
\bibitem [{\citenamefont {{Lewis Jr.}}\ \emph {et~al.}(2013)\citenamefont {{Lewis Jr.}}, \citenamefont {Courchet},\ and\ \citenamefont {Polleux}}]{LewisJr2013}%
  \BibitemOpen
  \bibfield  {author} {\bibinfo {author} {\bibfnamefont {T.~L.}\ \bibnamefont {{Lewis Jr.}}}, \bibinfo {author} {\bibfnamefont {J.}~\bibnamefont {Courchet}},\ and\ \bibinfo {author} {\bibfnamefont {F.}~\bibnamefont {Polleux}},\ }\bibfield  {title} {\bibinfo {title} {{{Cellular and molecular mechanisms underlying axon formation, growth, and branching}}},\ }\href {https://doi.org/10.1083/jcb.201305098} {\bibfield  {journal} {\bibinfo  {journal} {J. Cell Biol.}\ }\textbf {\bibinfo {volume} {202}},\ \bibinfo {pages} {837} (\bibinfo {year} {2013})},\ \bibinfo {note} {{\href{https://pubmed.ncbi.nlm.nih.gov/24043699/}{PMID: 24043699}}}\BibitemShut {NoStop}%
\bibitem [{\citenamefont {Yogev}\ \emph {et~al.}(2016)\citenamefont {Yogev}, \citenamefont {Cooper}, \citenamefont {Fetter}, \citenamefont {Horowitz},\ and\ \citenamefont {Shen}}]{Yogev2016}%
  \BibitemOpen
  \bibfield  {author} {\bibinfo {author} {\bibfnamefont {S.}~\bibnamefont {Yogev}}, \bibinfo {author} {\bibfnamefont {R.}~\bibnamefont {Cooper}}, \bibinfo {author} {\bibfnamefont {R.}~\bibnamefont {Fetter}}, \bibinfo {author} {\bibfnamefont {M.}~\bibnamefont {Horowitz}},\ and\ \bibinfo {author} {\bibfnamefont {K.}~\bibnamefont {Shen}},\ }\bibfield  {title} {\bibinfo {title} {{Microtubule Organization Determines Axonal Transport Dynamics}},\ }\href {https://doi.org/10.1016/j.neuron.2016.09.036} {\bibfield  {journal} {\bibinfo  {journal} {Neuron}\ }\textbf {\bibinfo {volume} {92}},\ \bibinfo {pages} {449} (\bibinfo {year} {2016})},\ \bibinfo {note} {{\href{https://pubmed.ncbi.nlm.nih.gov/27764672/}{PMID: 27764672}}}\BibitemShut {NoStop}%
\bibitem [{\citenamefont {Kelliher}\ \emph {et~al.}(2019)\citenamefont {Kelliher}, \citenamefont {Saunders},\ and\ \citenamefont {Wildonger}}]{KELLIHER201939}%
  \BibitemOpen
  \bibfield  {author} {\bibinfo {author} {\bibfnamefont {M.~T.}\ \bibnamefont {Kelliher}}, \bibinfo {author} {\bibfnamefont {H.~A.}\ \bibnamefont {Saunders}},\ and\ \bibinfo {author} {\bibfnamefont {J.}~\bibnamefont {Wildonger}},\ }\bibfield  {title} {\bibinfo {title} {{Microtubule control of functional architecture in neurons}},\ }\href {https://doi.org/10.1016/j.conb.2019.01.003} {\bibfield  {journal} {\bibinfo  {journal} {Curr. Opin. Neurobiol.}\ }\textbf {\bibinfo {volume} {57}},\ \bibinfo {pages} {39} (\bibinfo {year} {2019})},\ \bibinfo {note} {{Part of special issue: Molecular Neuroscience, \href{https://pubmed.ncbi.nlm.nih.gov/30738328/}{PMID: 30738328}}}\BibitemShut {NoStop}%
\bibitem [{\citenamefont {McMurray}(2000)}]{McMurray2000}%
  \BibitemOpen
  \bibfield  {author} {\bibinfo {author} {\bibfnamefont {C.}~\bibnamefont {McMurray}},\ }\bibfield  {title} {\bibinfo {title} {{Neurodegeneration: diseases of the cytoskeleton?}},\ }\href {https://doi.org/10.1038/sj.cdd.4400764} {\bibfield  {journal} {\bibinfo  {journal} {Cell Death Differ.}\ }\textbf {\bibinfo {volume} {7}},\ \bibinfo {pages} {861} (\bibinfo {year} {2000})},\ \bibinfo {note} {{\href{https://pubmed.ncbi.nlm.nih.gov/11279530/}{PMID: 11279530}}}\BibitemShut {NoStop}%
\bibitem [{\citenamefont {Jean}\ and\ \citenamefont {Baas}(2013)}]{Jean2013}%
  \BibitemOpen
  \bibfield  {author} {\bibinfo {author} {\bibfnamefont {D.~C.}\ \bibnamefont {Jean}}\ and\ \bibinfo {author} {\bibfnamefont {P.~W.}\ \bibnamefont {Baas}},\ }\bibfield  {title} {\bibinfo {title} {{It cuts two ways: microtubule loss during Alzheimer disease}},\ }\href {https://doi.org/10.1038/emboj.2013.219} {\bibfield  {journal} {\bibinfo  {journal} {The EMBO Journal}\ }\textbf {\bibinfo {volume} {32}},\ \bibinfo {pages} {2900} (\bibinfo {year} {2013})},\ \bibinfo {note} {{\href{https://pubmed.ncbi.nlm.nih.gov/24076651/}{PMID: 24076651}}}\BibitemShut {NoStop}%
\bibitem [{\citenamefont {Matamoros}\ and\ \citenamefont {Baas}(2016)}]{MATAMOROS2016217}%
  \BibitemOpen
  \bibfield  {author} {\bibinfo {author} {\bibfnamefont {A.~J.}\ \bibnamefont {Matamoros}}\ and\ \bibinfo {author} {\bibfnamefont {P.~W.}\ \bibnamefont {Baas}},\ }\bibfield  {title} {\bibinfo {title} {{Microtubules in health and degenerative disease of the nervous system}},\ }\href {https://doi.org/10.1016/j.brainresbull.2016.06.016} {\bibfield  {journal} {\bibinfo  {journal} {Brain Res. Bull.}\ }\textbf {\bibinfo {volume} {126}},\ \bibinfo {pages} {217} (\bibinfo {year} {2016})},\ \bibinfo {note} {{\href{https://pubmed.ncbi.nlm.nih.gov/27365230/}{PMID: 27365230}}}\BibitemShut {NoStop}%
\bibitem [{\citenamefont {Sferra}\ \emph {et~al.}(2020)\citenamefont {Sferra}, \citenamefont {Nicita},\ and\ \citenamefont {Bertini}}]{Sferra2020}%
  \BibitemOpen
  \bibfield  {author} {\bibinfo {author} {\bibfnamefont {A.}~\bibnamefont {Sferra}}, \bibinfo {author} {\bibfnamefont {F.}~\bibnamefont {Nicita}},\ and\ \bibinfo {author} {\bibfnamefont {E.}~\bibnamefont {Bertini}},\ }\bibfield  {title} {\bibinfo {title} {{Microtubule Dysfunction: A Common Feature of Neurodegenerative Diseases}},\ }\href {https://doi.org/10.3390/ijms21197354} {\bibfield  {journal} {\bibinfo  {journal} {Int. J. Mol. Sci.}\ }\textbf {\bibinfo {volume} {21}},\ \bibinfo {pages} {7354} (\bibinfo {year} {2020})},\ \bibinfo {note} {{\href{https://pubmed.ncbi.nlm.nih.gov/33027950/}{PMID: 33027950}}}\BibitemShut {NoStop}%
\bibitem [{\citenamefont {Fernandez-Valenzuela}\ \emph {et~al.}(2020)\citenamefont {Fernandez-Valenzuela}, \citenamefont {Sanchez-Varo}, \citenamefont {{n}oz Castro}, \citenamefont {Castro}, \citenamefont {Sanchez-Mejias}, \citenamefont {Navarro}, \citenamefont {Jimenez}, \citenamefont {{n}ez Diaz}, \citenamefont {Gomez-Arboledas}, \citenamefont {Moreno-Gonzalez},\ and\ \citenamefont {{et al.}}}]{Fernandez2020}%
  \BibitemOpen
  \bibfield  {author} {\bibinfo {author} {\bibfnamefont {J.~J.}\ \bibnamefont {Fernandez-Valenzuela}}, \bibinfo {author} {\bibfnamefont {R.}~\bibnamefont {Sanchez-Varo}}, \bibinfo {author} {\bibfnamefont {C.~M.}\ \bibnamefont {{n}oz Castro}}, \bibinfo {author} {\bibfnamefont {V.~D.}\ \bibnamefont {Castro}}, \bibinfo {author} {\bibfnamefont {E.}~\bibnamefont {Sanchez-Mejias}}, \bibinfo {author} {\bibfnamefont {V.}~\bibnamefont {Navarro}}, \bibinfo {author} {\bibfnamefont {S.}~\bibnamefont {Jimenez}}, \bibinfo {author} {\bibfnamefont {C.~N.}\ \bibnamefont {{n}ez Diaz}}, \bibinfo {author} {\bibfnamefont {A.}~\bibnamefont {Gomez-Arboledas}}, \bibinfo {author} {\bibfnamefont {I.}~\bibnamefont {Moreno-Gonzalez}},\ and\ \bibinfo {author} {\bibnamefont {{et al.}}},\ }\bibfield  {title} {\bibinfo {title} {{Enhancing microtubule stabilization rescues cognitive deficits and ameliorates pathological phenotype in an amyloidogenic Alzheimer’s disease model}},\ }\href {https://doi.org/10.1038/s41598-020-71767-4}
  {\bibfield  {journal} {\bibinfo  {journal} {Sci. Rep.}\ }\textbf {\bibinfo {volume} {10}},\ \bibinfo {pages} {14776} (\bibinfo {year} {2020})},\ \bibinfo {note} {{\href{https://pubmed.ncbi.nlm.nih.gov/32901091/}{PMID: 32901091}}}\BibitemShut {NoStop}%
\bibitem [{\citenamefont {Boiarska}\ and\ \citenamefont {Passarella}(2021)}]{BOIARSKA2021604}%
  \BibitemOpen
  \bibfield  {author} {\bibinfo {author} {\bibfnamefont {Z.}~\bibnamefont {Boiarska}}\ and\ \bibinfo {author} {\bibfnamefont {D.}~\bibnamefont {Passarella}},\ }\bibfield  {title} {\bibinfo {title} {{Microtubule-targeting agents and neurodegeneration}},\ }\href {https://doi.org/10.1016/j.drudis.2020.11.033} {\bibfield  {journal} {\bibinfo  {journal} {Drug Discovery Today}\ }\textbf {\bibinfo {volume} {26}},\ \bibinfo {pages} {604} (\bibinfo {year} {2021})},\ \bibinfo {note} {{\href{https://pubmed.ncbi.nlm.nih.gov/33279455/}{PMID: 33279455}}}\BibitemShut {NoStop}%
\bibitem [{\citenamefont {Peris}\ \emph {et~al.}(2022)\citenamefont {Peris}, \citenamefont {Parato}, \citenamefont {Qu}, \citenamefont {Soleilhac}, \citenamefont {Lant\'e}, \citenamefont {Kumar}, \citenamefont {Pero}, \citenamefont {Mart\'inez-Hern\'andez}, \citenamefont {Corrao}, \citenamefont {Falivelli},\ and\ \citenamefont {{et al.}}}]{Peris2022}%
  \BibitemOpen
  \bibfield  {author} {\bibinfo {author} {\bibfnamefont {L.}~\bibnamefont {Peris}}, \bibinfo {author} {\bibfnamefont {J.}~\bibnamefont {Parato}}, \bibinfo {author} {\bibfnamefont {X.}~\bibnamefont {Qu}}, \bibinfo {author} {\bibfnamefont {J.~M.}\ \bibnamefont {Soleilhac}}, \bibinfo {author} {\bibfnamefont {F.}~\bibnamefont {Lant\'e}}, \bibinfo {author} {\bibfnamefont {A.}~\bibnamefont {Kumar}}, \bibinfo {author} {\bibfnamefont {M.~E.}\ \bibnamefont {Pero}}, \bibinfo {author} {\bibfnamefont {J.}~\bibnamefont {Mart\'inez-Hern\'andez}}, \bibinfo {author} {\bibfnamefont {C.}~\bibnamefont {Corrao}}, \bibinfo {author} {\bibfnamefont {G.}~\bibnamefont {Falivelli}},\ and\ \bibinfo {author} {\bibnamefont {{et al.}}},\ }\bibfield  {title} {\bibinfo {title} {{{Tubulin tyrosination regulates synaptic function and is disrupted in Alzheimer’s disease}}},\ }\href {https://doi.org/10.1093/brain/awab436} {\bibfield  {journal} {\bibinfo  {journal} {Brain}\ }\textbf {\bibinfo {volume} {145}},\ \bibinfo {pages} {2486} (\bibinfo
  {year} {2022})},\ \bibinfo {note} {{\href{https://pubmed.ncbi.nlm.nih.gov/35148384/}{PMID: 35148384}}}\BibitemShut {NoStop}%
\bibitem [{\citenamefont {Saltmarche}\ \emph {et~al.}(2017)\citenamefont {Saltmarche}, \citenamefont {Naeser}, \citenamefont {Ho}, \citenamefont {Hamblin},\ and\ \citenamefont {Lim}}]{Saltmarche2017}%
  \BibitemOpen
  \bibfield  {author} {\bibinfo {author} {\bibfnamefont {A.~E.}\ \bibnamefont {Saltmarche}}, \bibinfo {author} {\bibfnamefont {M.~A.}\ \bibnamefont {Naeser}}, \bibinfo {author} {\bibfnamefont {K.~F.}\ \bibnamefont {Ho}}, \bibinfo {author} {\bibfnamefont {M.~R.}\ \bibnamefont {Hamblin}},\ and\ \bibinfo {author} {\bibfnamefont {L.}~\bibnamefont {Lim}},\ }\bibfield  {title} {\bibinfo {title} {{Significant Improvement in Cognition in Mild to Moderately Severe Dementia Cases Treated with Transcranial Plus Intranasal Photobiomodulation: Case Series Report}},\ }\href {https://doi.org/10.1089/pho.2016.4227} {\bibfield  {journal} {\bibinfo  {journal} {Photomed. Laser Surg.}\ }\textbf {\bibinfo {volume} {35}},\ \bibinfo {pages} {432} (\bibinfo {year} {2017})},\ \bibinfo {note} {{\href{https://pubmed.ncbi.nlm.nih.gov/28186867/}{PMID: 28186867}}}\BibitemShut {NoStop}%
\bibitem [{\citenamefont {Chao}(2019)}]{Chao2019}%
  \BibitemOpen
  \bibfield  {author} {\bibinfo {author} {\bibfnamefont {L.~L.}\ \bibnamefont {Chao}},\ }\bibfield  {title} {\bibinfo {title} {{Effects of Home Photobiomodulation Treatments on Cognitive and Behavioral Function, Cerebral Perfusion, and Resting-State Functional Connectivity in Patients with Dementia: A Pilot Trial}},\ }\href {https://doi.org/10.1089/photob.2018.4555} {\bibfield  {journal} {\bibinfo  {journal} {Photobiomodulation Photomed. Laser Surg.}\ }\textbf {\bibinfo {volume} {37}},\ \bibinfo {pages} {133} (\bibinfo {year} {2019})},\ \bibinfo {note} {{\href{https://pubmed.ncbi.nlm.nih.gov/31050950/}{PMID: 31050950}}}\BibitemShut {NoStop}%
\bibitem [{\citenamefont {Salehpour}\ \emph {et~al.}(2019)\citenamefont {Salehpour}, \citenamefont {Hamblin},\ and\ \citenamefont {DiDuro}}]{Salehpour2019}%
  \BibitemOpen
  \bibfield  {author} {\bibinfo {author} {\bibfnamefont {F.}~\bibnamefont {Salehpour}}, \bibinfo {author} {\bibfnamefont {M.~R.}\ \bibnamefont {Hamblin}},\ and\ \bibinfo {author} {\bibfnamefont {J.~O.}\ \bibnamefont {DiDuro}},\ }\bibfield  {title} {\bibinfo {title} {{Rapid Reversal of Cognitive Decline, Olfactory Dysfunction, and Quality of Life Using Multi-Modality Photobiomodulation Therapy: Case Report}},\ }\href {https://doi.org/10.1089/photob.2018.4569} {\bibfield  {journal} {\bibinfo  {journal} {Photobiomodulation Photomed. Laser Surg.}\ }\textbf {\bibinfo {volume} {37}},\ \bibinfo {pages} {159} (\bibinfo {year} {2019})},\ \bibinfo {note} {{\href{https://pubmed.ncbi.nlm.nih.gov/31050946/}{PMID: 31050946}}}\BibitemShut {NoStop}%
\bibitem [{\citenamefont {Liebert}\ \emph {et~al.}(2021)\citenamefont {Liebert}, \citenamefont {Bicknell}, \citenamefont {Laakso}, \citenamefont {Heller}, \citenamefont {Jalilitabaei}, \citenamefont {Tilley}, \citenamefont {Mitrofanis},\ and\ \citenamefont {Kiat}}]{Liebert2021}%
  \BibitemOpen
  \bibfield  {author} {\bibinfo {author} {\bibfnamefont {A.}~\bibnamefont {Liebert}}, \bibinfo {author} {\bibfnamefont {B.}~\bibnamefont {Bicknell}}, \bibinfo {author} {\bibfnamefont {E.-L.}\ \bibnamefont {Laakso}}, \bibinfo {author} {\bibfnamefont {G.}~\bibnamefont {Heller}}, \bibinfo {author} {\bibfnamefont {P.}~\bibnamefont {Jalilitabaei}}, \bibinfo {author} {\bibfnamefont {S.}~\bibnamefont {Tilley}}, \bibinfo {author} {\bibfnamefont {J.}~\bibnamefont {Mitrofanis}},\ and\ \bibinfo {author} {\bibfnamefont {H.}~\bibnamefont {Kiat}},\ }\bibfield  {title} {\bibinfo {title} {{{Improvements in clinical signs of Parkinson’s disease using photobiomodulation: a prospective proof-of-concept study}}},\ }\href {https://doi.org/10.1186/s12883-021-02248-y} {\bibfield  {journal} {\bibinfo  {journal} {{BMC} Neurol.}\ }\textbf {\bibinfo {volume} {21}},\ \bibinfo {pages} {256} (\bibinfo {year} {2021})},\ \bibinfo {note} {{\href{https://pubmed.ncbi.nlm.nih.gov/34215216/}{PMID: 34215216}}}\BibitemShut {NoStop}%
\bibitem [{\citenamefont {Kueper}\ \emph {et~al.}(2018)\citenamefont {Kueper}, \citenamefont {Speechley},\ and\ \citenamefont {Montero-Odasso}}]{Kueper2018}%
  \BibitemOpen
  \bibfield  {author} {\bibinfo {author} {\bibfnamefont {J.~K.}\ \bibnamefont {Kueper}}, \bibinfo {author} {\bibfnamefont {M.}~\bibnamefont {Speechley}},\ and\ \bibinfo {author} {\bibfnamefont {M.}~\bibnamefont {Montero-Odasso}},\ }\bibfield  {title} {\bibinfo {title} {{The Alzheimer’s Disease Assessment Scale--Cognitive Subscale (ADAS-Cog): Modifications and Responsiveness in Pre-Dementia Populations. A Narrative Review}},\ }\href {https://doi.org/10.3233/JAD-170991} {\bibfield  {journal} {\bibinfo  {journal} {J. Alzheimer's Dis.}\ }\textbf {\bibinfo {volume} {63}},\ \bibinfo {pages} {423} (\bibinfo {year} {2018})},\ \bibinfo {note} {{\href{https://pubmed.ncbi.nlm.nih.gov/29660938/}{PMID: 29660938}}}\BibitemShut {NoStop}%
\bibitem [{\citenamefont {Rogers}\ \emph {et~al.}(1998)\citenamefont {Rogers}, \citenamefont {Farlow}, \citenamefont {Doody}, \citenamefont {Mohs}, \citenamefont {Friedhoff},\ and\ \citenamefont {{Donepezil Study Group*}}}]{Rogers136}%
  \BibitemOpen
  \bibfield  {author} {\bibinfo {author} {\bibfnamefont {S.~L.}\ \bibnamefont {Rogers}}, \bibinfo {author} {\bibfnamefont {M.~R.}\ \bibnamefont {Farlow}}, \bibinfo {author} {\bibfnamefont {R.~S.}\ \bibnamefont {Doody}}, \bibinfo {author} {\bibfnamefont {R.}~\bibnamefont {Mohs}}, \bibinfo {author} {\bibfnamefont {L.~T.}\ \bibnamefont {Friedhoff}},\ and\ \bibinfo {author} {\bibnamefont {{Donepezil Study Group*}}},\ }\bibfield  {title} {\bibinfo {title} {{A 24-week, double-blind, placebo-controlled trial of donepezil in patients with Alzheimer{\textquoteright}s disease}},\ }\href {https://doi.org/10.1212/WNL.50.1.136} {\bibfield  {journal} {\bibinfo  {journal} {Neurology}\ }\textbf {\bibinfo {volume} {50}},\ \bibinfo {pages} {136} (\bibinfo {year} {1998})},\ \bibinfo {note} {{\href{https://pubmed.ncbi.nlm.nih.gov/9443470/}{PMID: 9443470}}}\BibitemShut {NoStop}%
\bibitem [{\citenamefont {Michalikova}\ \emph {et~al.}(2008)\citenamefont {Michalikova}, \citenamefont {Ennaceur}, \citenamefont {{van Rensburg}},\ and\ \citenamefont {Chazot}}]{MICHALIKOVA2008480}%
  \BibitemOpen
  \bibfield  {author} {\bibinfo {author} {\bibfnamefont {S.}~\bibnamefont {Michalikova}}, \bibinfo {author} {\bibfnamefont {A.}~\bibnamefont {Ennaceur}}, \bibinfo {author} {\bibfnamefont {R.}~\bibnamefont {{van Rensburg}}},\ and\ \bibinfo {author} {\bibfnamefont {P.}~\bibnamefont {Chazot}},\ }\bibfield  {title} {\bibinfo {title} {{Emotional responses and memory performance of middle-aged CD1 mice in a 3D maze: Effects of low infrared light}},\ }\href {https://doi.org/10.1016/j.nlm.2007.07.014} {\bibfield  {journal} {\bibinfo  {journal} {Neurobiol. Learn. Mem.}\ }\textbf {\bibinfo {volume} {89}},\ \bibinfo {pages} {480} (\bibinfo {year} {2008})},\ \bibinfo {note} {{\href{https://pubmed.ncbi.nlm.nih.gov/17855128/}{PMID: 17855128}}}\BibitemShut {NoStop}%
\bibitem [{\citenamefont {{De Taboada}}\ \emph {et~al.}(2011)\citenamefont {{De Taboada}}, \citenamefont {Yu}, \citenamefont {El-Amouri}, \citenamefont {Gattoni-Celli}, \citenamefont {Richieri}, \citenamefont {McCarthy}, \citenamefont {Streeter},\ and\ \citenamefont {Kindy}}]{DeTaboada2011}%
  \BibitemOpen
  \bibfield  {author} {\bibinfo {author} {\bibfnamefont {L.}~\bibnamefont {{De Taboada}}}, \bibinfo {author} {\bibfnamefont {J.}~\bibnamefont {Yu}}, \bibinfo {author} {\bibfnamefont {S.}~\bibnamefont {El-Amouri}}, \bibinfo {author} {\bibfnamefont {S.}~\bibnamefont {Gattoni-Celli}}, \bibinfo {author} {\bibfnamefont {S.}~\bibnamefont {Richieri}}, \bibinfo {author} {\bibfnamefont {T.}~\bibnamefont {McCarthy}}, \bibinfo {author} {\bibfnamefont {J.}~\bibnamefont {Streeter}},\ and\ \bibinfo {author} {\bibfnamefont {M.~S.}\ \bibnamefont {Kindy}},\ }\bibfield  {title} {\bibinfo {title} {{Transcranial Laser Therapy Attenuates Amyloid-$\upbeta$ Peptide Neuropathology in Amyloid-$\upbeta$ Protein Precursor Transgenic Mice}},\ }\href {https://doi.org/10.3233/JAD-2010-100894} {\bibfield  {journal} {\bibinfo  {journal} {J. Alzheimer's Dis.}\ }\textbf {\bibinfo {volume} {23}},\ \bibinfo {pages} {521} (\bibinfo {year} {2011})},\ \bibinfo {note} {{\href{https://pubmed.ncbi.nlm.nih.gov/21116053/}{PMID: 21116053}}}\BibitemShut
  {NoStop}%
\bibitem [{\citenamefont {Grillo}\ \emph {et~al.}(2013)\citenamefont {Grillo}, \citenamefont {Duggett}, \citenamefont {Ennaceur},\ and\ \citenamefont {Chazot}}]{GRILLO201313}%
  \BibitemOpen
  \bibfield  {author} {\bibinfo {author} {\bibfnamefont {S.}~\bibnamefont {Grillo}}, \bibinfo {author} {\bibfnamefont {N.}~\bibnamefont {Duggett}}, \bibinfo {author} {\bibfnamefont {A.}~\bibnamefont {Ennaceur}},\ and\ \bibinfo {author} {\bibfnamefont {P.}~\bibnamefont {Chazot}},\ }\bibfield  {title} {\bibinfo {title} {{Non-invasive infra-red therapy (1072nm) reduces $\upbeta$-amyloid protein levels in the brain of an Alzheimer’s disease mouse model, TASTPM}},\ }\href {https://doi.org/10.1016/j.jphotobiol.2013.02.015} {\bibfield  {journal} {\bibinfo  {journal} {J. Photochem. Photobiol. B: Biol.}\ }\textbf {\bibinfo {volume} {123}},\ \bibinfo {pages} {13} (\bibinfo {year} {2013})},\ \bibinfo {note} {{\href{https://pubmed.ncbi.nlm.nih.gov/23603448/}{PMID: 23603448}}}\BibitemShut {NoStop}%
\bibitem [{\citenamefont {Purushothuman}\ \emph {et~al.}(2014)\citenamefont {Purushothuman}, \citenamefont {Johnstone}, \citenamefont {Nandasena}, \citenamefont {Mitrofanis},\ and\ \citenamefont {Stone}}]{Purushothuman2014}%
  \BibitemOpen
  \bibfield  {author} {\bibinfo {author} {\bibfnamefont {S.}~\bibnamefont {Purushothuman}}, \bibinfo {author} {\bibfnamefont {D.~M.}\ \bibnamefont {Johnstone}}, \bibinfo {author} {\bibfnamefont {C.}~\bibnamefont {Nandasena}}, \bibinfo {author} {\bibfnamefont {J.}~\bibnamefont {Mitrofanis}},\ and\ \bibinfo {author} {\bibfnamefont {J.}~\bibnamefont {Stone}},\ }\bibfield  {title} {\bibinfo {title} {{Photobiomodulation with near infrared light mitigates Alzheimer’s disease-related pathology in cerebral cortex – evidence from two transgenic mouse models}},\ }\href {https://doi.org/10.1186/alzrt232} {\bibfield  {journal} {\bibinfo  {journal} {Alzheimer's Res. Ther.}\ }\textbf {\bibinfo {volume} {6}},\ \bibinfo {pages} {2} (\bibinfo {year} {2014})},\ \bibinfo {note} {{\href{https://pubmed.ncbi.nlm.nih.gov/24387311/}{PMID: 24387311}}}\BibitemShut {NoStop}%
\bibitem [{\citenamefont {Purushothuman}\ \emph {et~al.}(2015)\citenamefont {Purushothuman}, \citenamefont {Johnstone}, \citenamefont {Nandasena}, \citenamefont {{van Eersel}}, \citenamefont {Ittner}, \citenamefont {Mitrofanis},\ and\ \citenamefont {Stone}}]{PURUSHOTHUMAN2015155}%
  \BibitemOpen
  \bibfield  {author} {\bibinfo {author} {\bibfnamefont {S.}~\bibnamefont {Purushothuman}}, \bibinfo {author} {\bibfnamefont {D.~M.}\ \bibnamefont {Johnstone}}, \bibinfo {author} {\bibfnamefont {C.}~\bibnamefont {Nandasena}}, \bibinfo {author} {\bibfnamefont {J.}~\bibnamefont {{van Eersel}}}, \bibinfo {author} {\bibfnamefont {L.~M.}\ \bibnamefont {Ittner}}, \bibinfo {author} {\bibfnamefont {J.}~\bibnamefont {Mitrofanis}},\ and\ \bibinfo {author} {\bibfnamefont {J.}~\bibnamefont {Stone}},\ }\bibfield  {title} {\bibinfo {title} {{Near infrared light mitigates cerebellar pathology in transgenic mouse models of dementia}},\ }\href {https://doi.org/10.1016/j.neulet.2015.02.037} {\bibfield  {journal} {\bibinfo  {journal} {Neurosci. Lett.}\ }\textbf {\bibinfo {volume} {591}},\ \bibinfo {pages} {155} (\bibinfo {year} {2015})},\ \bibinfo {note} {{\href{https://pubmed.ncbi.nlm.nih.gov/25703226/}{PMID: 25703226}}}\BibitemShut {NoStop}%
\bibitem [{\citenamefont {{da Luz Eltchechem}}\ \emph {et~al.}(2017)\citenamefont {{da Luz Eltchechem}}, \citenamefont {{Salgado}}, \citenamefont {{Z\^angaro}}, \citenamefont {{da Silva Pereira}}, \citenamefont {{Kerppers}}, \citenamefont {{da Silva}},\ and\ \citenamefont {{Parreira}}}]{daLuzEltchechem2017}%
  \BibitemOpen
  \bibfield  {author} {\bibinfo {author} {\bibfnamefont {C.}~\bibnamefont {{da Luz Eltchechem}}}, \bibinfo {author} {\bibfnamefont {A.~S.~I.}\ \bibnamefont {{Salgado}}}, \bibinfo {author} {\bibfnamefont {R.~A.}\ \bibnamefont {{Z\^angaro}}}, \bibinfo {author} {\bibfnamefont {M.~C.}\ \bibnamefont {{da Silva Pereira}}}, \bibinfo {author} {\bibfnamefont {I.~I.}\ \bibnamefont {{Kerppers}}}, \bibinfo {author} {\bibfnamefont {L.~A.}\ \bibnamefont {{da Silva}}},\ and\ \bibinfo {author} {\bibfnamefont {R.~B.}\ \bibnamefont {{Parreira}}},\ }\bibfield  {title} {\bibinfo {title} {{Transcranial LED therapy on amyloid-$\upbeta$ toxin 25--35 in the hippocampal region of rats}},\ }\href {https://doi.org/10.1007/s10103-017-2156-3} {\bibfield  {journal} {\bibinfo  {journal} {Lasers Med. Sci.}\ }\textbf {\bibinfo {volume} {32}},\ \bibinfo {pages} {749} (\bibinfo {year} {2017})},\ \bibinfo {note} {{\href{https://pubmed.ncbi.nlm.nih.gov/28255783/}{PMID: 28255783}}}\BibitemShut {NoStop}%
\bibitem [{\citenamefont {Sommer}\ \emph {et~al.}(2012)\citenamefont {Sommer}, \citenamefont {Bieschke}, \citenamefont {Friedrich}, \citenamefont {Zhu}, \citenamefont {Wanker}, \citenamefont {Fecht}, \citenamefont {Mereles},\ and\ \citenamefont {Hunstein}}]{Sommer2012}%
  \BibitemOpen
  \bibfield  {author} {\bibinfo {author} {\bibfnamefont {A.~P.}\ \bibnamefont {Sommer}}, \bibinfo {author} {\bibfnamefont {J.}~\bibnamefont {Bieschke}}, \bibinfo {author} {\bibfnamefont {R.~P.}\ \bibnamefont {Friedrich}}, \bibinfo {author} {\bibfnamefont {D.}~\bibnamefont {Zhu}}, \bibinfo {author} {\bibfnamefont {E.~E.}\ \bibnamefont {Wanker}}, \bibinfo {author} {\bibfnamefont {H.~J.}\ \bibnamefont {Fecht}}, \bibinfo {author} {\bibfnamefont {D.}~\bibnamefont {Mereles}},\ and\ \bibinfo {author} {\bibfnamefont {W.}~\bibnamefont {Hunstein}},\ }\bibfield  {title} {\bibinfo {title} {{670 nm Laser Light and EGCG Complementarily Reduce Amyloid-$\upbeta$ Aggregates in Human Neuroblastoma Cells: Basis for Treatment of Alzheimer's Disease?}},\ }\href {https://doi.org/10.1089/pho.2011.3073} {\bibfield  {journal} {\bibinfo  {journal} {Photomed. Laser Surg.}\ }\textbf {\bibinfo {volume} {30}},\ \bibinfo {pages} {54} (\bibinfo {year} {2012})},\ \bibinfo {note} {{\href{https://pubmed.ncbi.nlm.nih.gov/22029866/}{PMID:
  22029866}}}\BibitemShut {NoStop}%
\bibitem [{\citenamefont {Salehpour}\ \emph {et~al.}(2021)\citenamefont {Salehpour}, \citenamefont {Khademi},\ and\ \citenamefont {Hamblin}}]{Salehpour2021}%
  \BibitemOpen
  \bibfield  {author} {\bibinfo {author} {\bibfnamefont {F.}~\bibnamefont {Salehpour}}, \bibinfo {author} {\bibfnamefont {M.}~\bibnamefont {Khademi}},\ and\ \bibinfo {author} {\bibfnamefont {M.~R.}\ \bibnamefont {Hamblin}},\ }\bibfield  {title} {\bibinfo {title} {{Photobiomodulation Therapy for Dementia: A Systematic Review of Pre-Clinical and Clinical Studies}},\ }\href {https://doi.org/10.3233/JAD-210029} {\bibfield  {journal} {\bibinfo  {journal} {J. Alzheimer's Dis.}\ }\textbf {\bibinfo {volume} {83}},\ \bibinfo {pages} {1431} (\bibinfo {year} {2021})},\ \bibinfo {note} {{\href{https://pubmed.ncbi.nlm.nih.gov/26642212/}{PMID: 26642212}}}\BibitemShut {NoStop}%
\bibitem [{NCT(2023{\natexlab{a}})}]{NCT03484143}%
  \BibitemOpen
  \href@noop {} {\bibinfo {title} {{Neuro RX Gamma - Pivotal Phase}}} (\bibinfo {year} {2023}{\natexlab{a}}),\ \bibinfo {note} {{\href{https://clinicaltrials.gov/study/NCT03484143}{NCT03484143}, accessed: 2023-08-01}}\BibitemShut {NoStop}%
\bibitem [{NCT(2023{\natexlab{b}})}]{NCT04018092}%
  \BibitemOpen
  \href@noop {} {\bibinfo {title} {{The Revitalize Study in Older Adults at Risk for Alzheimer's Disease}}} (\bibinfo {year} {2023}{\natexlab{b}}),\ \bibinfo {note} {{\href{https://clinicaltrials.gov/study/NCT04018092}{NCT04018092}, accessed: 2023-08-01}}\BibitemShut {NoStop}%
\bibitem [{NCT(2023{\natexlab{c}})}]{NCT03551392}%
  \BibitemOpen
  \href@noop {} {\bibinfo {title} {{Pilot Intervention With Near Infrared Stimulation}}} (\bibinfo {year} {2023}{\natexlab{c}}),\ \bibinfo {note} {{\href{https://clinicaltrials.gov/study/NCT03551392}{NCT03551392}, accessed: 2023-08-01}}\BibitemShut {NoStop}%
\bibitem [{\citenamefont {{de Freitas}}\ and\ \citenamefont {Hamblin}(2016)}]{Freitas2016}%
  \BibitemOpen
  \bibfield  {author} {\bibinfo {author} {\bibfnamefont {L.~F.}\ \bibnamefont {{de Freitas}}}\ and\ \bibinfo {author} {\bibfnamefont {M.~R.}\ \bibnamefont {Hamblin}},\ }\bibfield  {title} {\bibinfo {title} {{Proposed Mechanisms of Photobiomodulation or Low-Level Light Therapy}},\ }\href {https://doi.org/10.1109/JSTQE.2016.2561201} {\bibfield  {journal} {\bibinfo  {journal} {IEEE J. Sel. Top. Quantum Electron.}\ }\textbf {\bibinfo {volume} {22}},\ \bibinfo {pages} {348} (\bibinfo {year} {2016})},\ \bibinfo {note} {{\href{https://pubmed.ncbi.nlm.nih.gov/28070154/}{PMID: 28070154}}}\BibitemShut {NoStop}%
\bibitem [{\citenamefont {Miura}\ and\ \citenamefont {Thomas}(1995)}]{Miura1995}%
  \BibitemOpen
  \bibfield  {author} {\bibinfo {author} {\bibfnamefont {T.}~\bibnamefont {Miura}}\ and\ \bibinfo {author} {\bibfnamefont {G.~J.}\ \bibnamefont {Thomas}},\ }\bibinfo {title} {{{Raman Spectroscopy of Proteins and Their Assemblies}}},\ in\ \href {https://doi.org/10.1007/978-1-4899-1727-0_3} {\emph {\bibinfo {booktitle} {{{Proteins: Structure, Function, and Engineering}}}}},\ \bibinfo {editor} {edited by\ \bibinfo {editor} {\bibfnamefont {B.~B.}\ \bibnamefont {Biswas}}\ and\ \bibinfo {editor} {\bibfnamefont {S.}~\bibnamefont {Roy}}}\ (\bibinfo  {publisher} {Springer US},\ \bibinfo {address} {Boston, MA},\ \bibinfo {year} {1995})\ pp.\ \bibinfo {pages} {55--99},\ \bibinfo {note} {{\href{https://pubmed.ncbi.nlm.nih.gov/7900183/}{PMID: 7900183}}}\BibitemShut {NoStop}%
\bibitem [{\citenamefont {Lef\`evre}\ \emph {et~al.}(2007)\citenamefont {Lef\`evre}, \citenamefont {Rousseau},\ and\ \citenamefont {P\'ezolet}}]{Lefevre2007}%
  \BibitemOpen
  \bibfield  {author} {\bibinfo {author} {\bibfnamefont {T.}~\bibnamefont {Lef\`evre}}, \bibinfo {author} {\bibfnamefont {M.-E.}\ \bibnamefont {Rousseau}},\ and\ \bibinfo {author} {\bibfnamefont {M.}~\bibnamefont {P\'ezolet}},\ }\bibfield  {title} {\bibinfo {title} {{Protein Secondary Structure and Orientation in Silk as Revealed by Raman Spectromicroscopy}},\ }\href {https://doi.org/10.1529/biophysj.106.100339} {\bibfield  {journal} {\bibinfo  {journal} {Biophys. J.}\ }\textbf {\bibinfo {volume} {92}},\ \bibinfo {pages} {2885} (\bibinfo {year} {2007})},\ \bibinfo {note} {{\href{https://pubmed.ncbi.nlm.nih.gov/17277183/}{PMID: 17277183}}}\BibitemShut {NoStop}%
\bibitem [{\citenamefont {Barth}(2007)}]{Barth2007}%
  \BibitemOpen
  \bibfield  {author} {\bibinfo {author} {\bibfnamefont {A.}~\bibnamefont {Barth}},\ }\bibfield  {title} {\bibinfo {title} {{Infrared spectroscopy of proteins}},\ }\href {https://doi.org/10.1016/j.bbabio.2007.06.004} {\bibfield  {journal} {\bibinfo  {journal} {Biochim. Biophys. Acta - Bioenerg.}\ }\textbf {\bibinfo {volume} {1767}},\ \bibinfo {pages} {1073} (\bibinfo {year} {2007})},\ \bibinfo {note} {{\href{https://pubmed.ncbi.nlm.nih.gov/17692815/}{PMID: 17692815}}}\BibitemShut {NoStop}%
\bibitem [{\citenamefont {{De Meutter}}\ and\ \citenamefont {Goormaghtigh}(2021)}]{DeMeutter2021}%
  \BibitemOpen
  \bibfield  {author} {\bibinfo {author} {\bibfnamefont {J.}~\bibnamefont {{De Meutter}}}\ and\ \bibinfo {author} {\bibfnamefont {E.}~\bibnamefont {Goormaghtigh}},\ }\bibfield  {title} {\bibinfo {title} {{Evaluation of protein secondary structure from FTIR spectra improved after partial deuteration}},\ }\href {https://doi.org/10.1007/s00249-021-01502-y} {\bibfield  {journal} {\bibinfo  {journal} {Eur. Biophys. J.}\ }\textbf {\bibinfo {volume} {50}},\ \bibinfo {pages} {613} (\bibinfo {year} {2021})},\ \bibinfo {note} {{\href{https://pubmed.ncbi.nlm.nih.gov/33534058/}{PMID: 33534058}}}\BibitemShut {NoStop}%
\bibitem [{\citenamefont {Lee}\ \emph {et~al.}(1978)\citenamefont {Lee}, \citenamefont {Corfman}, \citenamefont {Frigon},\ and\ \citenamefont {Timasheff}}]{LEE19784}%
  \BibitemOpen
  \bibfield  {author} {\bibinfo {author} {\bibfnamefont {J.~C.}\ \bibnamefont {Lee}}, \bibinfo {author} {\bibfnamefont {D.}~\bibnamefont {Corfman}}, \bibinfo {author} {\bibfnamefont {R.~P.}\ \bibnamefont {Frigon}},\ and\ \bibinfo {author} {\bibfnamefont {S.~N.}\ \bibnamefont {Timasheff}},\ }\bibfield  {title} {\bibinfo {title} {{Conformational study of calf brain tubulin}},\ }\href {https://doi.org/10.1016/0003-9861(78)90137-6} {\bibfield  {journal} {\bibinfo  {journal} {Arch. Biochem. Biophys.}\ }\textbf {\bibinfo {volume} {185}},\ \bibinfo {pages} {4} (\bibinfo {year} {1978})},\ \bibinfo {note} {{\href{https://pubmed.ncbi.nlm.nih.gov/23729/}{PMID: 23729}}}\BibitemShut {NoStop}%
\bibitem [{\citenamefont {Greenfield}(2006)}]{Greenfield2006}%
  \BibitemOpen
  \bibfield  {author} {\bibinfo {author} {\bibfnamefont {N.~J.}\ \bibnamefont {Greenfield}},\ }\bibfield  {title} {\bibinfo {title} {{Using circular dichroism spectra to estimate protein secondary structure}},\ }\href {https://doi.org/10.1038/nprot.2006.202} {\bibfield  {journal} {\bibinfo  {journal} {Nat. Protoc.}\ }\textbf {\bibinfo {volume} {1}},\ \bibinfo {pages} {2876} (\bibinfo {year} {2006})},\ \bibinfo {note} {{\href{https://pubmed.ncbi.nlm.nih.gov/17406547/}{PMID: 17406547}}}\BibitemShut {NoStop}%
\bibitem [{\citenamefont {Li}\ \emph {et~al.}(2019)\citenamefont {Li}, \citenamefont {Lantz},\ and\ \citenamefont {Du}}]{Li2019}%
  \BibitemOpen
  \bibfield  {author} {\bibinfo {author} {\bibfnamefont {H.}~\bibnamefont {Li}}, \bibinfo {author} {\bibfnamefont {R.}~\bibnamefont {Lantz}},\ and\ \bibinfo {author} {\bibfnamefont {D.}~\bibnamefont {Du}},\ }\bibfield  {title} {\bibinfo {title} {{Vibrational Approach to the Dynamics and Structure of Protein Amyloids}},\ }\href {https://doi.org/10.3390/molecules24010186} {\bibfield  {journal} {\bibinfo  {journal} {Molecules}\ }\textbf {\bibinfo {volume} {24}},\ \bibinfo {pages} {186} (\bibinfo {year} {2019})},\ \bibinfo {note} {{\href{https://pubmed.ncbi.nlm.nih.gov/30621325/}{PMID: 30621325}}}\BibitemShut {NoStop}%
\bibitem [{\citenamefont {Kuhar}\ \emph {et~al.}(2021)\citenamefont {Kuhar}, \citenamefont {Sil},\ and\ \citenamefont {Umapathy}}]{Kuhar2021}%
  \BibitemOpen
  \bibfield  {author} {\bibinfo {author} {\bibfnamefont {N.}~\bibnamefont {Kuhar}}, \bibinfo {author} {\bibfnamefont {S.}~\bibnamefont {Sil}},\ and\ \bibinfo {author} {\bibfnamefont {S.}~\bibnamefont {Umapathy}},\ }\bibfield  {title} {\bibinfo {title} {{Potential of Raman spectroscopic techniques to study proteins}},\ }\href {https://doi.org/10.1016/j.saa.2021.119712} {\bibfield  {journal} {\bibinfo  {journal} {Spectrochim. Acta A Mol. Biomol. Spectrosc.}\ }\textbf {\bibinfo {volume} {258}},\ \bibinfo {pages} {119712} (\bibinfo {year} {2021})},\ \bibinfo {note} {{\href{https://pubmed.ncbi.nlm.nih.gov/33965670/}{PMID: 33965670}}}\BibitemShut {NoStop}%
\bibitem [{\citenamefont {Sadat}\ and\ \citenamefont {Joye}(2020)}]{Sadat_2020}%
  \BibitemOpen
  \bibfield  {author} {\bibinfo {author} {\bibfnamefont {A.}~\bibnamefont {Sadat}}\ and\ \bibinfo {author} {\bibfnamefont {I.~J.}\ \bibnamefont {Joye}},\ }\bibfield  {title} {\bibinfo {title} {{Peak Fitting Applied to Fourier Transform Infrared and Raman Spectroscopic Analysis of Proteins}},\ }\href {https://doi.org/10.3390/app10175918} {\bibfield  {journal} {\bibinfo  {journal} {Appl. Sci.}\ }\textbf {\bibinfo {volume} {10}},\ \bibinfo {pages} {5918} (\bibinfo {year} {2020})}\BibitemShut {NoStop}%
\bibitem [{\citenamefont {Gautam}\ \emph {et~al.}(2015)\citenamefont {Gautam}, \citenamefont {Vanga}, \citenamefont {Madan}, \citenamefont {Gayathri}, \citenamefont {Nongthomba},\ and\ \citenamefont {Umapathy}}]{doi:10.1021/ac503647x}%
  \BibitemOpen
  \bibfield  {author} {\bibinfo {author} {\bibfnamefont {R.}~\bibnamefont {Gautam}}, \bibinfo {author} {\bibfnamefont {S.}~\bibnamefont {Vanga}}, \bibinfo {author} {\bibfnamefont {A.}~\bibnamefont {Madan}}, \bibinfo {author} {\bibfnamefont {N.}~\bibnamefont {Gayathri}}, \bibinfo {author} {\bibfnamefont {U.}~\bibnamefont {Nongthomba}},\ and\ \bibinfo {author} {\bibfnamefont {S.}~\bibnamefont {Umapathy}},\ }\bibfield  {title} {\bibinfo {title} {{Raman Spectroscopic Studies on Screening of Myopathies}},\ }\href {https://doi.org/10.1021/ac503647x} {\bibfield  {journal} {\bibinfo  {journal} {Anal. Chem.}\ }\textbf {\bibinfo {volume} {87}},\ \bibinfo {pages} {2187} (\bibinfo {year} {2015})},\ \bibinfo {note} {{\href{https://pubmed.ncbi.nlm.nih.gov/25583313/}{PMID: 25583313}}}\BibitemShut {NoStop}%
\bibitem [{\citenamefont {Rygula}\ \emph {et~al.}(2013)\citenamefont {Rygula}, \citenamefont {Majzner}, \citenamefont {Marzec}, \citenamefont {Kaczor}, \citenamefont {Pilarczyk},\ and\ \citenamefont {Baranska}}]{Rygula2013}%
  \BibitemOpen
  \bibfield  {author} {\bibinfo {author} {\bibfnamefont {A.}~\bibnamefont {Rygula}}, \bibinfo {author} {\bibfnamefont {K.}~\bibnamefont {Majzner}}, \bibinfo {author} {\bibfnamefont {K.~M.}\ \bibnamefont {Marzec}}, \bibinfo {author} {\bibfnamefont {A.}~\bibnamefont {Kaczor}}, \bibinfo {author} {\bibfnamefont {M.}~\bibnamefont {Pilarczyk}},\ and\ \bibinfo {author} {\bibfnamefont {M.}~\bibnamefont {Baranska}},\ }\bibfield  {title} {\bibinfo {title} {{Raman spectroscopy of proteins: a review}},\ }\href {https://doi.org/10.1002/jrs.4335} {\bibfield  {journal} {\bibinfo  {journal} {J. Raman Spectrosc.}\ }\textbf {\bibinfo {volume} {44}},\ \bibinfo {pages} {1061} (\bibinfo {year} {2013})}\BibitemShut {NoStop}%
\bibitem [{\citenamefont {Krimm}\ and\ \citenamefont {Bandekar}(1986)}]{Krimm1986}%
  \BibitemOpen
  \bibfield  {author} {\bibinfo {author} {\bibfnamefont {S.}~\bibnamefont {Krimm}}\ and\ \bibinfo {author} {\bibfnamefont {J.}~\bibnamefont {Bandekar}},\ }\bibfield  {title} {\bibinfo {title} {{Vibrational Spectroscopy and Conformation of Peptides, Polypeptides, and Proteins}},\ }in\ \href {https://doi.org/10.1016/S0065-3233(08)60528-8} {\emph {\bibinfo {booktitle} {{Advances in Protein Chemistry, Volume 38}}}},\ \bibinfo {editor} {edited by\ \bibinfo {editor} {\bibfnamefont {C.}~\bibnamefont {Anfinsen}}, \bibinfo {editor} {\bibfnamefont {J.~T.}\ \bibnamefont {Edsall}},\ and\ \bibinfo {editor} {\bibfnamefont {F.~M.}\ \bibnamefont {Richards}}}\ (\bibinfo  {publisher} {Academic Press},\ \bibinfo {year} {1986})\ pp.\ \bibinfo {pages} {181--364},\ \bibinfo {note} {{\href{https://pubmed.ncbi.nlm.nih.gov/3541539/}{PMID: 3541539}}}\BibitemShut {NoStop}%
\bibitem [{\citenamefont {Simi{\'{c}}-Krsti{\'{c}}}\ \emph {et~al.}(1991)\citenamefont {Simi{\'{c}}-Krsti{\'{c}}}, \citenamefont {Jeremi{\'{c}}}, \citenamefont {Andjelkovi{\'{c}}},\ and\ \citenamefont {Koruga}}]{Simi_Krsti__1991}%
  \BibitemOpen
  \bibfield  {author} {\bibinfo {author} {\bibfnamefont {J.}~\bibnamefont {Simi{\'{c}}-Krsti{\'{c}}}}, \bibinfo {author} {\bibfnamefont {M.}~\bibnamefont {Jeremi{\'{c}}}}, \bibinfo {author} {\bibfnamefont {M.}~\bibnamefont {Andjelkovi{\'{c}}}},\ and\ \bibinfo {author} {\bibfnamefont {D.}~\bibnamefont {Koruga}},\ }\bibinfo {title} {{{A Raman Spectroscopic Study of Microtubule Protein}}},\ in\ \href {https://doi.org/10.1007/978-94-011-3392-0_9} {\emph {\bibinfo {booktitle} {{{Molecular Electronics: Materials and Methods}}}}},\ \bibinfo {editor} {edited by\ \bibinfo {editor} {\bibfnamefont {P.~I.}\ \bibnamefont {Lazarev}}}\ (\bibinfo  {publisher} {Springer Netherlands},\ \bibinfo {address} {Dordrecht},\ \bibinfo {year} {1991})\ pp.\ \bibinfo {pages} {79--86}\BibitemShut {NoStop}%
\bibitem [{\citenamefont {Salehpour}\ \emph {et~al.}(2020)\citenamefont {Salehpour}, \citenamefont {Gholipour-Khalili}, \citenamefont {Farajdokht}, \citenamefont {Kamari}, \citenamefont {Walski}, \citenamefont {Hamblin}, \citenamefont {DiDuro},\ and\ \citenamefont {Cassano}}]{Salehpour2020}%
  \BibitemOpen
  \bibfield  {author} {\bibinfo {author} {\bibfnamefont {F.}~\bibnamefont {Salehpour}}, \bibinfo {author} {\bibfnamefont {S.}~\bibnamefont {Gholipour-Khalili}}, \bibinfo {author} {\bibfnamefont {F.}~\bibnamefont {Farajdokht}}, \bibinfo {author} {\bibfnamefont {F.}~\bibnamefont {Kamari}}, \bibinfo {author} {\bibfnamefont {T.}~\bibnamefont {Walski}}, \bibinfo {author} {\bibfnamefont {M.~R.}\ \bibnamefont {Hamblin}}, \bibinfo {author} {\bibfnamefont {J.~O.}\ \bibnamefont {DiDuro}},\ and\ \bibinfo {author} {\bibfnamefont {P.}~\bibnamefont {Cassano}},\ }\bibfield  {title} {\bibinfo {title} {{Therapeutic potential of intranasal photobiomodulation therapy for neurological and neuropsychiatric disorders: a narrative review}},\ }\href {https://doi.org/10.1515/revneuro-2019-0063} {\bibfield  {journal} {\bibinfo  {journal} {Rev. Neurosci.}\ }\textbf {\bibinfo {volume} {31}},\ \bibinfo {pages} {269} (\bibinfo {year} {2020})},\ \bibinfo {note} {{\href{https://pubmed.ncbi.nlm.nih.gov/31812948/}{PMID:
  31812948}}}\BibitemShut {NoStop}%
\bibitem [{\citenamefont {Yoo}(2021)}]{Yoo2021}%
  \BibitemOpen
  \bibfield  {author} {\bibinfo {author} {\bibfnamefont {S.~H.}\ \bibnamefont {Yoo}},\ }\bibfield  {title} {\bibinfo {title} {{Intranasal Photobiomodulation Therapy for Brain Conditions: A Review}},\ }\href {https://doi.org/10.25289/ML.2021.10.3.132} {\bibfield  {journal} {\bibinfo  {journal} {Med. Lasers}\ }\textbf {\bibinfo {volume} {10}},\ \bibinfo {pages} {132} (\bibinfo {year} {2021})}\BibitemShut {NoStop}%
\bibitem [{\citenamefont {Eilers}\ and\ \citenamefont {Boelens}(2005)}]{Eilers2005}%
  \BibitemOpen
  \bibfield  {author} {\bibinfo {author} {\bibfnamefont {P.~H.~C.}\ \bibnamefont {Eilers}}\ and\ \bibinfo {author} {\bibfnamefont {H.~F.~M.}\ \bibnamefont {Boelens}},\ }\bibfield  {title} {\bibinfo {title} {{Baseline correction with asymmetric least squares smoothing}},\ }\href@noop {} {\bibfield  {journal} {\bibinfo  {journal} {Leiden University Medical Centre Report}\ }\textbf {\bibinfo {volume} {1}},\ \bibinfo {pages} {5} (\bibinfo {year} {2005})}\BibitemShut {NoStop}%
\bibitem [{\citenamefont {Savitzky}\ and\ \citenamefont {Golay}(1964)}]{Savitzky1964}%
  \BibitemOpen
  \bibfield  {author} {\bibinfo {author} {\bibfnamefont {A.}~\bibnamefont {Savitzky}}\ and\ \bibinfo {author} {\bibfnamefont {M.~J.~E.}\ \bibnamefont {Golay}},\ }\bibfield  {title} {\bibinfo {title} {{Smoothing and Differentiation of Data by Simplified Least Squares Procedures}},\ }\href {https://doi.org/10.1021/ac60214a047} {\bibfield  {journal} {\bibinfo  {journal} {Anal. Chem.}\ }\textbf {\bibinfo {volume} {36}},\ \bibinfo {pages} {1627} (\bibinfo {year} {1964})}\BibitemShut {NoStop}%
\bibitem [{\citenamefont {Maisuradze}\ and\ \citenamefont {Maisuradze}(2021)}]{Maisuradze2021}%
  \BibitemOpen
  \bibfield  {author} {\bibinfo {author} {\bibfnamefont {L.}~\bibnamefont {Maisuradze}}\ and\ \bibinfo {author} {\bibfnamefont {G.~G.}\ \bibnamefont {Maisuradze}},\ }\bibfield  {title} {\bibinfo {title} {{How Useful can the Voigt Profile be in Protein Folding Processes?}},\ }\href {https://doi.org/10.1007/s10930-020-09954-5} {\bibfield  {journal} {\bibinfo  {journal} {Protein J.}\ }\textbf {\bibinfo {volume} {40}},\ \bibinfo {pages} {140} (\bibinfo {year} {2021})},\ \bibinfo {note} {{\href{https://pubmed.ncbi.nlm.nih.gov/33398661/}{PMID: 33398661}}}\BibitemShut {NoStop}%
\bibitem [{\citenamefont {{Di Rocco}}\ \emph {et~al.}(2001)\citenamefont {{Di Rocco}}, \citenamefont {Iriarte},\ and\ \citenamefont {Pomarico}}]{DiRocco2001}%
  \BibitemOpen
  \bibfield  {author} {\bibinfo {author} {\bibfnamefont {H.~O.}\ \bibnamefont {{Di Rocco}}}, \bibinfo {author} {\bibfnamefont {D.~I.}\ \bibnamefont {Iriarte}},\ and\ \bibinfo {author} {\bibfnamefont {J.}~\bibnamefont {Pomarico}},\ }\bibfield  {title} {\bibinfo {title} {{General Expression for the Voigt Function that is of Special Interest for Applied Spectroscopy}},\ }\href {https://opg.optica.org/as/abstract.cfm?URI=as-55-7-822} {\bibfield  {journal} {\bibinfo  {journal} {Appl. Spectrosc.}\ }\textbf {\bibinfo {volume} {55}},\ \bibinfo {pages} {822} (\bibinfo {year} {2001})}\BibitemShut {NoStop}%
\bibitem [{\citenamefont {Levenberg}(1944)}]{LEVENBERG1944}%
  \BibitemOpen
  \bibfield  {author} {\bibinfo {author} {\bibfnamefont {K.}~\bibnamefont {Levenberg}},\ }\bibfield  {title} {\bibinfo {title} {{A Method for the Solution of Certain Non-Linear Problems in Least Squares}},\ }\href {https://doi.org/10.1090%2Fqam%2F10666} {\bibfield  {journal} {\bibinfo  {journal} {Quart. Appl. Math.}\ }\textbf {\bibinfo {volume} {2}},\ \bibinfo {pages} {164} (\bibinfo {year} {1944})}\BibitemShut {NoStop}%
\bibitem [{\citenamefont {Marquardt}(1963)}]{Marquardt1963}%
  \BibitemOpen
  \bibfield  {author} {\bibinfo {author} {\bibfnamefont {D.~W.}\ \bibnamefont {Marquardt}},\ }\bibfield  {title} {\bibinfo {title} {{An Algorithm for Least-Squares Estimation of Nonlinear Parameters}},\ }\href {https://doi.org/10.1137/0111030} {\bibfield  {journal} {\bibinfo  {journal} {J. Soc. Indust. Appl. Math.}\ }\textbf {\bibinfo {volume} {11}},\ \bibinfo {pages} {431} (\bibinfo {year} {1963})}\BibitemShut {NoStop}%
\bibitem [{\citenamefont {Ngarize}\ \emph {et~al.}(2004)\citenamefont {Ngarize}, \citenamefont {Herman}, \citenamefont {Adams},\ and\ \citenamefont {Howell}}]{Ngarize2004}%
  \BibitemOpen
  \bibfield  {author} {\bibinfo {author} {\bibfnamefont {S.}~\bibnamefont {Ngarize}}, \bibinfo {author} {\bibfnamefont {H.}~\bibnamefont {Herman}}, \bibinfo {author} {\bibfnamefont {A.}~\bibnamefont {Adams}},\ and\ \bibinfo {author} {\bibfnamefont {N.}~\bibnamefont {Howell}},\ }\bibfield  {title} {\bibinfo {title} {{Comparison of Changes in the Secondary Structure of Unheated, Heated, and High-Pressure-Treated $\upbeta$-Lactoglobulin and Ovalbumin Proteins Using Fourier Transform Raman Spectroscopy and Self-Deconvolution}},\ }\href {https://doi.org/10.1021/jf030649y} {\bibfield  {journal} {\bibinfo  {journal} {J. Agric. Food Chem.}\ }\textbf {\bibinfo {volume} {52}},\ \bibinfo {pages} {6470} (\bibinfo {year} {2004})},\ \bibinfo {note} {{\href{https://pubmed.ncbi.nlm.nih.gov/15479009/}{PMID: 15479009}}}\BibitemShut {NoStop}%
\bibitem [{\citenamefont {Surewicz}\ \emph {et~al.}(1993)\citenamefont {Surewicz}, \citenamefont {Mantsch},\ and\ \citenamefont {Chapman}}]{Surewicz1993}%
  \BibitemOpen
  \bibfield  {author} {\bibinfo {author} {\bibfnamefont {W.~K.}\ \bibnamefont {Surewicz}}, \bibinfo {author} {\bibfnamefont {H.~H.}\ \bibnamefont {Mantsch}},\ and\ \bibinfo {author} {\bibfnamefont {D.}~\bibnamefont {Chapman}},\ }\bibfield  {title} {\bibinfo {title} {{Determination of protein secondary structure by Fourier transform infrared spectroscopy: A critical assessment}},\ }\href {https://doi.org/10.1021/bi00053a001} {\bibfield  {journal} {\bibinfo  {journal} {Biochemistry}\ }\textbf {\bibinfo {volume} {32}},\ \bibinfo {pages} {389} (\bibinfo {year} {1993})},\ \bibinfo {note} {{\href{https://pubmed.ncbi.nlm.nih.gov/8422346/}{PMID: 8422346}}}\BibitemShut {NoStop}%
\bibitem [{\citenamefont {Sane}\ \emph {et~al.}(1999)\citenamefont {Sane}, \citenamefont {Cramer},\ and\ \citenamefont {Przybycien}}]{Sane1999}%
  \BibitemOpen
  \bibfield  {author} {\bibinfo {author} {\bibfnamefont {S.~U.}\ \bibnamefont {Sane}}, \bibinfo {author} {\bibfnamefont {S.~M.}\ \bibnamefont {Cramer}},\ and\ \bibinfo {author} {\bibfnamefont {T.~M.}\ \bibnamefont {Przybycien}},\ }\bibfield  {title} {\bibinfo {title} {{A Holistic Approach to Protein Secondary Structure Characterization Using Amide I Band Raman Spectroscopy}},\ }\href {https://doi.org/10.1006/abio.1999.4034} {\bibfield  {journal} {\bibinfo  {journal} {Anal. Biochem.}\ }\textbf {\bibinfo {volume} {269}},\ \bibinfo {pages} {255} (\bibinfo {year} {1999})},\ \bibinfo {note} {{\href{https://pubmed.ncbi.nlm.nih.gov/10221997/}{PMID: 10221997}}}\BibitemShut {NoStop}%
\bibitem [{\citenamefont {Wang}\ \emph {et~al.}(2020)\citenamefont {Wang}, \citenamefont {Li}, \citenamefont {Wang}, \citenamefont {Liu}, \citenamefont {Zhang},\ and\ \citenamefont {Zhang}}]{WANG2020105149}%
  \BibitemOpen
  \bibfield  {author} {\bibinfo {author} {\bibfnamefont {B.}~\bibnamefont {Wang}}, \bibinfo {author} {\bibfnamefont {Y.}~\bibnamefont {Li}}, \bibinfo {author} {\bibfnamefont {H.}~\bibnamefont {Wang}}, \bibinfo {author} {\bibfnamefont {X.}~\bibnamefont {Liu}}, \bibinfo {author} {\bibfnamefont {Y.}~\bibnamefont {Zhang}},\ and\ \bibinfo {author} {\bibfnamefont {H.}~\bibnamefont {Zhang}},\ }\bibfield  {title} {\bibinfo {title} {{In-situ analysis of the water distribution and protein structure of dough during ultrasonic-assisted freezing based on miniature Raman spectroscopy}},\ }\href {https://doi.org/10.1016/j.ultsonch.2020.105149} {\bibfield  {journal} {\bibinfo  {journal} {Ultrason. Sonochem.}\ }\textbf {\bibinfo {volume} {67}},\ \bibinfo {pages} {105149} (\bibinfo {year} {2020})},\ \bibinfo {note} {{\href{https://pubmed.ncbi.nlm.nih.gov/32371350/}{PMID: 32371350}}}\BibitemShut {NoStop}%
\bibitem [{\citenamefont {Welch}(1938)}]{Welch1938}%
  \BibitemOpen
  \bibfield  {author} {\bibinfo {author} {\bibfnamefont {B.~L.}\ \bibnamefont {Welch}},\ }\bibfield  {title} {\bibinfo {title} {{{The Significance of the Difference Between Two Means when the Population Variances are Unequal}}},\ }\href {https://doi.org/10.1093/biomet/29.3-4.350} {\bibfield  {journal} {\bibinfo  {journal} {Biometrika}\ }\textbf {\bibinfo {volume} {29}},\ \bibinfo {pages} {350} (\bibinfo {year} {1938})}\BibitemShut {NoStop}%
\bibitem [{\citenamefont {Welch}(1947)}]{Welch1947}%
  \BibitemOpen
  \bibfield  {author} {\bibinfo {author} {\bibfnamefont {B.~L.}\ \bibnamefont {Welch}},\ }\bibfield  {title} {\bibinfo {title} {{{The Generalization of `Student's' Problem When Several Different Population Variances Are Involved}}},\ }\href {https://doi.org/10.1093/biomet/34.1-2.28} {\bibfield  {journal} {\bibinfo  {journal} {Biometrika}\ }\textbf {\bibinfo {volume} {34}},\ \bibinfo {pages} {28} (\bibinfo {year} {1947})},\ \bibinfo {note} {{\href{https://pubmed.ncbi.nlm.nih.gov/20287819/}{PMID: 20287819}}}\BibitemShut {NoStop}%
\bibitem [{Ram()}]{RamanSuppMat}%
  \BibitemOpen
  \href@noop {} {}\bibinfo {note} {See Supplemental Material at \href{https://doi.org/10.6084/m9.figshare.24499825}{10.6084/m9.figshare.24499825} for a comprehensive presentation of our statistical hypothesis test results, measured amide I Raman spectra, and their corresponding spectral decompositions.}\BibitemShut {Stop}%
\bibitem [{\citenamefont {Ventilla}\ \emph {et~al.}(1972)\citenamefont {Ventilla}, \citenamefont {Cantor},\ and\ \citenamefont {Shelanski}}]{Ventilla1972}%
  \BibitemOpen
  \bibfield  {author} {\bibinfo {author} {\bibfnamefont {M.}~\bibnamefont {Ventilla}}, \bibinfo {author} {\bibfnamefont {C.~R.}\ \bibnamefont {Cantor}},\ and\ \bibinfo {author} {\bibfnamefont {M.}~\bibnamefont {Shelanski}},\ }\bibfield  {title} {\bibinfo {title} {{{A Circular Dichroism Study of Microtubule Protein}}},\ }\href {https://doi.org/10.1021/bi00759a002} {\bibfield  {journal} {\bibinfo  {journal} {Biochemistry}\ }\textbf {\bibinfo {volume} {11}},\ \bibinfo {pages} {1554} (\bibinfo {year} {1972})},\ \bibinfo {note} {{\href{https://pubmed.ncbi.nlm.nih.gov/5028102/}{PMID: 5028102}}}\BibitemShut {NoStop}%
\bibitem [{\citenamefont {{de Pereda}}\ \emph {et~al.}(1996)\citenamefont {{de Pereda}}, \citenamefont {Leynadier}, \citenamefont {Evangelio}, \citenamefont {Chac{\'{o}}n},\ and\ \citenamefont {Andreu}}]{de_Pereda_1996}%
  \BibitemOpen
  \bibfield  {author} {\bibinfo {author} {\bibfnamefont {J.~M.}\ \bibnamefont {{de Pereda}}}, \bibinfo {author} {\bibfnamefont {D.}~\bibnamefont {Leynadier}}, \bibinfo {author} {\bibfnamefont {J.~A.}\ \bibnamefont {Evangelio}}, \bibinfo {author} {\bibfnamefont {P.}~\bibnamefont {Chac{\'{o}}n}},\ and\ \bibinfo {author} {\bibfnamefont {J.~M.}\ \bibnamefont {Andreu}},\ }\bibfield  {title} {\bibinfo {title} {{Tubulin Secondary Structure Analysis, Limited Proteolysis Sites, and Homology to FtsZ}},\ }\href {https://doi.org/10.1021/bi961357b} {\bibfield  {journal} {\bibinfo  {journal} {Biochemistry}\ }\textbf {\bibinfo {volume} {35}},\ \bibinfo {pages} {14203} (\bibinfo {year} {1996})},\ \bibinfo {note} {{\href{https://pubmed.ncbi.nlm.nih.gov/8916905/}{PMID: 8916905}}}\BibitemShut {NoStop}%
\bibitem [{\citenamefont {Afrasiabi}\ \emph {et~al.}(2013)\citenamefont {Afrasiabi}, \citenamefont {Riazi}, \citenamefont {Dadras}, \citenamefont {Tavili}, \citenamefont {Ghalandari}, \citenamefont {Naghshineh}, \citenamefont {Mobasheri},\ and\ \citenamefont {Ahmadian}}]{Afrasiabi2013}%
  \BibitemOpen
  \bibfield  {author} {\bibinfo {author} {\bibfnamefont {A.}~\bibnamefont {Afrasiabi}}, \bibinfo {author} {\bibfnamefont {G.~H.}\ \bibnamefont {Riazi}}, \bibinfo {author} {\bibfnamefont {A.}~\bibnamefont {Dadras}}, \bibinfo {author} {\bibfnamefont {E.}~\bibnamefont {Tavili}}, \bibinfo {author} {\bibfnamefont {B.}~\bibnamefont {Ghalandari}}, \bibinfo {author} {\bibfnamefont {A.}~\bibnamefont {Naghshineh}}, \bibinfo {author} {\bibfnamefont {H.}~\bibnamefont {Mobasheri}},\ and\ \bibinfo {author} {\bibfnamefont {S.}~\bibnamefont {Ahmadian}},\ }\bibfield  {title} {\bibinfo {title} {{Electromagnetic fields with 217~Hz and 0.2~mT as hazardous factors for tubulin structure and assembly (in vitro study)}},\ }\href {https://doi.org/10.1007/s13738-013-0398-y} {\bibfield  {journal} {\bibinfo  {journal} {J. Iran. Chem. Soc.}\ }\textbf {\bibinfo {volume} {11}},\ \bibinfo {pages} {1295} (\bibinfo {year} {2013})}\BibitemShut {NoStop}%
\bibitem [{\citenamefont {Ackbarow}\ \emph {et~al.}(2007)\citenamefont {Ackbarow}, \citenamefont {Chen}, \citenamefont {Keten},\ and\ \citenamefont {Buehler}}]{Ackbarow2007}%
  \BibitemOpen
  \bibfield  {author} {\bibinfo {author} {\bibfnamefont {T.}~\bibnamefont {Ackbarow}}, \bibinfo {author} {\bibfnamefont {X.}~\bibnamefont {Chen}}, \bibinfo {author} {\bibfnamefont {S.}~\bibnamefont {Keten}},\ and\ \bibinfo {author} {\bibfnamefont {M.~J.}\ \bibnamefont {Buehler}},\ }\bibfield  {title} {\bibinfo {title} {Hierarchies, multiple energy barriers, and robustness govern the fracture mechanics of \textalpha-helical and \textbeta-sheet protein domains},\ }\href {https://doi.org/10.1073/pnas.0705759104} {\bibfield  {journal} {\bibinfo  {journal} {Proc. Natl. Acad. Sci. U.S.A.}\ }\textbf {\bibinfo {volume} {104}},\ \bibinfo {pages} {16410} (\bibinfo {year} {2007})},\ \bibinfo {note} {{\href{https://pubmed.ncbi.nlm.nih.gov/17925444/}{PMID: 17925444}}}\BibitemShut {NoStop}%
\bibitem [{\citenamefont {Perillo}\ \emph {et~al.}(2020)\citenamefont {Perillo}, \citenamefont {d’Apuzzo}, \citenamefont {Illario}, \citenamefont {Laino}, \citenamefont {{Di Spigna}}, \citenamefont {Lepore},\ and\ \citenamefont {Camerlingo}}]{s20020497}%
  \BibitemOpen
  \bibfield  {author} {\bibinfo {author} {\bibfnamefont {L.}~\bibnamefont {Perillo}}, \bibinfo {author} {\bibfnamefont {F.}~\bibnamefont {d’Apuzzo}}, \bibinfo {author} {\bibfnamefont {M.}~\bibnamefont {Illario}}, \bibinfo {author} {\bibfnamefont {L.}~\bibnamefont {Laino}}, \bibinfo {author} {\bibfnamefont {G.}~\bibnamefont {{Di Spigna}}}, \bibinfo {author} {\bibfnamefont {M.}~\bibnamefont {Lepore}},\ and\ \bibinfo {author} {\bibfnamefont {C.}~\bibnamefont {Camerlingo}},\ }\bibfield  {title} {\bibinfo {title} {{Monitoring Biochemical and Structural Changes in Human Periodontal Ligaments during Orthodontic Treatment by Means of Micro-Raman Spectroscopy}},\ }\href {https://doi.org/10.3390/s20020497} {\bibfield  {journal} {\bibinfo  {journal} {Sensors}\ }\textbf {\bibinfo {volume} {20}},\ \bibinfo {pages} {497} (\bibinfo {year} {2020})},\ \bibinfo {note} {{\href{https://pubmed.ncbi.nlm.nih.gov/31952367/}{PMID: 31952367}}}\BibitemShut {NoStop}%
\bibitem [{\citenamefont {Camerlingo}\ \emph {et~al.}(2014)\citenamefont {Camerlingo}, \citenamefont {d’Apuzzo}, \citenamefont {Grassia}, \citenamefont {Perillo},\ and\ \citenamefont {Lepore}}]{Camerlingo2014}%
  \BibitemOpen
  \bibfield  {author} {\bibinfo {author} {\bibfnamefont {C.}~\bibnamefont {Camerlingo}}, \bibinfo {author} {\bibfnamefont {F.}~\bibnamefont {d’Apuzzo}}, \bibinfo {author} {\bibfnamefont {V.}~\bibnamefont {Grassia}}, \bibinfo {author} {\bibfnamefont {L.}~\bibnamefont {Perillo}},\ and\ \bibinfo {author} {\bibfnamefont {M.}~\bibnamefont {Lepore}},\ }\bibfield  {title} {\bibinfo {title} {{Micro-Raman Spectroscopy for Monitoring Changes in Periodontal Ligaments and Gingival Crevicular Fluid}},\ }\href {https://doi.org/10.3390/s141222552} {\bibfield  {journal} {\bibinfo  {journal} {Sensors}\ }\textbf {\bibinfo {volume} {14}},\ \bibinfo {pages} {22552} (\bibinfo {year} {2014})},\ \bibinfo {note} {{\href{https://pubmed.ncbi.nlm.nih.gov/25436655/}{PMID: 25436655}}}\BibitemShut {NoStop}%
\bibitem [{\citenamefont {Paquin}\ and\ \citenamefont {Colomban}(2007)}]{Paquin2007}%
  \BibitemOpen
  \bibfield  {author} {\bibinfo {author} {\bibfnamefont {R.}~\bibnamefont {Paquin}}\ and\ \bibinfo {author} {\bibfnamefont {P.}~\bibnamefont {Colomban}},\ }\bibfield  {title} {\bibinfo {title} {{Nanomechanics of single keratin fibres: A Raman study of the \textalpha-helix →\textbeta-sheet transition and the effect of water}},\ }\href {https://doi.org/10.1002/jrs.1672} {\bibfield  {journal} {\bibinfo  {journal} {J. Raman Spectrosc.}\ }\textbf {\bibinfo {volume} {38}},\ \bibinfo {pages} {504} (\bibinfo {year} {2007})}\BibitemShut {NoStop}%
\bibitem [{\citenamefont {Beyer}\ \emph {et~al.}(2008)\citenamefont {Beyer}, \citenamefont {Jelezarov},\ and\ \citenamefont {Frohlich}}]{Beyer2008_2}%
  \BibitemOpen
  \bibfield  {author} {\bibinfo {author} {\bibfnamefont {C.}~\bibnamefont {Beyer}}, \bibinfo {author} {\bibfnamefont {I.}~\bibnamefont {Jelezarov}},\ and\ \bibinfo {author} {\bibfnamefont {J.}~\bibnamefont {Frohlich}},\ }\bibfield  {title} {\bibinfo {title} {{Real-time observation of potential conformational changes of proteins during electromagnetic field exposure}},\ }in\ \href {https://doi.org/10.1109/IEMBS.2008.4649309} {\emph {\bibinfo {booktitle} {{2008 30th Annual International Conference of the IEEE Engineering in Medicine and Biology Society}}}}\ (\bibinfo {year} {2008})\ pp.\ \bibinfo {pages} {939--942},\ \bibinfo {note} {{\href{https://pubmed.ncbi.nlm.nih.gov/19162812/}{PMID: 19162812}}}\BibitemShut {NoStop}%
\bibitem [{\citenamefont {Bekard}\ and\ \citenamefont {Dunstan}(2014)}]{Bekard2014}%
  \BibitemOpen
  \bibfield  {author} {\bibinfo {author} {\bibfnamefont {I.}~\bibnamefont {Bekard}}\ and\ \bibinfo {author} {\bibfnamefont {D.~E.}\ \bibnamefont {Dunstan}},\ }\bibfield  {title} {\bibinfo {title} {{Electric field induced changes in protein conformation}},\ }\href {https://doi.org/10.1039/C3SM52653D} {\bibfield  {journal} {\bibinfo  {journal} {Soft Matter}\ }\textbf {\bibinfo {volume} {10}},\ \bibinfo {pages} {431} (\bibinfo {year} {2014})},\ \bibinfo {note} {{\href{https://pubmed.ncbi.nlm.nih.gov/24652412/}{PMID: 24652412}}}\BibitemShut {NoStop}%
\bibitem [{\citenamefont {Lundholm}\ \emph {et~al.}(2015)\citenamefont {Lundholm}, \citenamefont {Rodilla}, \citenamefont {Wahlgren}, \citenamefont {Duelli}, \citenamefont {Bourenkov}, \citenamefont {Vukusic}, \citenamefont {Friedman}, \citenamefont {Stake}, \citenamefont {Schneider},\ and\ \citenamefont {Katona}}]{Lundholm2015}%
  \BibitemOpen
  \bibfield  {author} {\bibinfo {author} {\bibfnamefont {I.~V.}\ \bibnamefont {Lundholm}}, \bibinfo {author} {\bibfnamefont {H.}~\bibnamefont {Rodilla}}, \bibinfo {author} {\bibfnamefont {W.~Y.}\ \bibnamefont {Wahlgren}}, \bibinfo {author} {\bibfnamefont {A.}~\bibnamefont {Duelli}}, \bibinfo {author} {\bibfnamefont {G.}~\bibnamefont {Bourenkov}}, \bibinfo {author} {\bibfnamefont {J.}~\bibnamefont {Vukusic}}, \bibinfo {author} {\bibfnamefont {R.}~\bibnamefont {Friedman}}, \bibinfo {author} {\bibfnamefont {J.}~\bibnamefont {Stake}}, \bibinfo {author} {\bibfnamefont {T.}~\bibnamefont {Schneider}},\ and\ \bibinfo {author} {\bibfnamefont {G.}~\bibnamefont {Katona}},\ }\bibfield  {title} {\bibinfo {title} {{Terahertz radiation induces non-thermal structural changes associated with Fr\"{o}hlich condensation in a protein crystal}},\ }\href {https://doi.org/10.1063/1.4931825} {\bibfield  {journal} {\bibinfo  {journal} {Struct. Dyn.}\ }\textbf {\bibinfo {volume} {2}},\ \bibinfo {pages} {054702} (\bibinfo {year}
  {2015})},\ \bibinfo {note} {{\href{https://pubmed.ncbi.nlm.nih.gov/26798828/}{PMID: 26798828}}}\BibitemShut {NoStop}%
\bibitem [{\citenamefont {Maugeri}\ \emph {et~al.}(2018)\citenamefont {Maugeri}, \citenamefont {Griese}, \citenamefont {Branca}, \citenamefont {Miller}, \citenamefont {Smith}, \citenamefont {Eirich}, \citenamefont {H\"{o}gbom},\ and\ \citenamefont {Shafaat}}]{Maugeri2018}%
  \BibitemOpen
  \bibfield  {author} {\bibinfo {author} {\bibfnamefont {P.~T.}\ \bibnamefont {Maugeri}}, \bibinfo {author} {\bibfnamefont {J.~J.}\ \bibnamefont {Griese}}, \bibinfo {author} {\bibfnamefont {R.~M.}\ \bibnamefont {Branca}}, \bibinfo {author} {\bibfnamefont {E.~K.}\ \bibnamefont {Miller}}, \bibinfo {author} {\bibfnamefont {Z.~R.}\ \bibnamefont {Smith}}, \bibinfo {author} {\bibfnamefont {J.}~\bibnamefont {Eirich}}, \bibinfo {author} {\bibfnamefont {M.}~\bibnamefont {H\"{o}gbom}},\ and\ \bibinfo {author} {\bibfnamefont {H.~S.}\ \bibnamefont {Shafaat}},\ }\bibfield  {title} {\bibinfo {title} {{Driving Protein Conformational Changes with Light: Photoinduced Structural Rearrangement in a Heterobimetallic Oxidase}},\ }\href {https://doi.org/10.1021/jacs.7b11966} {\bibfield  {journal} {\bibinfo  {journal} {J. Am. Chem. Soc.}\ }\textbf {\bibinfo {volume} {140}},\ \bibinfo {pages} {1471} (\bibinfo {year} {2018})},\ \bibinfo {note} {{\href{https://pubmed.ncbi.nlm.nih.gov/29268610/}{PMID: 29268610}}}\BibitemShut {NoStop}%
\bibitem [{\citenamefont {Darwish}\ \emph {et~al.}(2020)\citenamefont {Darwish}, \citenamefont {Darwish},\ and\ \citenamefont {Darwish}}]{Darwish2020}%
  \BibitemOpen
  \bibfield  {author} {\bibinfo {author} {\bibfnamefont {S.~M.}\ \bibnamefont {Darwish}}, \bibinfo {author} {\bibfnamefont {A.~S.}\ \bibnamefont {Darwish}},\ and\ \bibinfo {author} {\bibfnamefont {D.~S.}\ \bibnamefont {Darwish}},\ }\bibfield  {title} {\bibinfo {title} {{An extremely low-frequency magnetic field can affect CREB protein conformation which may have a role in neuronal activities including memory}},\ }\href {https://doi.org/10.1088/2399-6528/ab66d2} {\bibfield  {journal} {\bibinfo  {journal} {J. Phys. Commun.}\ }\textbf {\bibinfo {volume} {4}},\ \bibinfo {pages} {015009} (\bibinfo {year} {2020})}\BibitemShut {NoStop}%
\bibitem [{\citenamefont {Zomorrodi}\ \emph {et~al.}(2021)\citenamefont {Zomorrodi}, \citenamefont {Karimpoor}, \citenamefont {Smith}, \citenamefont {Liburd}, \citenamefont {Loheswaran}, \citenamefont {Rashidi-Ranjbar}, \citenamefont {Hosseinkhah},\ and\ \citenamefont {Lim}}]{Zomorrodi2021}%
  \BibitemOpen
  \bibfield  {author} {\bibinfo {author} {\bibfnamefont {R.}~\bibnamefont {Zomorrodi}}, \bibinfo {author} {\bibfnamefont {M.}~\bibnamefont {Karimpoor}}, \bibinfo {author} {\bibfnamefont {A.}~\bibnamefont {Smith}}, \bibinfo {author} {\bibfnamefont {J.}~\bibnamefont {Liburd}}, \bibinfo {author} {\bibfnamefont {G.}~\bibnamefont {Loheswaran}}, \bibinfo {author} {\bibfnamefont {N.}~\bibnamefont {Rashidi-Ranjbar}}, \bibinfo {author} {\bibfnamefont {N.}~\bibnamefont {Hosseinkhah}},\ and\ \bibinfo {author} {\bibfnamefont {L.}~\bibnamefont {Lim}},\ }\bibfield  {title} {\bibinfo {title} {{Modulation of cortical oscillations using 10hz near-infrared transcranial and intranasal photobiomodulation: a randomized sham-controlled crossover study}},\ }\href {https://doi.org/10.1016/j.brs.2021.10.245} {\bibfield  {journal} {\bibinfo  {journal} {Brain Stimul.}\ }\textbf {\bibinfo {volume} {14}},\ \bibinfo {pages} {1665} (\bibinfo {year} {2021})}\BibitemShut {NoStop}%
\bibitem [{\citenamefont {Cooper}(2000)}]{Cooper2000}%
  \BibitemOpen
  \bibfield  {author} {\bibinfo {author} {\bibfnamefont {G.~M.}\ \bibnamefont {Cooper}},\ }\href {https://www.ncbi.nlm.nih.gov/books/NBK9932/} {\emph {\bibinfo {title} {{The Cell: A Molecular Approach}}}},\ \bibinfo {edition} {2nd}\ ed.\ (\bibinfo  {publisher} {Sunderland (MA): Sinauer Associates},\ \bibinfo {year} {2000})\ Chap.\ \bibinfo {chapter} {Microtubules}\BibitemShut {NoStop}%
\bibitem [{\citenamefont {Muhia}\ \emph {et~al.}(2016)\citenamefont {Muhia}, \citenamefont {Thies}, \citenamefont {Labont\'e}, \citenamefont {Ghiretti}, \citenamefont {Gromova}, \citenamefont {Xompero}, \citenamefont {Lappe-Siefke}, \citenamefont {Hermans-Borgmeyer}, \citenamefont {Kuhl}, \citenamefont {Schweizer},\ and\ \citenamefont {{et al.}}}]{MUHIA2016}%
  \BibitemOpen
  \bibfield  {author} {\bibinfo {author} {\bibfnamefont {M.}~\bibnamefont {Muhia}}, \bibinfo {author} {\bibfnamefont {E.}~\bibnamefont {Thies}}, \bibinfo {author} {\bibfnamefont {D.}~\bibnamefont {Labont\'e}}, \bibinfo {author} {\bibfnamefont {A.~E.}\ \bibnamefont {Ghiretti}}, \bibinfo {author} {\bibfnamefont {K.~V.}\ \bibnamefont {Gromova}}, \bibinfo {author} {\bibfnamefont {F.}~\bibnamefont {Xompero}}, \bibinfo {author} {\bibfnamefont {C.}~\bibnamefont {Lappe-Siefke}}, \bibinfo {author} {\bibfnamefont {I.}~\bibnamefont {Hermans-Borgmeyer}}, \bibinfo {author} {\bibfnamefont {D.}~\bibnamefont {Kuhl}}, \bibinfo {author} {\bibfnamefont {M.}~\bibnamefont {Schweizer}},\ and\ \bibinfo {author} {\bibnamefont {{et al.}}},\ }\bibfield  {title} {\bibinfo {title} {{The Kinesin KIF21B Regulates Microtubule Dynamics and Is Essential for Neuronal Morphology, Synapse Function, and Learning and Memory}},\ }\href {https://doi.org/10.1016/j.celrep.2016.03.086} {\bibfield  {journal} {\bibinfo  {journal} {Cell Rep.}\ }\textbf
  {\bibinfo {volume} {15}},\ \bibinfo {pages} {968} (\bibinfo {year} {2016})},\ \bibinfo {note} {{\href{https://pubmed.ncbi.nlm.nih.gov/27117409/}{PMID: 27117409}}}\BibitemShut {NoStop}%
\bibitem [{\citenamefont {Ashraf}\ \emph {et~al.}(2014)\citenamefont {Ashraf}, \citenamefont {Greig}, \citenamefont {Khan}, \citenamefont {Hassan}, \citenamefont {Tabrez}, \citenamefont {Shakil}, \citenamefont {Sheikh}, \citenamefont {Zaidi}, \citenamefont {Akram}, \citenamefont {Jabir},\ and\ \citenamefont {{et al.}}}]{Ashraf2014}%
  \BibitemOpen
  \bibfield  {author} {\bibinfo {author} {\bibfnamefont {G.~M.}\ \bibnamefont {Ashraf}}, \bibinfo {author} {\bibfnamefont {N.~H.}\ \bibnamefont {Greig}}, \bibinfo {author} {\bibfnamefont {T.~A.}\ \bibnamefont {Khan}}, \bibinfo {author} {\bibfnamefont {I.}~\bibnamefont {Hassan}}, \bibinfo {author} {\bibfnamefont {S.}~\bibnamefont {Tabrez}}, \bibinfo {author} {\bibfnamefont {S.}~\bibnamefont {Shakil}}, \bibinfo {author} {\bibfnamefont {I.~A.}\ \bibnamefont {Sheikh}}, \bibinfo {author} {\bibfnamefont {S.~K.}\ \bibnamefont {Zaidi}}, \bibinfo {author} {\bibfnamefont {M.}~\bibnamefont {Akram}}, \bibinfo {author} {\bibfnamefont {N.~R.}\ \bibnamefont {Jabir}},\ and\ \bibinfo {author} {\bibnamefont {{et al.}}},\ }\bibfield  {title} {\bibinfo {title} {{Protein Misfolding and Aggregation in Alzheimer's Disease and Type 2 Diabetes Mellitus}},\ }\href {https://doi.org/10.2174/1871527313666140917095514} {\bibfield  {journal} {\bibinfo  {journal} {CNS Neurol. Disord. Drug Targets}\ }\textbf {\bibinfo {volume} {13}},\
  \bibinfo {pages} {1280} (\bibinfo {year} {2014})},\ \bibinfo {note} {{\href{https://pubmed.ncbi.nlm.nih.gov/25230234/}{PMID: 25230234}}}\BibitemShut {NoStop}%
\bibitem [{\citenamefont {Koster}(2022)}]{Koster2022}%
  \BibitemOpen
  \bibfield  {author} {\bibinfo {author} {\bibfnamefont {P.~M.}\ \bibnamefont {Koster}},\ }\emph {\bibinfo {title} {{{Near Infrared Light Penetration in Human Tissue: An Analysis of Tissue Structure and Heterogeneities}}}},\ \href {https://epublications.marquette.edu/theses_open/739} {Master's thesis},\ \bibinfo  {school} {Marquette University} (\bibinfo {year} {2022})\BibitemShut {NoStop}%
\bibitem [{\citenamefont {Iqbal}\ \emph {et~al.}(2005)\citenamefont {Iqbal}, \citenamefont {{del C. Alonso}}, \citenamefont {Chen}, \citenamefont {Chohan}, \citenamefont {El-Akkad}, \citenamefont {Gong}, \citenamefont {Khatoon}, \citenamefont {Li}, \citenamefont {Liu}, \citenamefont {Rahman},\ and\ \citenamefont {{et al.}}}]{IQBAL2005}%
  \BibitemOpen
  \bibfield  {author} {\bibinfo {author} {\bibfnamefont {K.}~\bibnamefont {Iqbal}}, \bibinfo {author} {\bibfnamefont {A.}~\bibnamefont {{del C. Alonso}}}, \bibinfo {author} {\bibfnamefont {S.}~\bibnamefont {Chen}}, \bibinfo {author} {\bibfnamefont {M.~O.}\ \bibnamefont {Chohan}}, \bibinfo {author} {\bibfnamefont {E.}~\bibnamefont {El-Akkad}}, \bibinfo {author} {\bibfnamefont {C.-X.}\ \bibnamefont {Gong}}, \bibinfo {author} {\bibfnamefont {S.}~\bibnamefont {Khatoon}}, \bibinfo {author} {\bibfnamefont {B.}~\bibnamefont {Li}}, \bibinfo {author} {\bibfnamefont {F.}~\bibnamefont {Liu}}, \bibinfo {author} {\bibfnamefont {A.}~\bibnamefont {Rahman}},\ and\ \bibinfo {author} {\bibnamefont {{et al.}}},\ }\bibfield  {title} {\bibinfo {title} {{Tau pathology in Alzheimer disease and other tauopathies}},\ }\href {https://doi.org/10.1016/j.bbadis.2004.09.008} {\bibfield  {journal} {\bibinfo  {journal} {Biochim. Biophys. Acta - Mol. Basis Dis.}\ }\textbf {\bibinfo {volume} {1739}},\ \bibinfo {pages} {198} (\bibinfo {year}
  {2005})},\ \bibinfo {note} {{Part of special issue: The Biology and Pathobiology of Tau, \href{https://pubmed.ncbi.nlm.nih.gov/15615638/}{PMID: 15615638}}}\BibitemShut {NoStop}%
\bibitem [{\citenamefont {Iqbal}\ \emph {et~al.}(2010)\citenamefont {Iqbal}, \citenamefont {Liu}, \citenamefont {Gong},\ and\ \citenamefont {Grundke-Iqbal}}]{Iqbal2010}%
  \BibitemOpen
  \bibfield  {author} {\bibinfo {author} {\bibfnamefont {K.}~\bibnamefont {Iqbal}}, \bibinfo {author} {\bibfnamefont {F.}~\bibnamefont {Liu}}, \bibinfo {author} {\bibfnamefont {C.-X.}\ \bibnamefont {Gong}},\ and\ \bibinfo {author} {\bibfnamefont {I.}~\bibnamefont {Grundke-Iqbal}},\ }\bibfield  {title} {\bibinfo {title} {{Tau in Alzheimer Disease and Related Tauopathies}},\ }\href {https://doi.org/10.2174/156720510793611592} {\bibfield  {journal} {\bibinfo  {journal} {Curr. Alzheimer Res.}\ }\textbf {\bibinfo {volume} {7}},\ \bibinfo {pages} {656} (\bibinfo {year} {2010})},\ \bibinfo {note} {{\href{https://pubmed.ncbi.nlm.nih.gov/20678074/}{PMID: 20678074}}}\BibitemShut {NoStop}%
\bibitem [{\citenamefont {Lyros}\ \emph {et~al.}(2020)\citenamefont {Lyros}, \citenamefont {Ragoschke-Schumm}, \citenamefont {Kostopoulos}, \citenamefont {Sehr}, \citenamefont {Backens}, \citenamefont {Kalampokini}, \citenamefont {Decker}, \citenamefont {Lesmeister}, \citenamefont {Liu}, \citenamefont {Reith},\ and\ \citenamefont {{et al.}}}]{LYROS2020}%
  \BibitemOpen
  \bibfield  {author} {\bibinfo {author} {\bibfnamefont {E.}~\bibnamefont {Lyros}}, \bibinfo {author} {\bibfnamefont {A.}~\bibnamefont {Ragoschke-Schumm}}, \bibinfo {author} {\bibfnamefont {P.}~\bibnamefont {Kostopoulos}}, \bibinfo {author} {\bibfnamefont {A.}~\bibnamefont {Sehr}}, \bibinfo {author} {\bibfnamefont {M.}~\bibnamefont {Backens}}, \bibinfo {author} {\bibfnamefont {S.}~\bibnamefont {Kalampokini}}, \bibinfo {author} {\bibfnamefont {Y.}~\bibnamefont {Decker}}, \bibinfo {author} {\bibfnamefont {M.}~\bibnamefont {Lesmeister}}, \bibinfo {author} {\bibfnamefont {Y.}~\bibnamefont {Liu}}, \bibinfo {author} {\bibfnamefont {W.}~\bibnamefont {Reith}},\ and\ \bibinfo {author} {\bibnamefont {{et al.}}},\ }\bibfield  {title} {\bibinfo {title} {{Normal brain aging and Alzheimer's disease are associated with lower cerebral pH: an in vivo histidine $^{1}$H-MR spectroscopy study}},\ }\href {https://doi.org/10.1016/j.neurobiolaging.2019.11.012} {\bibfield  {journal} {\bibinfo  {journal} {Neurobiol. Aging}\ }\textbf
  {\bibinfo {volume} {87}},\ \bibinfo {pages} {60} (\bibinfo {year} {2020})},\ \bibinfo {note} {{\href{https://pubmed.ncbi.nlm.nih.gov/31902521/}{PMID: 31902521}}}\BibitemShut {NoStop}%
\bibitem [{\citenamefont {Decker}\ \emph {et~al.}(2021)\citenamefont {Decker}, \citenamefont {N\'emeth}, \citenamefont {Schomburg}, \citenamefont {Chemla}, \citenamefont {F{\"u}l{\"o}p}, \citenamefont {Menger}, \citenamefont {Liu},\ and\ \citenamefont {Fassbender}}]{DECKER2021}%
  \BibitemOpen
  \bibfield  {author} {\bibinfo {author} {\bibfnamefont {Y.}~\bibnamefont {Decker}}, \bibinfo {author} {\bibfnamefont {E.}~\bibnamefont {N\'emeth}}, \bibinfo {author} {\bibfnamefont {R.}~\bibnamefont {Schomburg}}, \bibinfo {author} {\bibfnamefont {A.}~\bibnamefont {Chemla}}, \bibinfo {author} {\bibfnamefont {L.}~\bibnamefont {F{\"u}l{\"o}p}}, \bibinfo {author} {\bibfnamefont {M.~D.}\ \bibnamefont {Menger}}, \bibinfo {author} {\bibfnamefont {Y.}~\bibnamefont {Liu}},\ and\ \bibinfo {author} {\bibfnamefont {K.}~\bibnamefont {Fassbender}},\ }\bibfield  {title} {\bibinfo {title} {{Decreased pH in the aging brain and Alzheimer's disease}},\ }\href {https://doi.org/10.1016/j.neurobiolaging.2020.12.007} {\bibfield  {journal} {\bibinfo  {journal} {Neurobiol. Aging}\ }\textbf {\bibinfo {volume} {101}},\ \bibinfo {pages} {40} (\bibinfo {year} {2021})},\ \bibinfo {note} {{\href{https://pubmed.ncbi.nlm.nih.gov/33578193/}{PMID: 33578193}}}\BibitemShut {NoStop}%
\bibitem [{\citenamefont {Schwartz}\ \emph {et~al.}(2020)\citenamefont {Schwartz}, \citenamefont {Peres}, \citenamefont {Jolicoeur},\ and\ \citenamefont {{da Veiga Moreira}}}]{Schwartz2020}%
  \BibitemOpen
  \bibfield  {author} {\bibinfo {author} {\bibfnamefont {L.}~\bibnamefont {Schwartz}}, \bibinfo {author} {\bibfnamefont {S.}~\bibnamefont {Peres}}, \bibinfo {author} {\bibfnamefont {M.}~\bibnamefont {Jolicoeur}},\ and\ \bibinfo {author} {\bibfnamefont {J.}~\bibnamefont {{da Veiga Moreira}}},\ }\bibfield  {title} {\bibinfo {title} {{Cancer and Alzheimer’s disease: intracellular pH scales the metabolic disorders}},\ }\href {https://doi.org/10.1007/s10522-020-09888-6} {\bibfield  {journal} {\bibinfo  {journal} {Biogerontology}\ }\textbf {\bibinfo {volume} {21}},\ \bibinfo {pages} {683} (\bibinfo {year} {2020})},\ \bibinfo {note} {{\href{https://pubmed.ncbi.nlm.nih.gov/32617766/}{PMID: 32617766}}}\BibitemShut {NoStop}%
\end{thebibliography}%

\end{document}